\documentclass[oldversion]{aa}  
\usepackage{graphicx}
\usepackage{savesymbol}
\usepackage{amsmath}
\savesymbol{iint}
\usepackage{txfonts}
\restoresymbol{TXF}{iint}
\usepackage{txfonts}
\usepackage{natbib}
\bibpunct{(}{)}{;}{a}{}{,}
\DeclareGraphicsExtensions{.eps}
\graphicspath{{./pub_graphics/}}
\renewcommand{\deg}{\hbox{$^\circ$}}
\newcommand{\cng}[1]{{#1}}
\newcommand{\cngs}[1]{{#1}}
\newcommand{\scc}{ For symbol- and colour coding see the introduction to this section.}
\begin{document}
   \title{Kinematic modelling of disk galaxies}

   \subtitle{I. A new method to fit tilted rings to data cubes}

   \author{G. I. G. J\'ozsa\thanks{\email{gjozsa@astro.uni-bonn.de}}
          \inst{1}
          \and
          F. Kenn\thanks{\email{fkenn@astro.uni-bonn.de}}
          \inst{1}
          \and
          U. Klein\thanks{\email{uklein@astro.uni-bonn.de}}
          \inst{1}
\and
T. A. Oosterloo\thanks{\email{oosterloo@astron.nl}}
\inst{2,3}
          }

   \offprints{G. J\'ozsa}

   \institute{Argelander Institute for Astronomy (AIfA), Univ. Bonn,
              Auf dem H\"ugel 71, D-53121 Bonn\\        
         \and 
              Netherlands Foundation for Research in Astronomy, Postbus 2, 7990 AA Dwingeloo, The Netherlands\\
\and
Kapteyn Astronomical Institute, Univ. Groningen, Postbus 800, 9700 AV Groningen, The Netherlands}
   \date{Received <date> / Accepted <date>}

 
  \abstract
%
   {This is the first of a series of papers in which the kinematics of disk galaxies over a range of scales is scrutinised employing spectroscopy.

 A fundamental aspect of these studies is
   presented here:  the new publicly available software tool
%
%
   TiRiFiC
   (http://www.astro.uni-bonn.de/{\~{}}gjozsa/tirific.html) enables a
   direct fit of a ``tilted-ring model'' to spectroscopic data cubes. 
%
%
   The algorithm generates model data cubes from the tilted-ring
   parametrisation of a rotating disk, which are automatically
   adjusted to reach an optimum fit  
   via a chi-squared minimisation method to an observed data cube. The structure of the new
   software, the shortcomings of the \cng{previously} available
   programs to produce a tilted-ring model, and the performance of TiRiFiC are discussed.
%
%

   Our method is \cng{less} affected by the well-known problem of
   beam smearing that occurs when fitting to the velocity field. \cng{Since with our method we fit many data points in a data cube simultaneously, TiRiFiC is sensitive to very faint structures and can hence be used to derive tilted-ring models significantly extending \cng{in radius beyond} those derived from a velocity field.}
  The software is able to parametrise {H\,{\small
   I}} disks of galaxies that are intersected by the
   line-of-sight twice or more, i.e. if the disks are heavily warped\cng{, and/or with a significant shift of the projected centre of rotation, and/or if} seen
   edge-on. Furthermore, our method delivers the surface-brightness
   profile of the examined galaxy in addition to the orientational
   parameters and the rotation curve.
%
%

   In order to derive kinematic and morphological models of disk
   galaxies, especially reliable rotation curves, a direct-fit method
   as implemented in our code should be \cng{the tool of choice}.}


   \maketitle %
\section{Introduction} 
\label{Sect_2.1} 

Kinematic analyses based on spectroscopy are an important tool to constrain
the dynamical structure and hence the distribution of matter in
galaxies. \cng{The d}iscovery and description of global features in the
kinematics of galaxies, such as the flatness of rotation curves
\citep[][]{Rubin70,Bosma78}, directly influenced cosmology.  Since then, the increase of computational power enabled
theorists to provide testable predictions on the mass distribution on
sub-galaxy scales for given cosmological models \citep[e.g. 
][]{Navarro97,Moore99,Reed05}, such that kinematical studies can be
utilised as immediate tests for cosmology.  While in recent years
the debate concentrated mainly on the spherical distribution of Dark
Matter in relaxed systems
\citep[e.g.][]{Swaters03,Blok03,Gentile04,Navarro04a}, deviations from this, 
which are evident from lopsidedness \citep[e.g. 
][]{Schoenmakers97} or warping \citep[e.g.  ][]{Bosma78}, now come into
the focus of theoretical research in the cosmological context
\citep[][]{Sharma05,Gao06}.  The aim to constrain the structure of
anisotropies in the DM distribution imposes an observational challenge,
requiring both observations with high sensitivity as well as
appropriate analysis tools. 

This is the first paper in a series \cng{with the} aim \cng{to} \cng{push} forward the
observational limits in order to put further constraints on theory, 
 to gain insight in the large-scale structure of the disk-halo
system in disk galaxies, and \cng{to} further test proposed
mass-density profiles.  One of the major ingredients is the necessary
development of improved analysis methods.  A new software tool to
analyse the kinematics of disk galaxies is presented in this paper. 

It is known from observations that the orbits of the disk material in
spiral galaxies without large bars have a comparably low ellipticity
\citep[e.g.][]{Bosma81b,Schoenmakers97}, such that it is a good
approximation to treat them as being circular.  This means that to first
order the kinematics of a galactic disk at a certain galactocentric
radius can be described by a set of three parameters, viz.  the
rotation velocity, and two parameters that describe the local disk
orientation with respect to some reference system.  Such a ``tilted-ring
model'' was first constructed for M83 by \citet[][]{Rogstad74}. 

In some cases the tilted-ring model is an oversimplification, for
example when the orbits are significantly non-circular due to
the presence of a bar \citep[e.g.][]{Bosma78,Simon03}.  Nevertheless,
\cng{in} many cases it serves as a good approximation and hence is
the standard kinematic model for galaxies.  Several algorithms exist
to fit such a model to spectroscopic data. The
most extensively used is the ROTCUR routine
\citep[][]{Albada85,Begemann87}, which is a generalisation of the
original method of \citet[][]{Warner73}.  It is implemented in several
data analysis packages, e.g.  GIPSY \citep[][]{Hulst92}, NEMO
\citep[][]{Teuben95} and AIPS \citep[][]{Fomalont81}.  ROTCUR fits
a set of inclined rings to a velocity field.  ROTCUR derivatives and extensions exist.  The GIPSY
routine RESWRI takes into account that non-axisymmetric potentials have
a characteristic imprint on the velocity field
\citep[][]{Binney78,Teuben91} and allows for any azimuthal variations of
the rotation velocity \citep[see also][]{Franx94,Schoenmakers97}. 
Similarly, RINGFIT \citep[][]{Simon03} allows for the same variation,
while at the same time only treating a flat disk.  \citet[][]{Simon05}
use\cng{d} this routine to test the reliability of their rotation curves. 
ROTCURSHAPE \citep[implemented in NEMO,][]{Teuben95} performs a global
fit of the velocity field, in which the ring parameters are not fitted
independently, but ``in one go''.  A reduction of the higher order terms
in a harmonic expansion is used by \cng{KINEMETRY} \citep[][]{Krajnovic06} in
order to find the best fitting solution of a tilted-ring model. 

All these routines have in common that they are based on an analysis
of the velocity field, which itself is  derived from a data cube, so
it is an intermediate step when going from the observed data cube to a
tilted ring model.  Various methods exist to extract a velocity field
by analysing  the spectra in a data cube.  All of these
methods have their shortcomings.  First, for galaxies with large warps
or galaxies seen close to edge-on, the derivation of a single
representative velocity field is impossible, because even if the disk
geometry is known, the line-of-sight  intersects the disk twice or
more.  This means that for some positions on the sky more than one
velocity has to be inferred.  Second, a velocity field is contaminated by beam smearing
effects \citep[see below, also][]{Teuben02}.  The integration of
emission along the line-of-sight in a thickened galaxy causes a
similar effect.  This leads to a situation where, in principle, a
tilted-ring model resulting from a fit to a velocity field has to be
cross-checked with a model data cube as can be produced by the
GIPSY routine GALMOD \citep[][ originally designed by T.S.  van
Albada]{Hulst92}. \cng{Some authors have fitted such model data
cubes by adjusting the model parameters and
successively} improving the model data cube by comparing it with the
original data cube ``by eye'' \citep[][]{Arnaboldi93,Swaters99,Gentile04}. 

The drawbacks of just using the velocity field fits led to the
development of a number of software tools that perform direct fits
to the data cube \citep[][]{Irwin91} or to a position-velocity
diagram \citep[][]{Simard99,Takamiya02,Boehm04}, thus circumventing the
beam-smearing or smoothing problem.  These fit routines, which simulate
an observation from a model parametrisation, do not allow,
however, for an intrinsic change of the orientational parameters and
hence are  only limited implementations of the tilted-ring model. 
Especially for the analysis of {H\,{\small I}} kinematics, full 3D
fitting is most suitable, and has first been implemented by
\citet[][]{Corbelli97} and applied to the spiral galaxy
M33.  In this case, however, the fit was performed on single spectra
resulting from a single-dish observation.  Thus, besides not being
available publicly, this algorithm would not suit to perform a fit to a
data cube produced by synthesis or integral-field spectroscopic
observations. 

In this paper, \cng{the} new, publicly available software called
TiRiFiC (``\emph{Ti}lted-\emph{Ri}ng-\emph{Fi}tting-\emph{C}ode'') is
presented, which performs an automated fit of a tilted-ring model to a
data cube.  This software was originally developed to construct
tilted-ring models of heavily warped galaxies observed in the
{H\,{\small I}} emission line.  It can, however, be used for all kinds of
spectroscopic data cubes of translucent objects that can be approximated
by the tilted-ring model.  While the software will be under development
for quite some time in order to improve the performance and to 
extend functionality beyond the classic tilted-ring model,
TiRiFiC is well usable in its current form. 

In this paper, we try to answer two basic issues: does a method to
fit a tilted-ring model directly to the data cube lead to more reliable
results than a fit to the velocity field and is it thus possible to
reach a reasonable fit within a reasonable computational time?
The paper
is laid out as follows: in Sect.~\ref{Sect_2.2} the model layout and
the fitting procedure are introduced, in Sect.~\ref{Sect_2.3} the
smoothing effect of velocity field-based fitting algorithms is
discussed, in Sect.~\ref{Sect_2.4} the results of \cng{a few tests}
of TiRiFiC \cng{are} presented. 
In Sect.~\ref{Sect_2.6}
the results are summarised. First applications of TiRiFiC to purposely selected (warped) galaxies will be published in forthcoming papers.
\section{TiRiFiC layout}
\label{Sect_2.2}
\subsection{Tilted-ring model}
\label{Sect_2.2.1} 
\begin{figure*} \parbox[c]{8.5cm}
{
  \begin{center}
\includegraphics[width=8.5cm]{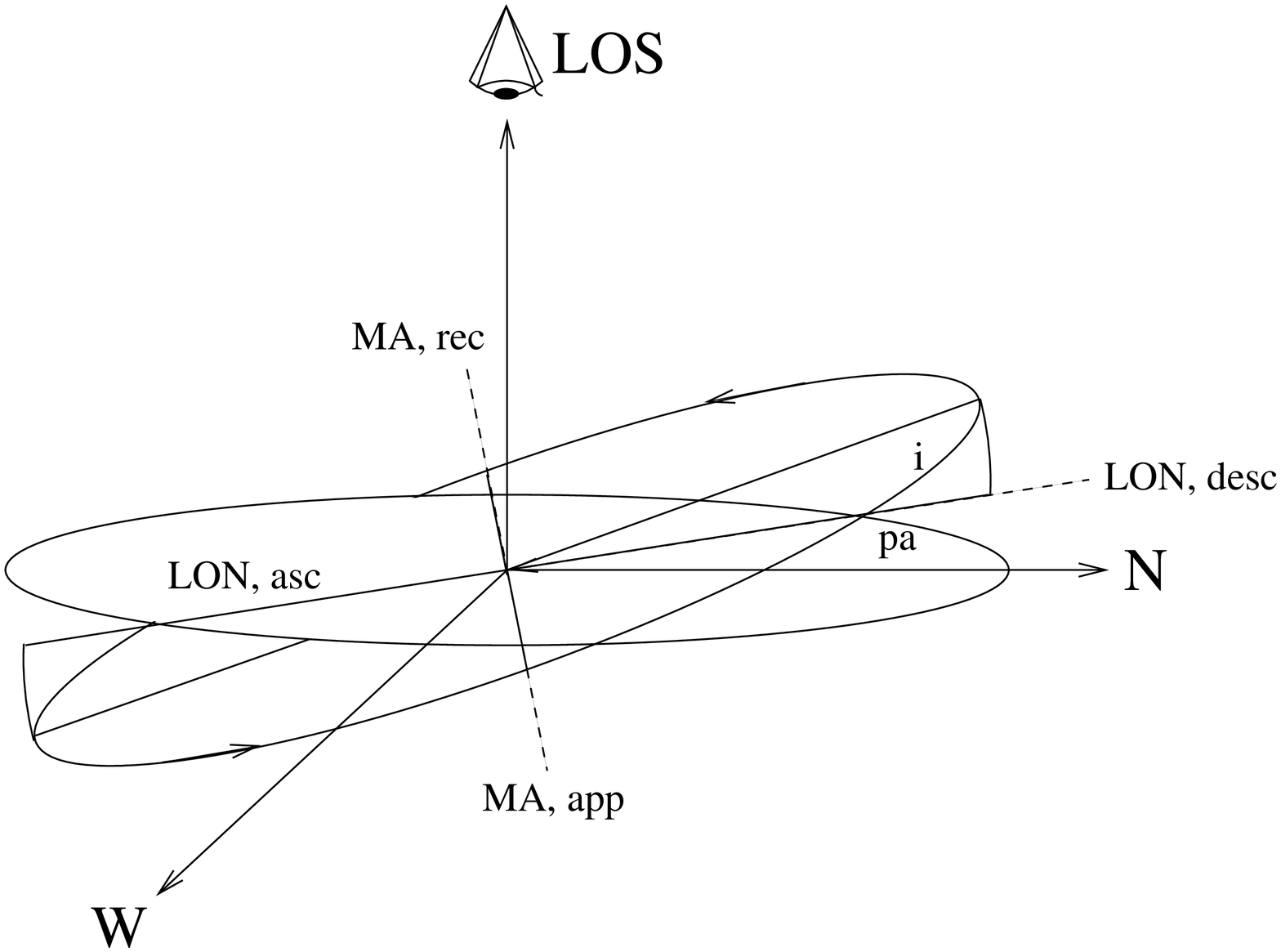}
  \end{center}
}
\parbox[c]{8.5cm} {
{
    \begin{center}
\includegraphics[width=8.5cm]{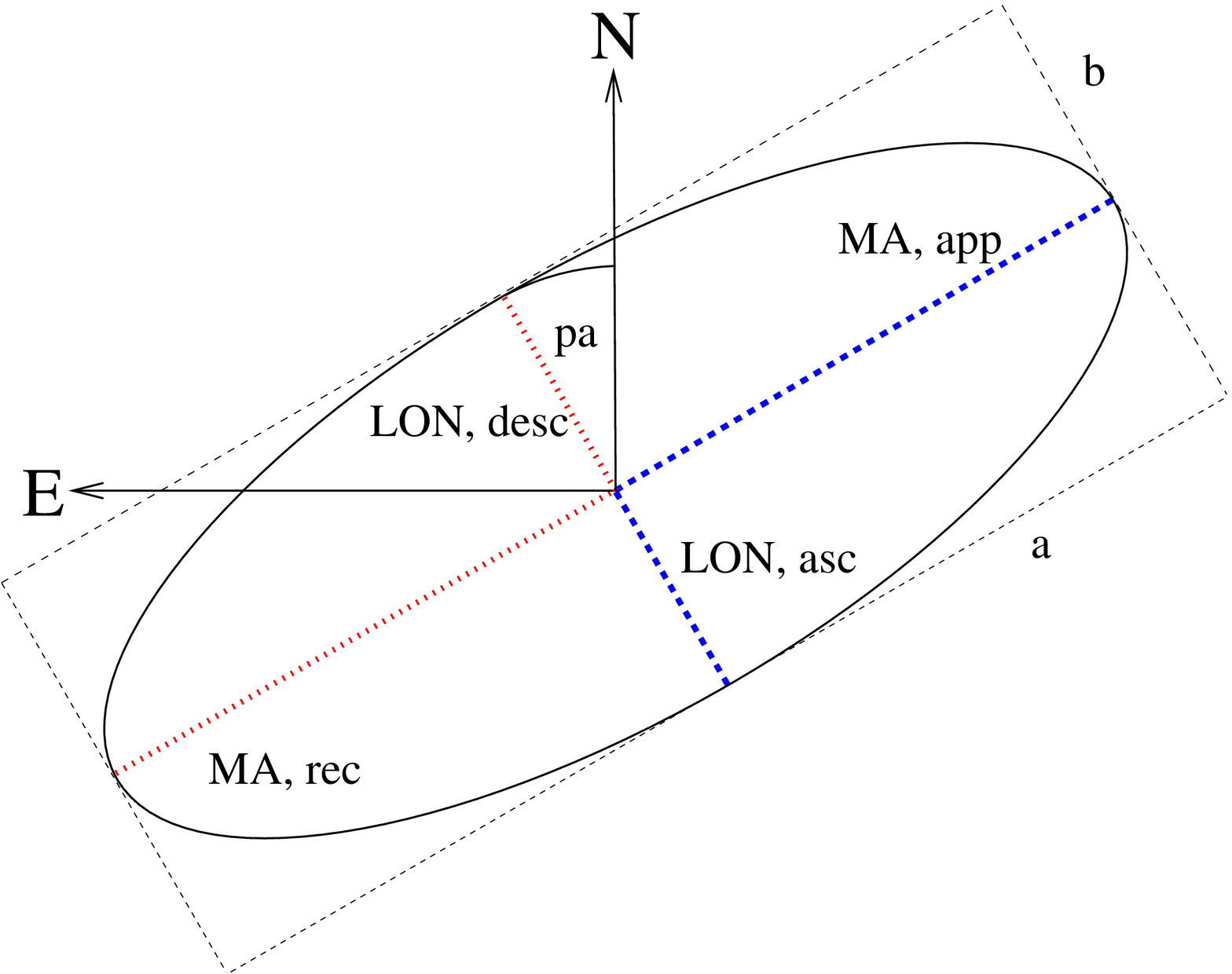}
    \end{center}
}
}
\caption{Definition of the orientational parameters of TiRiFiC, viz. 
inclination $i$ and position angle $pa$.  Left: view of a circular orbit
of the tracer material at an arbitrary position.  A circular orbit
appears to the observer as an ellipse as shown in the right graph.  The
three-dimensional orientation of the orbit circle is parametrised by the
position angle enclosed by the north axis and the ``descending''
line-of-nodes (LON,desc), and the inclination.  The descending
line-of-nodes is the half-minor axis of the projected ellipse defined by
the centre of the ellipse and the point where the galaxy material
switches from having a higher recession velocity (than systemic) to a
lower one, moving anti-clockwise along the ellipse.  With the
inclination being the angle between the celestial plane and the orbital
plane, the orientation of the orbit is fixed.  LOS is the line-of-sight,
MA,app is the approaching half-major axis, MA,rec is the receding
half-major axis, and LON,asc is the ascending line-of-nodes.}
\label{Fig_2.1} \end{figure*}

A TiRiFiC tilted-ring model is specified by a set of parameters
that vary with radius\cng{,} plus a set of global parameters.  We refer to the
set of parameters belonging to a given radius as a "ring" in order to
stick to the traditional terminology.  A model is calculated in a very
similar, but not identical, way as in the GIPSY routine GALMOD.  To
calculate a model, a number of ``sub-rings'' with a user-specified width
is created by linear interpolation of the ring-specific parameters. 
\cng{These} sub-rings are then modelled by a Monte-Carlo integration to rotate
with the same tangential velocity with a certain orientation w.r.t.  the
observer, determined by the position angle and inclination.  The
resulting velocities are then projected onto a cube with dimensions
set by the input data cube.  The orientational parametrisation used by
TiRiFiC is shown in Fig.~\ref{Fig_2.1}.  The final step to obtain a
model representing an observation consists of a convolution with a
3D-Gaussian, representing the instrumental effects of a finite observing
beam and the finite resolution in the frequency
domain, as well as the internal dispersion within the tracer material of
a galaxy.  

The user specifies the number of rings and for each ring (see also
Fig.~\ref{Fig_2.1}):
\begin{itemize}
\item the radius.
\item the circular velocity.
\item the scale height (for the specification of the vertical density distribution see
below).
\item the surface brightness.
\item the inclination.
\item the position angle. Contrary to the usual convention (e.g. GALMOD,
ROTCUR), the position angle is defined as the angle between the meridian
and the minor axis of each ring from north through east (see Fig.~\ref{Fig_2.1}). 
In comparison to GALMOD the TiRiFiC position angle is $90 \deg$ larger. 
 TiRiFiC uses the same  definition as used by \citet[][]{Briggs90}. 
\item the right ascension of the central positions of the rings, which
is allowed to vary from ring to ring.  \item the declination of the
central positions of the rings, which is allowed to vary from ring to
ring.  \item the systemic velocity, which is also allowed to vary from
ring to ring.  \end{itemize}

The user specifies the following global parameters:
\begin{itemize}

\item a global isotropic velocity dispersion, which includes the
instrumental dispersion.  It is clear that such a treatment is not
entirely physical, as it is known that the dispersion changes with
radius in spiral galaxies and is also not the same in \cng{the} vertical and
horizontal direction of the disk.  On the other hand, one has to keep in
mind that the tilted-ring model itself is in fact a geometric parametrisation of the galaxy kinematics,
which still needs to be interpreted, even if it resembles the
true kinematics quite well.  The reason to keep a velocity dispersion
as a global parameter is to improve the computational speed.  It
provides the possibility to shift the modelling of the dispersion to a
convolution instead of a (time-expensive) Monte-Carlo integration. 

\item the second global parameter accounts for the vertical distribution
of the gas density.  The user can chose between a Gaussian, $sech^2$,
exponential, Lorentzian and a box layer. 

\item the constant total flux of a single point source (a ``cloud''). 
The number of Monte-Carlo point sources is approximately the total flux
of the model divided by this number.  When calculating the point sources
for a subring, however, the cloud flux is changed by a tiny fraction in
order to keep the flux of the subring accurate.\end{itemize}

After calculation of a point source model and gridding of the
Monte-Carlo point sources on\cng{to} a model cube, the cube is convolved with a 3D-Gaussian,
a product of a 2D (anisotropic) Gaussian in the $x$-$y$-plane, and a 1D
Gaussian determined by the global velocity dispersion.  The form of the
Gaussian in the $x$-$y$-plane is determined by the \cng{observing} beam of
the input cube \cng{(the CLEAN beam)}, which can be redefined by the user.
\subsection{$\chi^2$ evaluation} \label{Sect_2.2.3} The convolution
routine implemented in TiRiFiC \cng{aims} at computational speed. 
The only possibility to reach the required speed is by an
FFT-convolution, for which the FFTW \citep[][]{Frigo05} library is used. 
It seems to possess the highest flexibility and quality (in terms of
computational speed) of all freely available FFT libraries. 

TiRiFiC calculates the $\chi^2$ and, along with this, the relative
probability of two models.  Usually, $\chi^2$ is calculated via
\begin{equation}
\label{eq_2.1}
\begin{aligned}
\chi^2 & = \sum_k \frac{(M_k-O_k)^2}{\sigma_k^2} & = \sum_k
\frac{(M_k-O_k)^2}{w_k}
\end{aligned}
\qquad,
\end{equation}
where $k$ is an index running over all pixels in the cube, $M$ is
the model data cube, $O$ the original, and $\sigma_k$ the noise of
the original data cube in the $k$th pixel.  If the quantisation noise is
treated correctly, the weight $w_k$ of a pixel should be
\label{eq_2.6}
\begin{equation}
w(k) = \sigma_{\mathrm{rms}}^2+(\sigma^{\mathrm{q}}_k)^2 \rm ,
\end{equation}

where $\sigma_{\mathrm{rms}}$ is the rms noise of the original data cube
and $\sigma^{\mathrm{q}}_k$ the quantisation noise of the convolved
artificial data cube.  TiRiFiC is able to calculate
$\sigma^{\mathrm{q}}_k$ \citep[for a mathematical treatment
see][]{Jozsa06}.  It is, however, a larger computational effort to do
so, and the user might be inclined to modify the goodness-of-fit
evaluation.  With a ``weight parameter'' $W$ a weight map is calculated
via

\begin{equation}
\label{eq_2.7}
w_k = \frac{\sigma_{\mathrm{rms}}^2\cdot W^2+(\sigma^{\mathrm{q}}_k)^2}{W^2} \qquad .
\end{equation}
If $W \rightarrow \infty$, the weight map becomes a
constant, $\sigma_{\mathrm{rms}}^2$, giving TiRiFiC the possibility to save one
convolution. If $W=1$, the noise will be estimated correctly,
including the quantisation noise (which can in principle exceed the rms
noise of the data cube). The parameter is kept continuous to give the
user kind of a weighting scheme at hand. With an increasing weight
parameter, the emphasis will be more and more taken away from regions
with high quantisation noise (and thus high surface density)\cng{,} towards regions
of low surface density.

\subsection{$\chi^2$ minimisation}
\label{Sect_2.2.4}

With the type of model generation in TiRiFiC, the usual $\chi^2$
minimisation algorithms employing partial derivatives of the model
function with respect to the fitting parameters like the
Levenberg-Marquardt algorithm
\citep[][]{Levenberg44,Marquardt63,Press86} are not easily realised,
since the analytic form of the fitting function is unknown.  The
Monte-Carlo method  \cng{was chosen} as a way to produce a model cube because
an analytic form of the tilted-ring model does not exist.  Furthermore,
the model itself, for which a Jacobian would have to be supplied, is the
\emph{convolved} tilted-ring model in three dimensions. 

For TiRiFiC the simplest solution without the need for derivatives was
implemented first hand.  The golden-section search algorithm \citep[][
Chap.  10.1]{Press86} is an uneconomic but \cng{simple} method (see below)
to find local $\chi^2-$minima in parameter space without the necessity
to normalise the $\chi^2$ evaluated with TiRiFiC, i.e.  for this
algorithm the knowledge of the rms noise in the analysed data cube does
not play a major role. 

A tentatively implemented Metropolis Monte-Carlo-Markov-chain algorithm
\citep[][]{Metropolis53} proved not to be applicable in tests using
real H{\,{\small I}} data cubes, mainly because a smooth
disk as fitted to the data cubes is not close enough to a realistic, clumpy
galaxy disk. The Metropolis algorithm requires an ideal
parameter-set to be able to reproduce the data perfectly. For more
details on the fitting procedure see \citet[][ Chap.~2]{Jozsa06}.

Parameter fitting in TiRiFiC is rather flexible.  The user can specify
single parameters to be fitted, or groups of parameters, which will be
fitted at once as one parameter.  As an example, the user has the
possibility to fit a changing projected centre from ring to ring as is,
e.g., expected for a galaxy disk with a bowl-shaped warp.  Such
bowl-shaped warps result in an asymmetric projected appearance.  For
rather symmetric galaxies this might not be desirable, because in this
case one would desire to keep the number of free parameters low.  With
TiRiFiC the user has the possibility to fit a single centre for all
rings, defining their centres as a group and fitting the value of this
group as a whole. 

\section{Beam smearing and smoothing inherent to velocity fields}
\label{Sect_2.3}
\cngs{
The success of a tilted-ring fit to a velocity field depends on how
well the velocity field is able to represent the mid-plane recession
velocity of the rotating disk. In practice, to derive a velocity field from a data cube one usually tries to
determine the peak positions of the velocity profiles \citep[e.g.][ Chap. 3]{Swaters99}.
The simplest way to extract this peak velocity from a noisy velocity profile is to compute the intensity-weighted mean
\citep[the first moment, e.g.][]{Rogstad71,Simon03,Simon05}, while a
fit of full Gaussians \citep[e.g.][]{Begemann87,Swaters99} or
half-Gaussians \citep[e.g.][]{Garcia-Ruiz02a} is considered a
better approach. Even more sophisticated methods exist to
reliably derive the peak velocities from the line profiles
\citep[][]{Gentile04}. With artificial data cubes, however, one is neither
restricted by noise nor by the velocity resolution. Therefore, by
producing a noiseless artificial observation with a very
high spectral resolution and reading out the peak velocities from the
profiles, one gets a velocity field which would ideally be derived from a measured data cube.}

\cngs{It can be shown in a simple experiment that biases will occur in a velocity field-based tilted-ring analysis even when using such an idealised case of a velocity field.}

\begin{figure*}
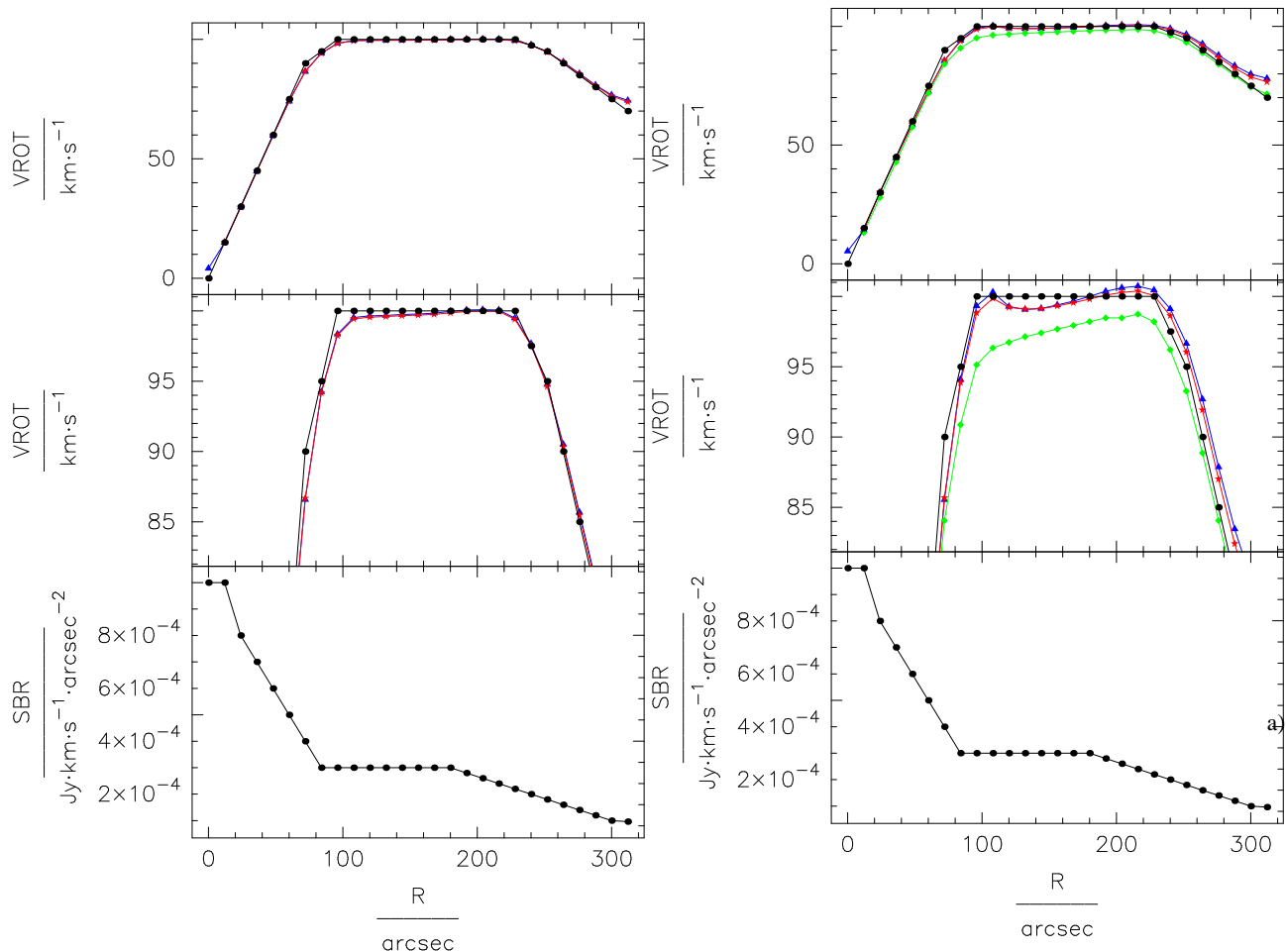

 \parbox[c]{8.5cm}
{
  \begin{center}
\includegraphics[width=8.5cm]{fig_2a.eps}
  \end{center}
}
\parbox[c]{8.5cm} {
{
    \begin{center}
\includegraphics[width=8.5cm]{fig_2b.eps}
    \end{center}
}
\parbox[c]{8.5cm} 
{
  \begin{flushright}
  \vspace{-200pt}
  \hspace{-200pt}
  a)
  \end{flushright}
}
}
\caption{Testing the extraction of rotation \cng{curves} from velocity fields. Filled circles and connecting lines (black) show the parametrisation with which two \cng{artificial} observations have been generated, a velocity field and a data cube (see text). Top panel: rotation curve. Centre panel: blow-up of the rotation curve (same lay-out as above). Bottom panel: {H\,{\small I}} surface brightness. The \cng{inclinations are $60\deg$ (left) and $75\deg$ (right) respectively.} \cng{Velocity fields were} generated as described in the text. Triangles and connecting lines (blue in online version): results from a ROTCUR fit with a free-angle of $0\deg$ and a uniform weighting. Stars and connecting lines (red in online version): results from a ROTCUR fit with a free angle of $30\deg$ and a cosine weighting. Diamonds and connecting lines \cng{in the right panel} (green in online-version): results from a cut along the galaxy major axis, corrected for inclination.
\cng{Filled circles and connecting lines (black) also denote the results of a TiRiFiC fit; they are identical to the parametrisations of the artificial observation.}
}
\label{Fig_2.2}
\end{figure*}

A Gaussian beam with a HPBW of $12\,\arcsec$ and a (total) velocity
dispersion of $10\, {\mathrm{km}}\,{\mathrm{s}}^{-1}$ was used to
produce with TiRiFiC \cng{two} low-noise \cng{artificial}
observation\cng{s} of a galaxy with $17.6\cdot10^6$ point sources in a
data cube with a spatial pixel separation of $4\arcsec$ (for the
parametrisation see Fig.~\ref{Fig_2.2}, a detailed description can be
found in \citealt{Jozsa06}). \cng{The parameters were sampled with a
separation of $12\arcsec$.}
\cng{Inclinations}
of $60\deg$ \cng{and} $75\deg$ \cng{were} chosen, and the position angle was kept fixed at all
radii. 

In order to produce a ``perfect'' velocity field, the cube\cng{s were} \cng{generated with} an unrealistic channel separation of $0.26\,{\mathrm{km\,s}}^{-1}$. 
With \cng{these data cubes} peak-velocity map\cng{s were} created. 
Using \cng{these velocity fields}, rotation curve\cng{s were} produced using the GIPSY routine
ROTCUR, employing two slightly different approaches with an input guess identical to the model parameters\cng{, using a sampling of the rings identical to the input model}.
 The results of both fits can be seen in
Fig.~\ref{Fig_2.2} \citep[see also][]{Teuben02}.

The fitted rotation velocities are biased following a certain pattern
that depends on the shape of the surface brightness profile, which in
this test case is falling monotonically.  In those regions where the
rotation curve is rising, the rotation velocities are underestimated,
while in regions where the rotation curve is falling the rotation
velocities are overestimated.  Hence, the fitted rotation velocity tends
to be biased towards regions of higher surface brightness.  The reason
for this lies in the fact that the measured velocity profile at a
certain position of the galaxy is a convolution of the beam with the
true intensity distribution.  Thus, if the intensity is not constant
over the beam area, the resulting ``smeared'' profile receives more
contribution from areas with higher intensity and hence becomes
asymmetric, even if it would be symmetric \cng{when observed with} a pencil-beam.  The
position of its maximum is dragged towards velocities with a higher
associated surface brightness.  This effect is \cng{cannot be treated} in a
straightforward manner  because it also depends on how velocities
change within a beam. It is also visible at radii where both the
rotation curve and surface brightness are flat over a large range.  This
means that ``smoothing'' affects the rotation curve at all radii.  The
reason for this lies in the fact that in projection the galaxy becomes
\cng{fore shortened} along the kinematical minor axis. \cng{Here, t}he gradient of the
velocity field becomes largest and the \cng{observing} beam covers the
largest range of velocities along the kinematical minor axis. Points in
the vicinity of the kinematical minor axis are thus more strongly
affected by beam smearing. \cng{This is why u}sually a section of the velocity field is
thus not regarded (by specifying a ``free-angle\footnote{angle defining
cones around the minor axis within which radial velocities are discarded
from the computation}'') in a rotation curve analysis and a weighting of
data points is applied, while our experiment shows that this treatment
does not help much for the analysis of the inner regions, but improves
matters in the outer ones.  An extreme choice for a free angle would be
to only read out the velocities along the major axis of the galaxy and
to correct the velocities for the inclination.  This is equivalent to an
extraction of a rotation curve from a long-slit observation.  The
results from such a ``fit'' are shown in \cng{the right hand panel of} Fig.~\ref{Fig_2.2}.  It can be
seen that with such a choice the results get slightly better in the
outskirts of the galaxy at the expense of far worse results in \cng{its inner parts}. 

\cng{In order to demonstrate how a velocity field analysis is influenced by the a-priori unknown distribution of the \ion{H}{i}, we chose a rather high inclination of at least $i=60\deg$ for our artificial observations. Since this covers only one third of all observable galaxies, we emphasise that not necessarily all velocity field analyses are affected by beam smearing. Furthermore, the clumpiness of the distribution of the interstellar medium (ISM) introduces further errors in a tilted-ring analysis, which might well exceed the ones arising from the shown variant of beam smearing.}

\cng{However\cngs{, using a velocity field as would be ideally derived in a noiseless observation, }we demonstrated how \cngs{in analysing} velocity fields of galaxies with an inclination higher than $i=60\deg$ systematic errors might be introduced by beam smearing.}
Even if the maximum deviation from the true rotation curve is not very
large (\cng{about $3.3\,{\mathrm{km\,s}}^{-1}$ for $i=60\deg$ and}
about $3.9\,{\mathrm{km\,s}}^{-1}$ for $i=75\deg$ in this experiment),
this variant of beam smearing can introduce a significant
misinterpretation of the measurement, as the resulting rotation curve
is \emph{systematically} biased. \cng{Moreover, second order treatments of the kinematics of gaseous disks, like a harmonic analysis, require a precision in that range.} An assignment of a (large)
statistical error to compensate for beam smearing is thus not a
solution. Results from mass decompositions in which (unknown) density
profiles are fitted to rotation curves derived from velocity fields
are very vulnerable in that respect. The smoothing effect can lead to
a preference of rotation curves with a smaller curvature
${\partial^2 v}/{\partial r^2}$.
\section{Testing TiRiFiC}
\label{Sect_2.4}
\subsection{The trivial test} \label{Sect_2.4.1}
\cng{Figure \ref{Fig_2.2} also shows the results of a
fit with TiRiFiC to artificial data cubes that are identical to the
data cubes used to produce the velocity fields, except for a larger
channel width of $2.06\,{\rm km\,s}^{-1}$. As for ROTCUR, the
first-guess-parameters were identical to the model
parameters and sampled at identical radii. Naturally, TiRiFiC makes a perfect fit and the resulting
model is identical to the input model. It is however noticeable that, in
principle, with TiRiFiC it is thus possible to produce an artificial
observation that can be reproduced by a fitting process over and
over. For routines working on the velocity field this is not the
case. First, the surface brightness distribution, which is a free
extra output of TiRiFiC, has to be inferred using other information
(e.g. by fitting a tilted-ring model to the total intensity map). Second, an artificial
data cube will always have to be subject to a convolution with an
artificial beam, hence introducing beam smearing; therefore, the artificial cube never reproduces a velocity field that fits
to the tilted-ring model found by the fitting routine.}

\subsection{Rotation curve of an edge-on disk}\label{Sect_2.4.2}
\begin{figure}
\resizebox{\hsize}{!}{\includegraphics{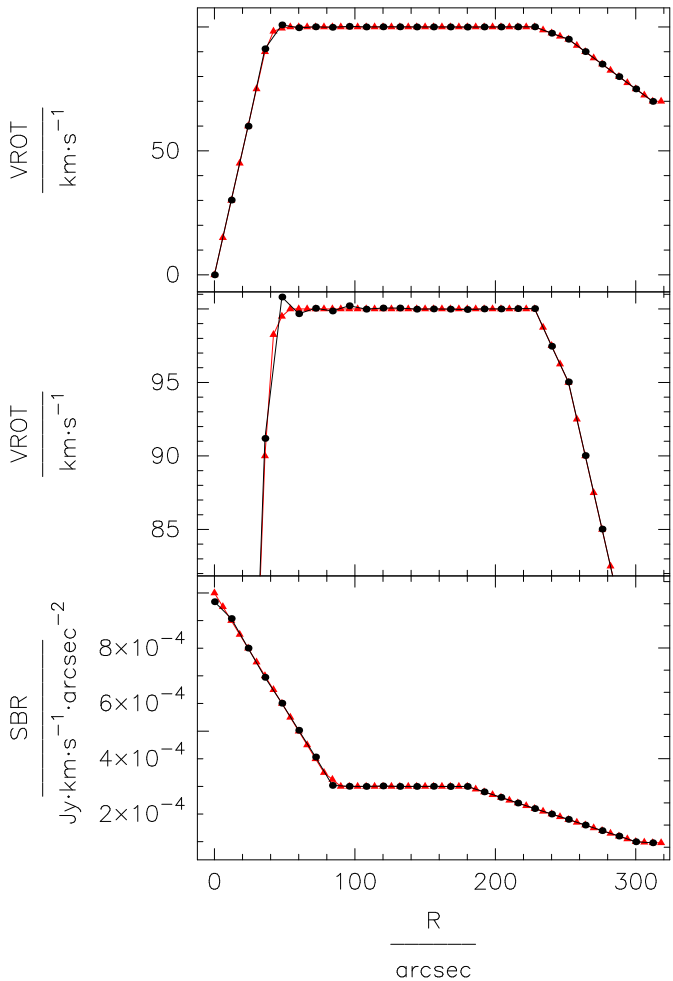}}
\caption{Testing rotation curve extraction with TiRiFiC. Triangles and connecting lines (red in online version) show the parametrisation with which a artificial observation of a flat galaxy with $90\deg$ inclination has been generated (edge-on case). Top panel: rotation velocity. Middle panel: rotation velocity, blow-up. Bottom panel: surface brightness. The parametrisation for the artificial observation is twice as dense as for the fit of the artificial observation performed with TiRiFiC. The fitted parameters are represented by dots (black in online version) and connecting lines.}
\label{fig_2.4}
\end{figure}

\cng{One of the strengths of TiRiFiC is supposed to lie in the ability to
perform reliable fits where velocity field methods are bound to fail
by definition. This is in particular the case when the line-of-sight
crosses the galaxy disk twice or more often. An extreme case of such a
situation is encountered when a galaxy is observed (nearly) edge-on, because
then the galaxy disk is crossed not only twice but an infinite number
of times. We simulated a galaxy observation using the rotation
curve and the surface brightness profile shown in Fig.~\ref{fig_2.4}, a template cube and an artificial beam as in Sect.~\ref{Sect_2.4.1}, putting it in an edge-on orientation. The parameters for the artificial observation were sampled in steps of $6\arcsec$. Then,
the artificial observation was fitted with TiRiFiC, taking the input model
parameters as a first guess, but leaving the rotation velocity, the
surface density profile, and the orientational parameters, inclination
and position angle as free parameters. In order to prevent a trivial solution, the fitting was performed with a parameter sampling in steps of $12\arcsec$.
TiRiFiC
reaches a nearly perfect fit, interpolating in the regions where the
fit results cannot reproduce the artificial observation by construction (see
Fig.~\ref{fig_2.4}). Hence, TiRiFiC is in principle able to figure out reliable
rotation curves from observations of edge-on galaxies. A slight ``beam-smearing'' effect takes place at radii where it is not possible to follow the high curvature of the parametrisation of the artificial observation, owing to the fact that in the fitting process the parameters are sampled with half the rate as for the artificial observation.

A caveat is that
TiRiFiC assumes a galaxy to be translucent. Observations of edge-on
galaxies may suffer significantly from
self-absorption and scattering processes at certain wavelengths \citep[e.g.][]{Baes03}. This
has to be taken into account when interpreting TiRiFiC results,
especially when fitting to optical spectroscopic data.}

\subsection{An edge-on warp} \label{Sect_2.4.3} 
TiRiFiC has been
constructed to be able to fit data cubes of observations of warped
galaxies, which is one of the main applications of the tilted-ring
model. In order to construct a case tailored to put TiRiFiC to the
test and for which no appropriate \cng{automated} software exists yet, a warp in an
edge-on galaxy was constructed. This is a situation where it is
impossible to construct an unambiguous velocity field, as the
line-of-sight crosses the disk several times. Again, the model observation was generated using a template data cube and an artificial beam as described in the previous subsections, sampling the model parameters in steps of $6\arcsec$.

\begin{figure}
\resizebox{\hsize}{!}{\includegraphics{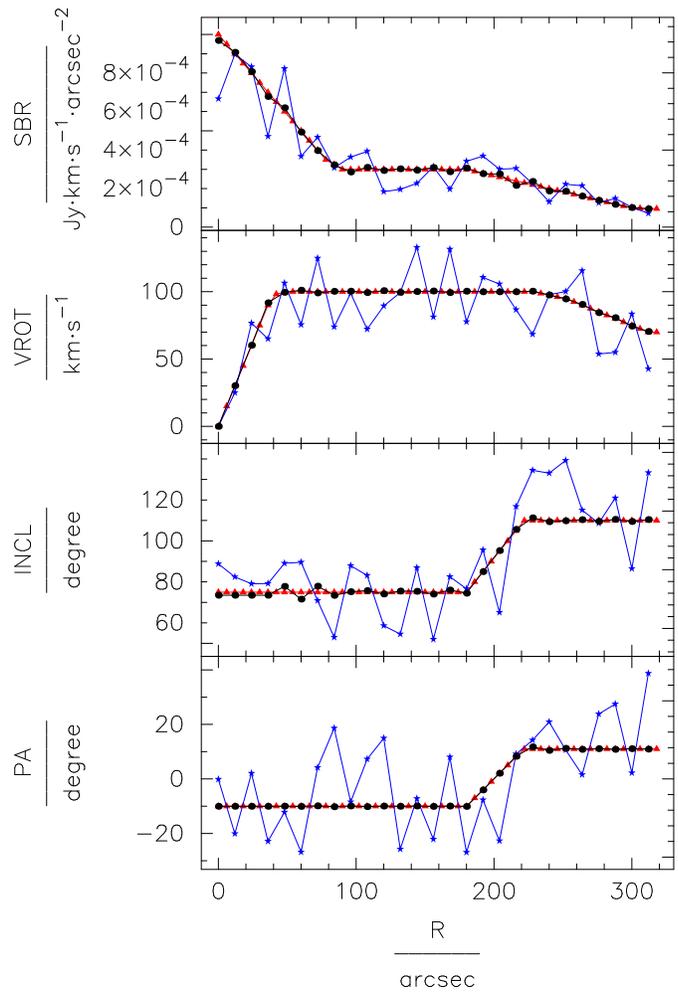}}
\caption{Testing tilted-ring fitting with TiRiFiC. Triangles and connecting lines (red in online version) show the parametrisation with which a \cng{artificial} observation of a heavily warped galaxy has been generated. stars and connecting lines (blue in online version) show the first-guess input to TiRiFiC. The fitted parameters are represented by black filled circles and connecting lines.}
\label{Fig_2.5}
\end{figure}

The input parameters to TiRiFiC were chosen to randomly deviate from the
input model parameters (see Fig.~\ref{Fig_2.5}). \cng{Again, in fitting the data the model parameters were sampled in steps of $12\arcsec$ in order to prevent the possibility of a trivial solution.}

In fitting the data cube, the rotation velocity, surface brightness,
inclination and position angle were fitted independently for every
single ring; the other parameters were fitted by letting them vary
together. An exception was made for the innermost four rings
where the orientation parameters were fitted \cng{with the constraint} to be
the same for each ring.  A number of $3.5\cdot 10^6$ point-sources was
chosen for the fitting process, while the \cng{artificial} observation was created
with $17.6\cdot10^6$ point sources.  

The result of the fit process
(Fig.~\ref{Fig_2.5}) shows that with TiRiFiC, it is possible to fit to a
data cube of an edge-on galaxy with a large warp with a very high
accuracy.  
\cng{
\subsection{Sensitivity}
\label{Sect_2.4.3a}
Since a galactic disk (or any gaseous astronomical object with disk symmetry) becomes fainter with increasing radius, the extent of a 
tilted-ring analysis is either limited by the observational
noise or by the fact that the ring symmetry breaks down. For a conventional velocity field analysis the radial
velocities have to be derived from the spectra in the single pixels in
a data cube. Hence, the extent of the velocity field and, with this, the
maximal radius that can be addressed in a tilted-ring analysis is
determined by the pixels where the signal-to-noise ratio
significantly exceeds unity. Since with the TiRiFiC approach one fits
a global model to the data cube, the number of pixels that determine
the parametrisation of a ring is large. Therefore, even if the
emission from an object is well below the nominal noise per pixel,
one should still be able to derive a reliable
parametrisation with TiRiFiC . Provided the object follows a tilted-ring symmetry the detection limit should in theory scale with ${1}/{\sqrt(N)}$, where $N$
is the number of pixels with emission in a rotating ring projected
onto the 3D data cube. 

In order to get an impression of the realistic performance of TiRiFiC
when fitting faint structures, we generated a large number of artificial
observations of gaseous rings with varying uniform surface brightness
and varying surface solid angle. The rings were projected onto a data
cube with a size of $512\arcsec\times 512\arcsec$ and a pixel size of $4\arcsec$, 64 channels with a separation of $4.12\,{\rm km}\,{\rm s}^{-1}$ using an observational
beam of $12\arcsec\times14\arcsec$ (Half Power Beam Width), and
a global velocity dispersion of $7\,{\rm km}\,{\rm s}^{-1}$. With
these specifications, an \ion{H}{i} observation with the Westerbork
Radio Synthesis Telescope (WSRT) is simulated. In order to add
realistic noise to the model, we utilised line-free channels from a
$2\times 12\,\rm hrs$ observation with the WSRT having an rms noise
level of $3.6\,{\rm mJy}/{\rm beam}$. The rings were chosen to have an
inclination of $60\deg$, a rotation velocity of $75\,{\rm km}\,{\rm
s}^{-1}$, and a scale height of $2\arcsec$ using a $sech^2$-law for
the vertical distribution. While the maximal extent of each ring was
fixed at $200\arcsec$, the inner radius was varied between $45\arcsec$
and $195\arcsec$, varying linearly the face-on solid angle occupied by
the ring, which is roughly proportional to the number of pixels containing emission in the
projection of the ring onto the cube.

We fitted the artificial observations with rings of the same extent, with an
initial guess deviating significantly from the artificial parametrisation
(surface brightness $1\cdot 10^{-5}\,{\rm Jy}\,{\rm km}\,{\rm s}^{-1}\,{\rm arcsec}^-2$,
rotation velocity $65\,{\rm km}\,{\rm s}^{-1}$, inclination and
position angle deviating by $10\deg$, all other input parameters
identical to the parametrisation of the artificial observation) in a two
step process, first fitting all parameters except the surface
brightness, then fitting all parameters. In order to distinguish a
detection of such a ring from a non-detection we also fitted a data
cube containing only noise and no emission. Calculating the
differences of the fitted parameters and the parameters used for the
artificial observation, one can clearly distinguish between cases where the
fit was successful and where it failed (Fig.~\ref{fig_thresh_pa} and
Figs.~\ref{fig_senso} in the online version of this paper). 
\begin{figure}
\resizebox{\hsize}{!}{\includegraphics[angle=270]{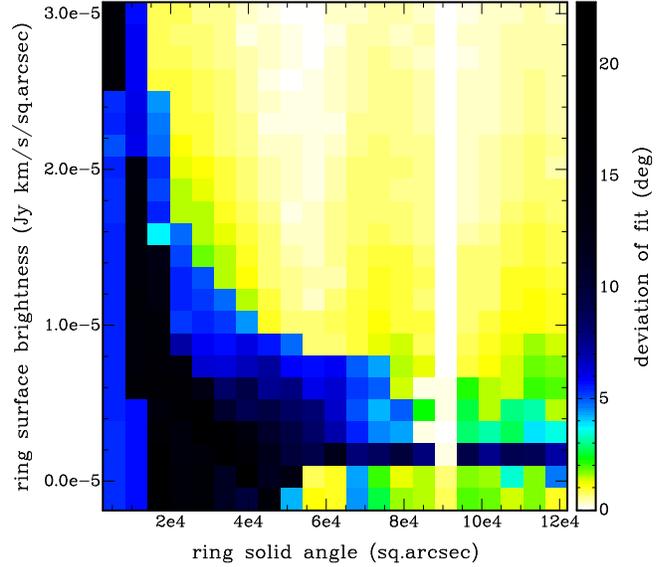}}
\caption{Testing the sensitivity of TiRiFiC, example. Each pixel represents a fit to a artificial observation (including realistic noise) of an inclined rotating ring with a given solid angle and surface brightness, the lowest row representing a fit to a data cube containing noise only. The pixel values represent the deviation of the fitted position angle from the true position angle of the ring. Clearly distinguishable are regions where the fit succeeded (bright regions) and where it failed (dark regions). Similar plots can be generated for other fitting parameters (see Fig.~\ref{fig_senso} in the online version of the paper).}
\label{fig_thresh_pa}
\end{figure}
By setting a
threshold for the deviations of the parameters, viz. rotation velocity
($4\,{\rm km}\,{\rm s}^{-1}$), surface brightness ($7\cdot
10^{-7}\,{\rm Jy}\,{\rm km}\,{\rm s}^{-1}\,{\rm arcsec}^-2$), position angle ($2\deg$), and
inclination ($2\deg$), a binary diagram could be constructed in which
a detection for combinations of ring surface brightness and ring solid
angle was marked (see lower right panel in Fig.~\ref{fig_senso}). Finally, the detection limits
as a function of the ring solid angle could be extracted
(Fig.~\ref{fig_sens}).
\begin{figure}
\resizebox{\hsize}{!}{\includegraphics[angle=270]{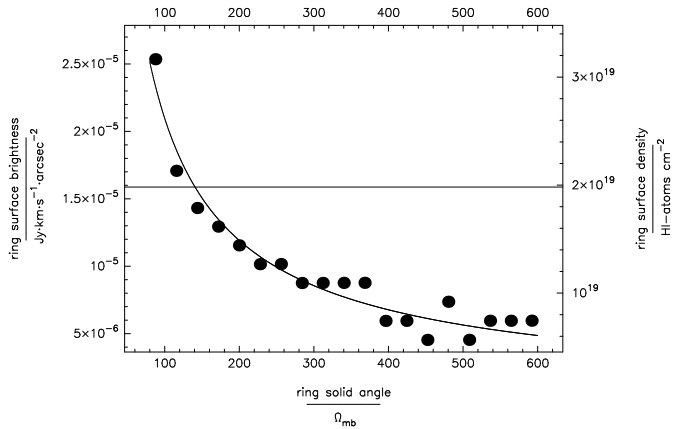}}
\caption{Testing the sensitivity of TiRiFiC. Detection limits as a function of ring-surface solid angle in units of the main beam solid angle ($\Omega_{\rm mb}=202\,{\rm arcsec}^2$, see text). The dependence is fairly well fitted by a power law (curved line). The vertical line marks the surface brightness where all pixels in the noiseless data cube lie below the 1-$\sigma_{\rm rms}$ level of the noisy data cube. The right hand axis is scaled such that the surface brightness of a ring is converted to surface column density of an \ion{H}{i} observation.}
\label{fig_sens}
\end{figure}
It turned out that while the detection limit determined that way shows
a monotonic decrease with increasing ring solid angle, the expected
proportionality did not show up. Also when trying to estimate the
number of independent data points which the projected ring occupies in the
artificial observation we could not establish a simple law that connects the
ring solid angle to the detection limits.

However, Fig.~\ref{fig_sens} clearly shows how the sensitivity in
detecting faint symmetric structures in gaseous disks is enhanced
using the TiRiFiC approach. With a power law
\begin{equation}
\label{eqn}
\frac{\sigma_{\rm tir}}{{\rm Jy}\,{\rm km}\,{\rm s}^{-1}\,{\rm arcsec}^{-2}} = 9.0\cdot 10^{-4}\cdot \frac{\Omega_{\rm ring}^{-0.82}}{\Omega_{\rm mb}} \qquad ,
\end{equation}
where $\sigma_{\rm tir}$ is the detection limit and $\Omega_{\rm
ring}$ is the ring surface solid angle, and $\Omega_{\rm mb} = 202\,
{\rm arcsec}^2$ the beam solid angle, we achieve a reasonable fit. With this we can estimate the ring solid angle of $141\,\Omega_{\rm mb}$ that is needed to detect a ring that in projection onto
the cube has an intensity just below the noise $\sigma_{\rm rms}$ in the data cube ($1.6\cdot 10^{-5}\,{\rm Jy}\,{\rm km}\,{\rm s}^{-1}$ ($\,\hat{\approx}
2\cdot 10^{19} \,{\rm atoms}\,{\rm cm}^{-2}$ in an \ion{H}{i} observation, horizontal
line in Fig.~\ref{fig_sens}). At a radius of $200\arcsec$, this
corresponds to a ring width of $24\arcsec$.

A caveat arises from the fact that even when no real emission is
present, a fit with TiRiFiC will result in a
detection a faint structures in the presence of noise: in a noisy data set a residual offset in
surface brightness can be surely found for a specific optimal ring
parametrisation even if averaging over many pixels. This can be tested
when fitting to a data cube that contains only noise, like it has been
done in the experiment discussed here. Figure~\ref{fig_sens_noem}
shows the detected surface brightnesses as a function of ring width when fitting a data cube that contains only noise. The erroneously detected surface brightnesses lie well below the detection limit as quoted in Eq.~\ref{eqn}, the detection limit exceedig the erraneously detected surface brightnesses by a factor of a few. 
However in analysing observations with TiRiFiC, one has to be cautious when accepting a faint structure detected by TiRiFiC as real, since then residual calibration and CLEANing errors become more and more important; in particular cases one should possibly apply additional tests such as the analysis of independent observations.
\begin{figure}
\resizebox{\hsize}{!}{\includegraphics[angle=270]{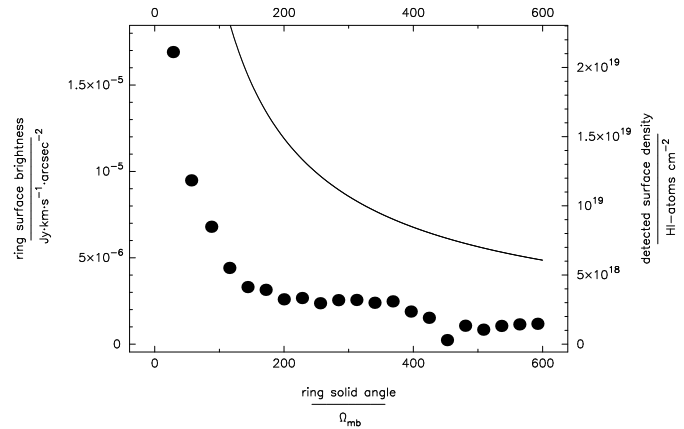}}
\caption{Resulting surface brightness from a fit of a ring to a cube that contains only noise as a function of ring surface solid angle in units of the beam solid angle ($\Omega_{\rm mb}=202\,{\rm arcsec}^2$, see text). Over-plotted is the detection limit determined from fits to faint rings (curved line, see text). The erroneously detected surface brightness lies well below that line.}
\label{fig_sens_noem}
\end{figure}

Compared to e.g. \ion{H}{i} observations of nearby spiral galaxies,
the size of the test data cube is rather small. Since the surface solid
angle of a ring with constant width scales with radius, the detection limits will scale with radius for larger objects.
We thus
conclude that with TiRiFiC, one is able to detect faint structures
well below the noise -- provided the observed object shows the
symmetry inherent to the tilted-ring model. The detection limits can
be estimated as shown above. 
\subsection{Realistic tests}
\label{Sect_2.4.3b}
In order to test how TiRiFiC works in practice, we constructed a
number of artificial \ion{H}{i} observations of galaxies that had to be
analysed with TiRiFiC by a (variable) author of this paper, the
``observer''. The template cube and the \cng{observing} beam had the
same specifications as given in Sect.~{\ref{Sect_2.4.3a}}. We
performed fits to artificial observations with and without realistic noise
added to the data cube as described above, and with varying degrees of
symmetry. The observer was informed about the symmetry of the object
(like ``flat disk'' or ``constant centre and systemic velocity''), but
strictly not about the specific parametrisation of the artificial galaxy
disk. We also included cases where \emph{all} parameters were kept
variable and hence a lopsided or a U-shaped galaxy with a possible
flare was analysed. Finally, we constructed a case where the cloud
flux was drastically enhanced in order to simulate the clumpiness of
the gas in the observed object.

All fits were performed following the same pattern. First, the
observer estimated the rough geometry of the object from the data
cube, in order to then run a first iteration process with TiRiFiC
using only a few parameters (fitting a common centre and systemic
velocity, a common scale height, a common orientation, and a surface
brightness distribution and a rotation curve parametrised by a few
data points). After that, the sampling of the parameters was refined
and, if necessary, the parameters were fitted independently. It turns
out that a fitting process with TiRiFiC needs the personal attention
of an observer. In order to achieve a good fit, it is necessary to
inspect the results and to correct single data points that are
obviously ``outliers'' in order to restart the fitting process. After
a few such iterations (in the range of 2-10) no improvement is achieved
and the last fit is the final result. 

In order to compare TiRiFiC to methods, \cngs{where a velocity field is derived and fitted}, i.e. a ROTCUR \cngs{application together with an idealised realistic construction of a velocity field}, we additionally generated data
cubes with a ten times higher velocity resolution \cngs{(a channel width of $0.41\,{\rm km}\,{\rm s}^{-1}$)} and (if the cube
given to the observer contained noise) a ten times lower noise
level. The data cubes were flagged where the intensity in the
noiseless data cube was below $0.5 \sigma_{\rm rms}$. Both a peak intensity and a first moment velocity field was
generated from the flagged
data cube using the GIPSY routine MOMENTS. \cngs{This way we simulated the construction of a velocity field as would ideally be derived by an observer \cngs{(see Sect.~\ref{Sect_2.3}) using common methods for an extraction of a velocity field. The velocity fields were} analysed
with ROTCUR, the input guesses either deviating slightly from
the optimal guess or being optimal. 

A full documentation of all specific fitting results is given in the
online version of the paper. Here, we provide a summary.
  
Independent of the specific iteration method preferred by a single observer,
TiRiFiC
bears reliable results at galactocentric radii greater than 1-1.5 HPBWs (in our example greater than $20\arcsec$). Below that radius, the
number of independent data points is too low to allow a fit of the
full parameter set. In a statistical sense, TiRiFiC is thus not free
from beam smearing but by far less affected than the traditional
method: for the bulk of the data points, the ROTCUR results are worse
than those of TiRiFiC, and the minimal radius where ROTCUR produces
reliable results (under idealised circumstances) lies beyond 4-5 HPBWs
of the observing beam. 

From Eq.~\ref{eqn} we estimate the detection limit for a ring width of
$15\arcsec$ at a radius of $200\arcsec$ to be $2.3\cdot 10^{-5}\,{\rm
Jy}\,{\rm km}\,{\rm s}^{-1}$ ($\,\hat{\approx}\, 2.9\cdot 10^{19} \,{\rm
atoms}\,{\rm cm}^{-2}$ for an \ion{H}{i} observation). This value is a
conservative upper limit. In most cases the detection limits are
significantly lower, especially when the observer could make use of a
certain symmetry, e.g. a common centre of all rings.

TiRiFiC fits reliably disks that possess shifting centres and systemic
velocities. Hence, with TiRiFiC one is able to fit asymmetric galaxies
that maintain the tilted-ring symmetry.

A clumpy \ion{H}{i} distribution simulated as described above does
not hamper the performance of TiRiFiC. TiRiFiC averages over
azimuth, hence a random distribution of a still sizeable number of
clouds results in a larger $\chi^2$, but not in a different result. We did
not simulate the case where the \ion{H}{i} distribution is distinctly
asymmetric. It is to be suspected that in some cases an m=2 harmonic
in the distribution of the tracer material can have an effect on the
results.

Above a limit of $25\deg$ in inclination (which is close to \cngs(face-on}) TiRiFiC disentangles reliably
rotation velocity and inclination. In this respect, ROTCUR made a
better performance \cng{, due to to the fact that our derived velocity fields become very accurate at low inclination. It was shown by \citet[][]{Begemann87}, that for an inclination lower than $60\deg$, inclination and rotation velocity become significantly degenerate in a tilted-ring analysis of the velocity field. For a velocity field analysis with ROTCUR, to derive reliable rotation curves (and with that, a reliable estimate for the inclination) of galaxies with an inclination of less than $40\deg$ it is necessary to apply the method on a velocity field with a statistical error of less than $1\,{rm km}\,{s}^{-1}$. Such an accuracy is hard to reach using real observational material.} In cases with low inclination, the surface density
profile determined with TiRiFiC had the smallest errors.

For most parameters a stable solution is found in the first iteration
process. The fact that nevertheless a few iterations are needed to
reach a final model shows that for some singular parameters the
algorithm used for the $\chi^2$ minimisation does not perform an
effective scann of the parameter space. Then, the user takes the role
of stirring up the parameters in order to enable the localisation of
the better minimum (for a few data points). The success and the
quality of the fits in our tests performed was independent of identity
of the observers, who used slightly differing approaches with regards
to the fitting parameters and the choice how to approach the best-fit
parametrisation from start on. Thus, it seems that the actual way the
user shifts singular data points in order to start a next iteration
process has no big influence on the quality of results. 

We fail to achieve good results in one
particular case with both, TiRiFiC and ROTCUR. In the case of a galaxy with a thick disk
(a vertical scale height of $33\arcsec$ in a disk with an
\ion{H}{i} radius of about $200\arcsec$, see Fig.~\ref{Fig02_add} and Fig.~\ref{Fig02b_add} in the online version of the paper), a TiRiFiC fit will converge
extremely slowly, since in
this case the degeneracy of inclination and rotation velocity is
enhanced. This slow convergence then leads to a misinterpretation of the results as being already the final solution. Also ROTCUR fails with extremely large errors, since the
kinematical information is smeared out along the kinematical minor
axis, making the $\chi^2$ minimum in parameter space shallower. In
other words, the iso-velocity contours get stretched artificially along
the minor axis. In case of TiRiFiC we have the hope to overcome
this problem by implementing a more effective minimisation method. The
user will in any case be able to detect a thickened disk, and be
aware of the problem. We expressively caution, however, against the
over-interpretation of rotation curves regardless of the analysis
method in the case of galaxies likely to have a thickened disk like
it is the case for dIrr galaxies \citep[see
e.g. ][]{Staveley-Smith92,Bottema86,Walter99}. Without showing a
galactic disk to be thin enough, these rotation curves are extremely
unreliable.

\subsection{Solid-body rotation}
\label{Sect_2.4.3c}
Another problem occurring when extracting rotation curves using
velocity fields occurs in the case of a solid-body rotation, as can be
found again in dIrr galaxies \citep[e.g. ][]{Walter99}. In that case,
rotation velocity and inclination are completely degenerate. Since
with TiRiFiC one fits not only the kinematics but also the radial surface brightness profile, this degeneracy is broken, provided the gas distribution
is symmetric enough and shows a radial gradient, as we can show in a further
experiment (last viewgraph in the online version of the paper) that we performed with
the same specifications for the artificial observation as given in
Sect.~\ref{Sect_2.4.3b} using a noiseless data cube. TiRiFiC fits very
reliably a flat disk with solid-body rotation.
}
\subsection{TiRiFiC performance}
\label{Sect_2.4.4}
TiRiFiC is slow. \cngs{Nearly} 160 hours computation time needed to
perform the fit \cng{described in Sect.~\ref{Sect_2.4.3}}, working with
an AMD Athlon XP 2400+ CPU, can serve as an orientation.
\cngs{In practice, however, the user has the possibility to reduce the computing time considerably. In a first iteration process, the size of the data cube can be reduced by binning data points. The model can be ported easily from a low-resolution data cube to a high-resolution data cube, since the model parameters are independent of the specifications of the data cube, unless the reference coordinate system (the epoch and the velocity definition) changes. Assumptions about symmetries can be made in order to reduce the number of parameters. E.g., if a disk is not warped, all position angles and inclinations can be fitted as one single parameter. A-priori known parameters like a surface brightness profile can be excluded from the fitting process.}

Up to now, we concentrated on an implementation of a working version of
the software in order to demonstrate its applicability.  Future
development, however, will include a parallelisation of the code in
order to make the use of computer clusters possible.  Furthermore, the
possibilities to use more sensible fitting procedures will be
explored. With this we certainly expect a jump in the computing
performance of TiRiFiC.
\section{Results and discussion}
\label{Sect_2.6}
We introduce TiRiFiC as a program that fits a tilted-ring model
directly to spectroscopic data cubes \cng{and we provide a number of tests of the program}. We show that the TiRiFiC
tilted-ring model has a high flexibility and even allows modelling
of \cng{galaxy disks that are asymmetric in projection.} We are able to 
show that TiRiFiC is not affected by the drawbacks of methods that work on
velocity fields. \cng{For galaxies with an inclination above $60\deg$, we show that velocity} fields are affected by systematics
that depend on a-priori unknown factors, such as the surface density
distribution. \cng{While the deviations of a rotation curve from the true one derived with fits to the velocity field are small, they occur in a systematic manner, possibly biasing the results of a rotation curve analysis.}
Using a \cng{direct} fit to the data cube,
TiRiFiC reduces these systematics significantly, delivering the
unknown surface density profile as an extra. \cng{A finite observing beam affects the reliability of the TiRiFiC results at the smallest radii in a statistical sense, since the number of fit parameters becomes too large in comparison to the number of independent data points. In general, TiRiFiC delivers more reliable results than a fit to the velocity field.}  

The biggest advantage of TiRiFiC lies in its applicability to edge-on
and warped galaxies. Several tests show that TiRiFiC is able to reach
a best-fit model that comes very close to artificial input models
having a large \cng{edge-on} warp.

\cng{Since the approach of TiRiFiC is to fit a global model to the data
cube, a detection and quantification of a faint outer disk is possible,
even if the bulk of the emission of that disk is fainter than the rms
noise of the data cube, provided the outer disk follows the tilted-ring symmetry.

TiRiFiC fails to disentangle inclination and rotation velocity below a
disk inclination of $25\deg$ (i.e. close to face-on). In that case, to estimate a more
reliable rotation curve, a velocity field approach is to be
preferred.

The determination of rotation curves for galaxies with very thick disks like
dIrr galaxies is unreliable, regardless of the fitting method, since
the degeneracy of inclination and rotation velocity is significantly
enhanced. Since the surface-brightness distribution is also fitted by
TiRiFiC, it is nevertheless possible with this approach to derive reliable rotation curves in the case
of solid-body rotation that occurs predominantly in dIrr galaxies.}

The major \cng{current} drawback of TiRiFiC is its slowness. Performing a
single fit when using a conventional PC takes a long time. Hence,
TiRiFiC is not (yet) usable to perform many fits in a short time
unless a certain amount of computational power can be employed.

\cng{We present a series of practical tests in the course of which an ``observer'' had to parametrise an unknown galactic disk with TiRiFiC, and we are able to show that such a parametrisation is reliable even when analysing data cubes with realistic noise. The success in the limits stated above is independent of the specific approach of the observer.}
It is of course the application \cng{to} real observational data as will be
presented in subsequent papers of this series that will show how
useful TiRiFiC is in \cng{gaining} new insights in the kinematic
structure of rotating disks, especially galaxy disks.  TiRiFiC can be
considered as the most straightforward realisation of a tilted-ring
fitting \cng{method, since with its approach one does not require any careful analysis of the single spectra in a data cube in order to extract a velocity field}. Hence, if TiRiFiC fails to produce a model data cube
that fits to an observation of a galactic disk, the implication is
that its kinematics cannot be represented by a tilted-ring model.  The
development of fitting routines similar to TiRiFiC is a good step
\cng{towards} improving analysis methods of observational data containing
kinematical information, as will be demonstrated in the consecutive
papers of this series.  As secondary kinematical features (and
their influence on the analysis) not compliant with the tilted-ring
model \cng{have received} more and more attention in the recent literature, e.g. by applying harmonical analyses, the
aim must be to extend the analysis tools in order to put theory \cng{of galactic structure formation that includes non-circular motions} to the
test. This work is already in progress.
\begin{acknowledgements}
The underlying PhD thesis project was partly financed by the Deutsche
Forschungsgemeinschaft in the framework of the Graduiertenkolleg 787
``Galaxy Groups as Labor atories for Baryonic and Dark Matter''. 
The underlying PhD thesis project was also partly financed by the University of Bonn in the framework of the ``Research Group Bonn: Dark Matter and Dark Energy''.
We thank the referee Rob Swaters for his numerous and valuable suggestions that enhanced the quality of the paper. 
\end{acknowledgements}
\bibliographystyle{aa}
\bibliography{aa.bib}
\Online
\section{TiRiFiC sensitivity: additional plots}
\label{appendix_A}
\begin{figure*}[htbp]
 \parbox[c]{8.5cm}
{
  \begin{center}
\includegraphics[angle=270,width=8.5cm]{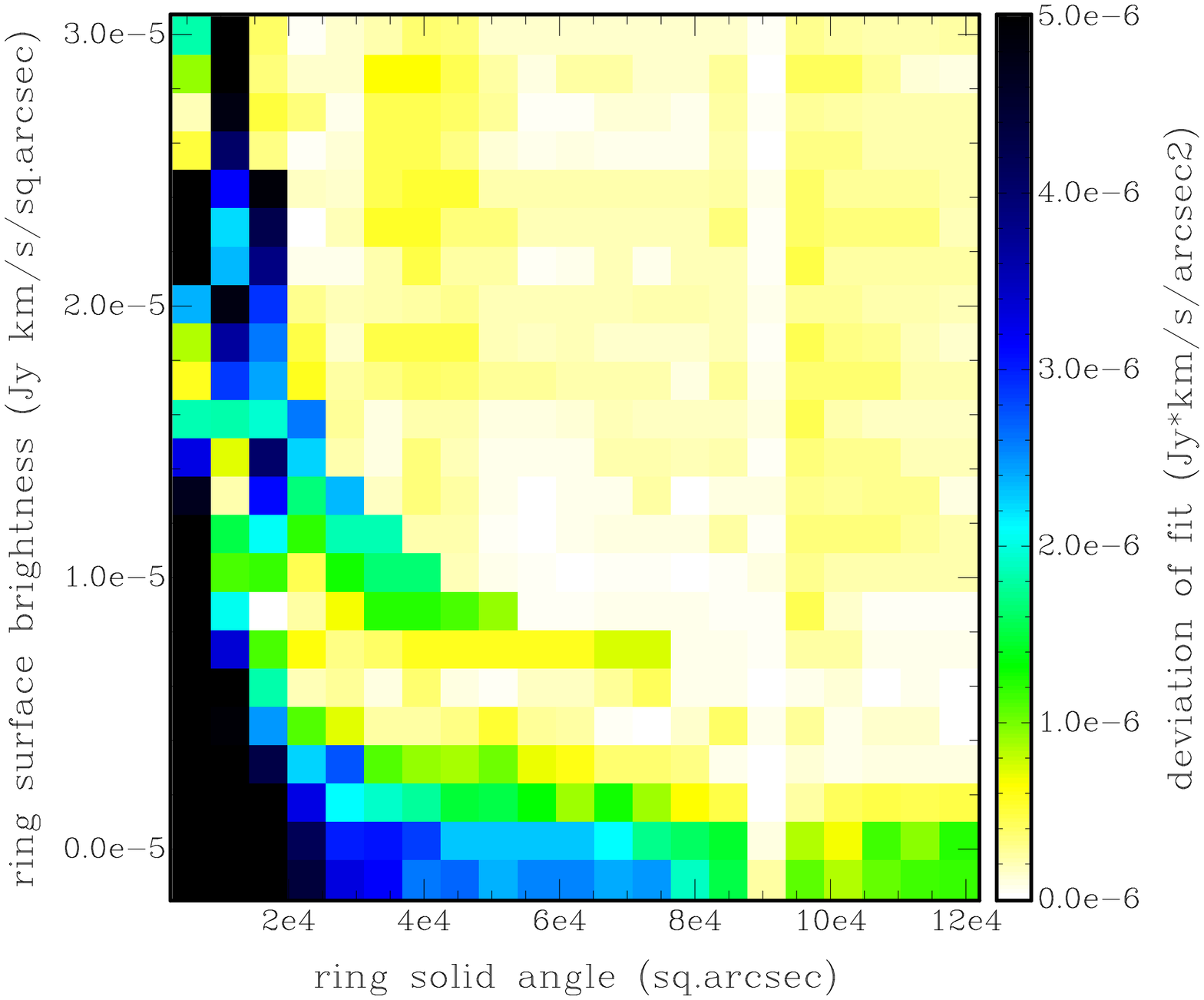}
  \end{center}
}
\parbox[c]{8.5cm}
{
    \begin{center}
\includegraphics[angle=270,width=8.5cm]{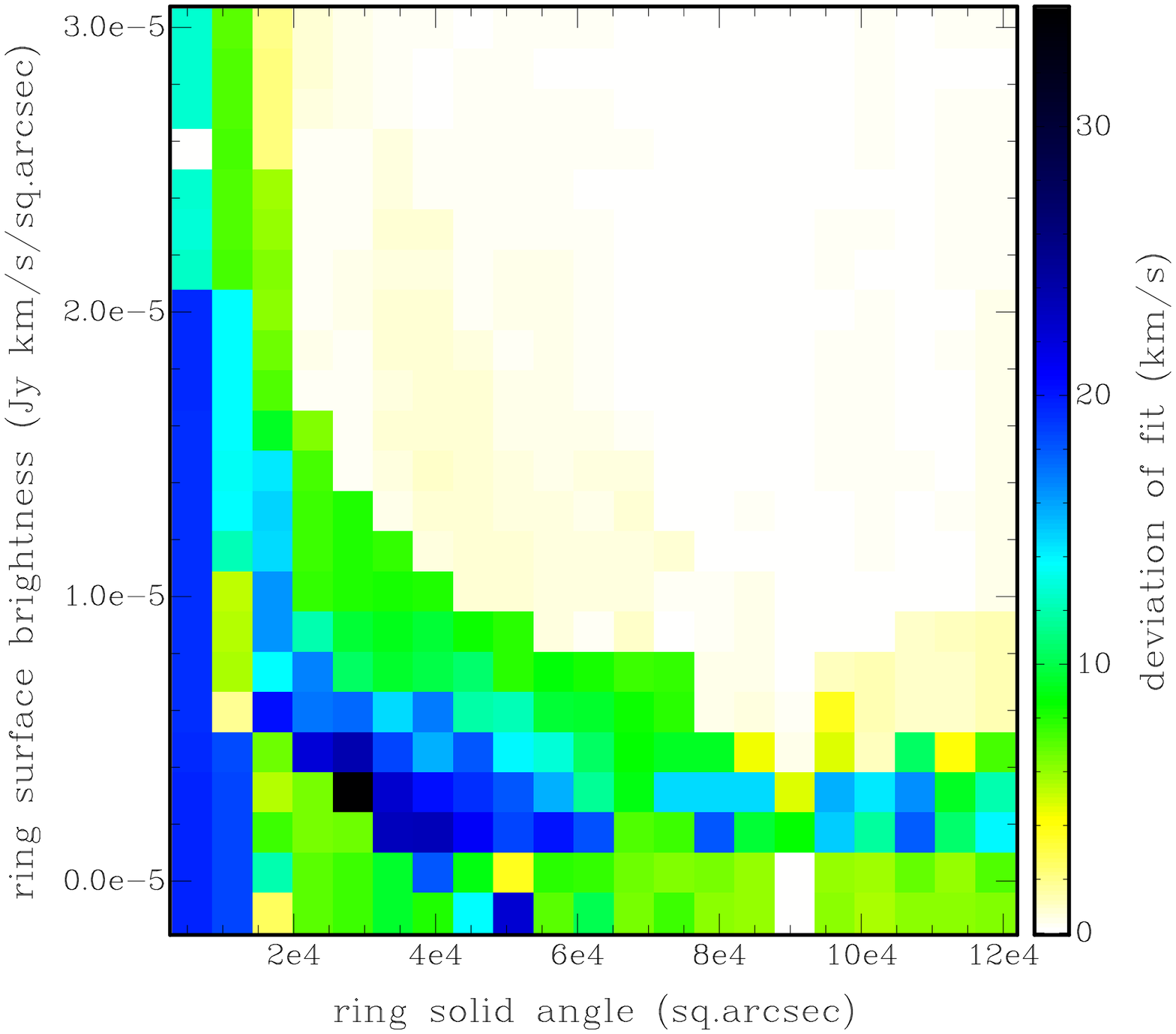}
    \end{center}
}
 \parbox[c]{8.5cm}
{
  \begin{center}
\includegraphics[angle=270,width=8.5cm]{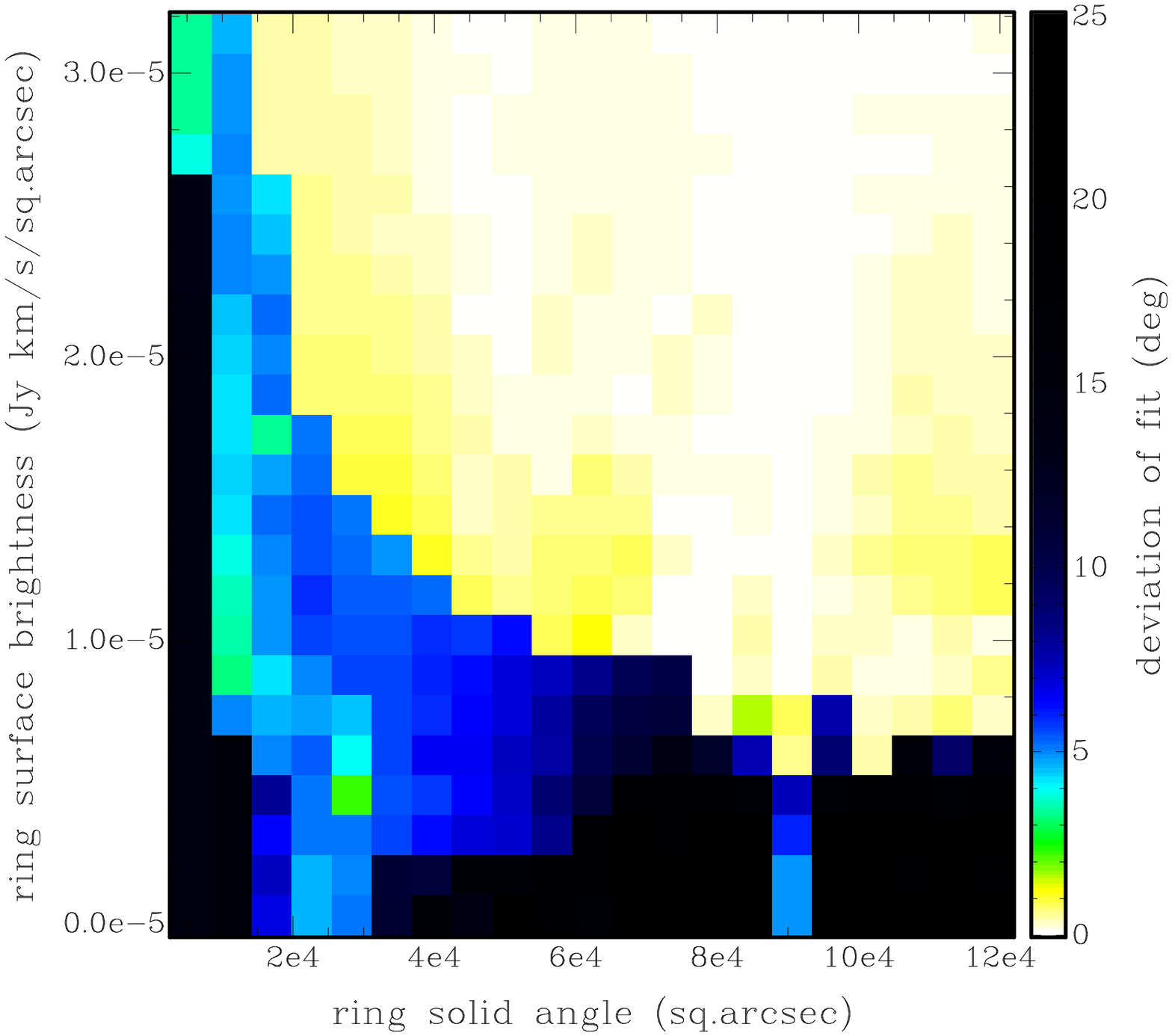}
  \end{center}
}
~~~~~~~~~~~~\parbox[c]{8.5cm}
{
    \begin{center}
\includegraphics[angle=270,width=8.5cm]{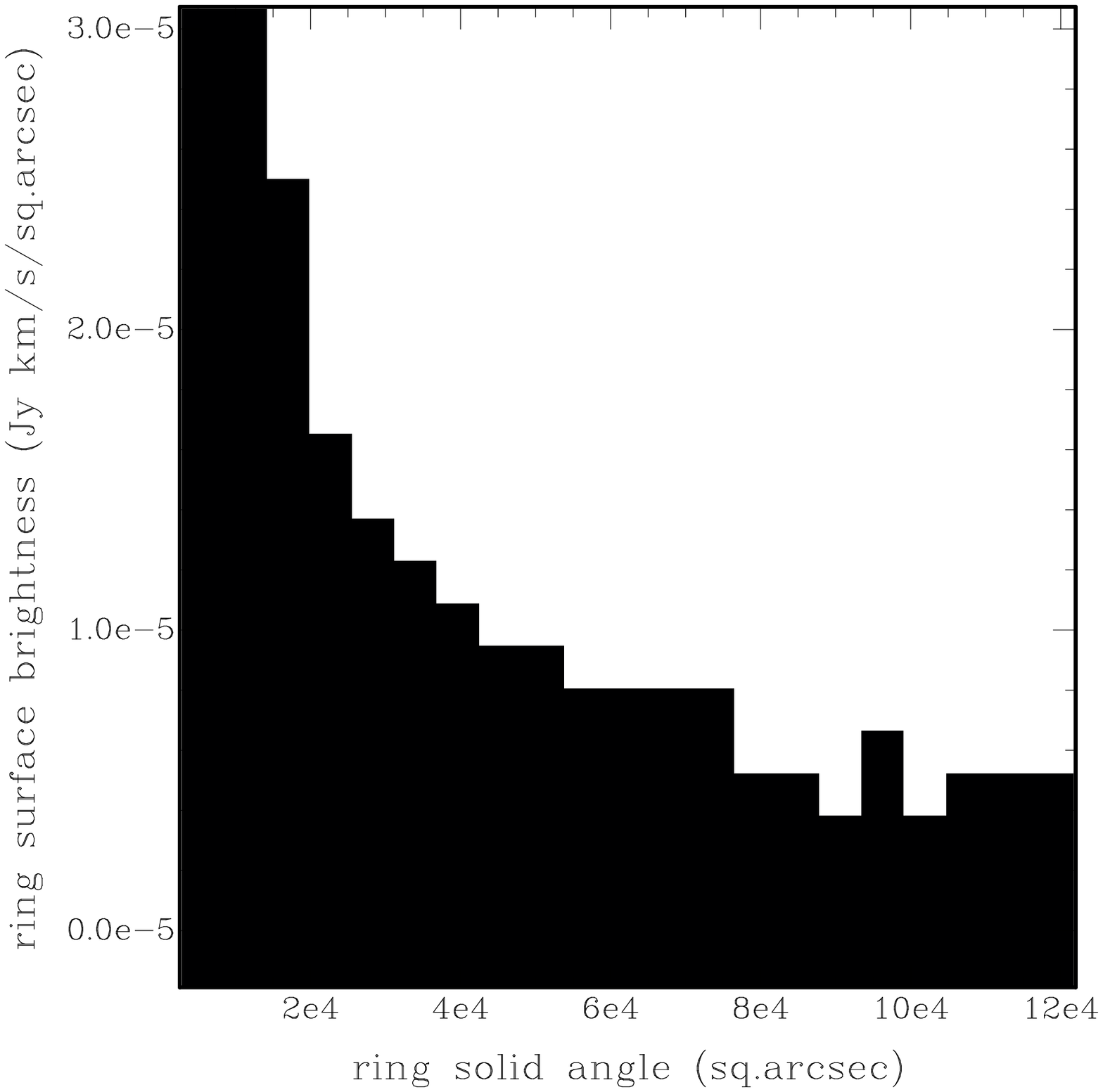}
    \end{center}
}
\caption{Testing the sensitivity of TiRiFiC. Each pixel represents a fit to a artificial observation (including realistic noise) of an inclined rotating ring with a given solid angle and surface brightness, the lowest row representing a fit to a data cube containing noise only. The pixel values represent the deviation of the fitted surface brightness (upper left panel), the rotation velocity (upper right panel), and the inclination (lower left panel) from the known true values. Clearly distinguishable are the regions where the fit succeeded (bright regions) and where the fit failed (dark regions). By setting a threshold a binary diagram can be generated (lower right panel), from which the detection limits in dependence of the ring surface solid angle can be estimated (Fig.~\ref{fig_sens}).}
\label{fig_senso}
\end{figure*}
\clearpage
\section{Testing TiRiFiC: individual tests}
\label{appendix_B}
TiRiFiC was tested by constructing a number of artificial observations of a
number of \ion{H}{i} disks of a galaxy at a distance of $4\,\rm
Mpc$. Here we show the results of the fits, which were performed by an
``observer'' using TiRiFiC. In addition, we present the results derived with ROTCUR on idealised velocity fields, which were
extracted from a artificial observation of the same disk \emph{with ten times
lower noise and ten times higher velocity resolution}. ROTCUR was run
on a peak-velocity field and a first-moment velocity field, the input guesses reported in the captions. The fit was performed using a free
angle of $45\deg$, fixing the expansion velocity of the rings to $0\,{\rm
km}\,{\rm s}^{-1}$. For
each test we provide a plot (left) containing the parametrisation
of the artificial disk (black dots and connecting black lines), the results
of the TiRiFiC fit (red triangles and connecting lines), and the
results of both ROTCUR fits (blue diamonds and connecting lines: fit
to first-moment velocity field, stars in magenta and connecting
lines: fit to the peak velocity field). From top to
bottom the panels contain: SD: \ion{H}{i} face-on column density and surface density in
${\rm atoms}\,{\rm cm}^{-2}$ and ${\rm M}_\odot\,{\rm pc}^{-2}$,
respectively. VROT: rotation velocity in ${\rm km}\,{\rm
s}^{-1}$. INCL: inclination in degrees and radians, respectively. PA:
position angle in degrees and radians. RA: the right ascension of the
central position in hh:mm:ss. RASH: the projected shift in right ascension in $\rm
kpc$ with respect to the centre of ring 1. DEC: the
declination of the central position in dd:mm:ss. DESH: the projected shift in
declination in $\rm kpc$ with respect to the centre of ring
1. VSYS: the systemic velocity in ${\rm km}\,{\rm s}^{-1}$. SCHT: the
scale height in arcseconds and $\rm pc$, respectively. The black
vertical line marks the end of the artificial disk, while the red vertical
line marks the end of the disk modelled by the observer. The blue
vertical line marks a radius of 1.5 HPBWs, beyond which TiRiFiC produces
reliable results in nearly all cases. The blue horizontal line marks
the detection limit for a ring of $15\arcsec$ width at a radius of
$200\arcsec$ as derived in Sect.~\ref{Sect_2.4.3a}. The right-hand
graph shows the deviations of the TiRiFiC fits from the model
parametrisations. Black dots and lines represent the deviation of the black
dots (the true parametrisation of the artificial galaxy) in the left
viewgraph from the red lines (the parametrisation fitted with TiRiFiC)
in the left panel, in red the deviation of the red
triangles (the parametrisation fitted with TiRiFiC) in the left
panel from the black lines (the true parametrisation of the artificial
galaxy) in the left panel. Blue diamonds show the deviation of the
ROTCUR fit to the first-moment velocity field from the parametrisation
of the artificial galaxy, stars in magenta show the deviation of the ROTCUR fit
to the peak-velocity field from the parametrisation of the artificial
galaxy. The vertical lines in the right-hand plot have the same
meaning as in the left-hand plot.

We analysed the quality of the TiRiFiC fits by means of
statistics. To this end, we extracted the deviations of the fits from the
parametrisation of the artificial galaxy and built the mean and rms, and
calculated the maximal error. The tables contain: \\
{\bf (1)} The
quantity listed. $\overline{\Delta}$: mean of
deviations. $\sqrt{\overline{\Delta^2}}$: rms of
deviations. $\Delta_{\rm max}$: Maximum deviation in the range
described by (2).\\ {\bf (2)} The range in which the deviations were
determined. all: taking all data points into account. 70\%: taking the
70\% lowest deviations into account. $n-\sigma$: taking those data
points into account where the expected surface brightness is below the
$n-\sigma$- level of the artificial observation. The $1-\sigma$ level is
determined by the face-on column density of a ring constructed as
described in Sect.~\ref{Sect_2.4.3a}, which in projection onto a data cube
contains no emission above the $1-\sigma_{\rm rms}$ level in the noisy
data cube ($\sigma = 2\cdot10^{19} \,{\rm atoms}\,{\rm
cm}^{-2}$). excs: only for the surface column density: deviations
where the expected surface column density is 0, i.e. where the artificial
disk has already ended.\\ {\bf (3)} Method to determine the
quantities. tir: use of TiRiFiC. m1: use of ROTCUR, fitting to the
first-moment velocity field. pk: use of ROTCUR, fitting to the peak-velocity field.\\ {\bf (4)} Error in face-on {\ion{H}{i}} column
density ($10^{18}\,{\rm atoms}\,{\rm cm}^{-2}$).\\ {\bf (5)} Ratio of
error in face-on {\ion{H}{i}} column density and expected face-on
{\ion{H}{i}} column density.\\ {\bf (6)} Error in rotation velocity
(${\rm km}\,{\rm s}^{-1}$).\\ {\bf (7)} Error in inclination
($\,\deg$).\\ {\bf (8)} Error in position angle ($\,\deg$).\\ {\bf
(9)} Error in right ascension of ring centre ($\,\arcsec$).\\ {\bf
(10)} Error in declination of ring centre ($\,\arcsec$).\\ {\bf (11)}
Error in systemic velocity (${\rm km}\,{\rm s}^{-1}$).\\ {\bf (12)}
Error in scale height ($\,\arcsec$).
\begin{figure*}
\parbox[c]{8.5cm}{
  \begin{center}
\includegraphics[width=8.5cm,height=22cm]{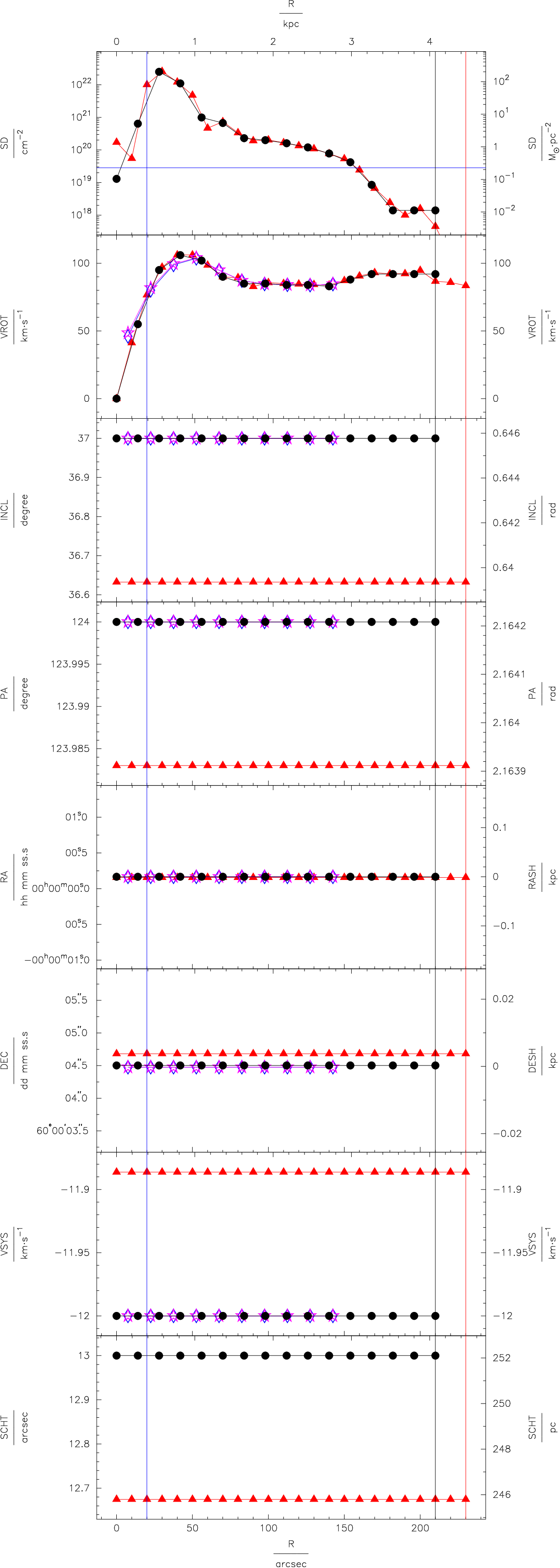}
  \end{center}
}
\parbox[c]{7.5cm} {
{
    \begin{center}
\includegraphics[width=8.5cm,height=22cm]{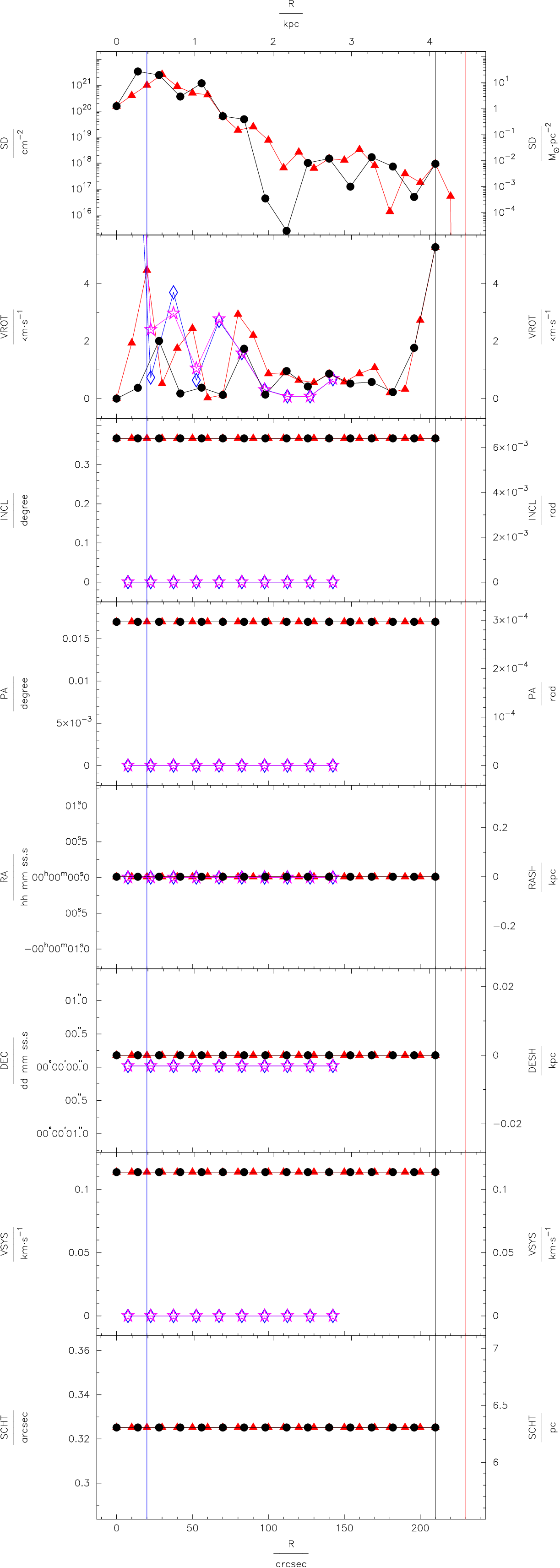}
    \end{center}
}
}
\input{fig_cap_09}
\end{figure*}\begin{center}
\begin{table*}

\caption{Test 1: Flat disk with constant scale height without noise. The
orientation parameters, centre, systemic velocity and scale height were fitted "as one" with TiRiFiC. Deviation in global dispersion: 0.01 ${\rm km}\,{\rm s}^{-1}$. ROTCUR: input model with optimal guesses. Only the rotation velocity was fitted.}
\begin{center}
\begin{tabular}{rrrrrrrrrrrr}
\hline
\hline
Quantity & Range & method & $ N_{\ion{H}{i}} $ & $ \frac{N_{\ion{H}{i}}}{N_{\ion{H}{i}},exp}$ & $v_{\rm rot}$ & $ i $ & $ pa $ & $ {\rm RA} $ & $ {\rm DEC} $ & $ v_{\rm sys} $ & $ z_0 $ \\
 (1)     & (2)   & (3) & (4)                & (5)                                          & (6)           & (7)   & (8)    & (9)          & (10)          & (11)            & (12)\\ 
\hline\\
$ \overline{\Delta} $ & all & tir &  257.32 &0.710 &1.4 &0.4 &0.0 &0.2 &0.2 &0.1 &0.3 \\
$\sqrt{\overline{\Delta^2}}$ & all & tir &  625.61 &2.610 &2.0 &0.4 &0.0 &0.2 &0.2 &0.1 &0.3 \\
$\Delta_{\rm max}$ & all & tir &  2629.47 &12.176 &5.3 &0.4 &0.0 &0.2 &0.2 &0.1 &0.3 \\
\hline
$ \overline{\Delta} $ & 70\% & tir &  4.00 &0.058 &0.6 &0.4 &0.0 &0.2 &0.2 &0.1 &0.3 \\
$\sqrt{\overline{\Delta^2}}$ & 70\% & tir &  8.17 &0.071 &0.8 &0.4 &0.0 &0.2 &0.2 &0.1 &0.3 \\
$\Delta_{\rm max}$ & 70\% & tir &  25.31 &0.117 &1.8 &0.4 &0.0 &0.2 &0.2 &0.1 &0.3 \\
\hline
$ \overline{\Delta} $ & hi & tir &  353.39 &0.088 &1.1 &0.4 &0.0 &0.2 &0.2 &0.1 &0.3 \\
$\sqrt{\overline{\Delta^2}}$ & hi & tir &  793.24 &0.149 &1.4 &0.4 &0.0 &0.2 &0.2 &0.1 &0.3 \\
$\Delta_{\rm max}$ & hi & tir &  2629.47 &0.481 &2.9 &0.4 &0.0 &0.2 &0.2 &0.1 &0.3 \\
\hline
$ \overline{\Delta} $ & $8-\sigma$ & tir &  0.99 &0.140 &1.3 &0.4 &0.0 &0.2 &0.2 &0.1 &0.3 \\
$\sqrt{\overline{\Delta^2}}$ & $8-\sigma$ & tir &  1.43 &0.242 &2.0 &0.4 &0.0 &0.2 &0.2 &0.1 &0.3 \\
$\Delta_{\rm max}$ & $8-\sigma$ & tir &  3.35 &0.678 &5.3 &0.4 &0.0 &0.2 &0.2 &0.1 &0.3 \\
\hline
$ \overline{\Delta} $ & $2-\sigma$ & tir &  0.72 &0.221 &1.7 &0.4 &0.0 &0.2 &0.2 &0.1 &0.3 \\
$\sqrt{\overline{\Delta^2}}$ & $2-\sigma$ & tir &  1.27 &0.312 &2.5 &0.4 &0.0 &0.2 &0.2 &0.1 &0.3 \\
$\Delta_{\rm max}$ & $2-\sigma$ & tir &  3.35 &0.678 &5.3 &0.4 &0.0 &0.2 &0.2 &0.1 &0.3 \\
\hline
$ \overline{\Delta} $ & $0.5-\sigma$ & tir &  0.34 &0.241 &1.9 &0.4 &0.0 &0.2 &0.2 &0.1 &0.3 \\
$\sqrt{\overline{\Delta^2}}$ & $0.5-\sigma$ & tir &  0.50 &0.337 &2.7 &0.4 &0.0 &0.2 &0.2 &0.1 &0.3 \\
$\Delta_{\rm max}$ & $0.5-\sigma$ & tir &  0.95 &0.678 &5.3 &0.4 &0.0 &0.2 &0.2 &0.1 &0.3 \\
\hline
$ \overline{\Delta} $ & excs  & tir &  0.03 \\
$\sqrt{\overline{\Delta^2}}$ & excs & tir &  0.04 \\
$\Delta_{\rm max}$ & excs & tir &  0.05 \\
\hline
$ \overline{\Delta} $ & all & pk &  & & 3.1 &0.0 &0.0 &0.0 &0.0 &0.0 & \\
$\sqrt{\overline{\Delta^2}}$ & all & pk &  & & 6.2 &0.0 &0.0 &0.0 &0.0 &0.0 & \\
$\Delta_{\rm max}$ & all & pk &  & & 0.0 &0.0 &0.0 &0.0 &0.0 &0.0 & \\
\hline
$ \overline{\Delta} $ & 70\% & pk &  & & 0.9 &0.0 &0.0 &0.0 &0.0 &0.0 & \\
$\sqrt{\overline{\Delta^2}}$ & 70\% & pk &  & & 1.2 &0.0 &0.0 &0.0 &0.0 &0.0 & \\
$\Delta_{\rm max}$ & 70\% & pk &  & & 2.4 &0.0 &0.0 &0.0 &0.0 &0.0 & \\
\hline
$ \overline{\Delta} $ & hi & pk &  & & 1.3 &0.0 &0.0 &0.0 &0.0 &0.0 & \\
$\sqrt{\overline{\Delta^2}}$ & hi & pk &  & & 1.7 &0.0 &0.0 &0.0 &0.0 &0.0 & \\
$\Delta_{\rm max}$ & hi & pk &  & & 3.0 &0.0 &0.0 &0.0 &0.0 &0.0 & \\
\hline
$ \overline{\Delta} $ & all & m1 &  & & 2.7 &0.0 &0.0 &0.0 &0.0 &0.0 & \\
$\sqrt{\overline{\Delta^2}}$ & all & m1 &  & & 5.5 &0.0 &0.0 &0.0 &0.0 &0.0 & \\
$\Delta_{\rm max}$ & all & m1 &  & & 0.0 &0.0 &0.0 &0.0 &0.0 &0.0 & \\
\hline
$ \overline{\Delta} $ & 70\% & m1 &  & & 0.6 &0.0 &0.0 &0.0 &0.0 &0.0 & \\
$\sqrt{\overline{\Delta^2}}$ & 70\% & m1 &  & & 0.8 &0.0 &0.0 &0.0 &0.0 &0.0 & \\
$\Delta_{\rm max}$ & 70\% & m1 &  & & 1.6 &0.0 &0.0 &0.0 &0.0 &0.0 & \\
\hline
$ \overline{\Delta} $ & hi & m1 &  & & 1.2 &0.0 &0.0 &0.0 &0.0 &0.0 & \\
$\sqrt{\overline{\Delta^2}}$ & hi & m1 &  & & 1.7 &0.0 &0.0 &0.0 &0.0 &0.0 & \\
$\Delta_{\rm max}$ & hi & m1 &  & & 3.7 &0.0 &0.0 &0.0 &0.0 &0.0 & \\
\hline
\hline
\end{tabular}
\end{center}
\end{table*}
\end{center}
\clearpage
\begin{figure*}
\parbox[c]{8.5cm}{
  \begin{center}
\includegraphics[width=8.5cm,height=22cm]{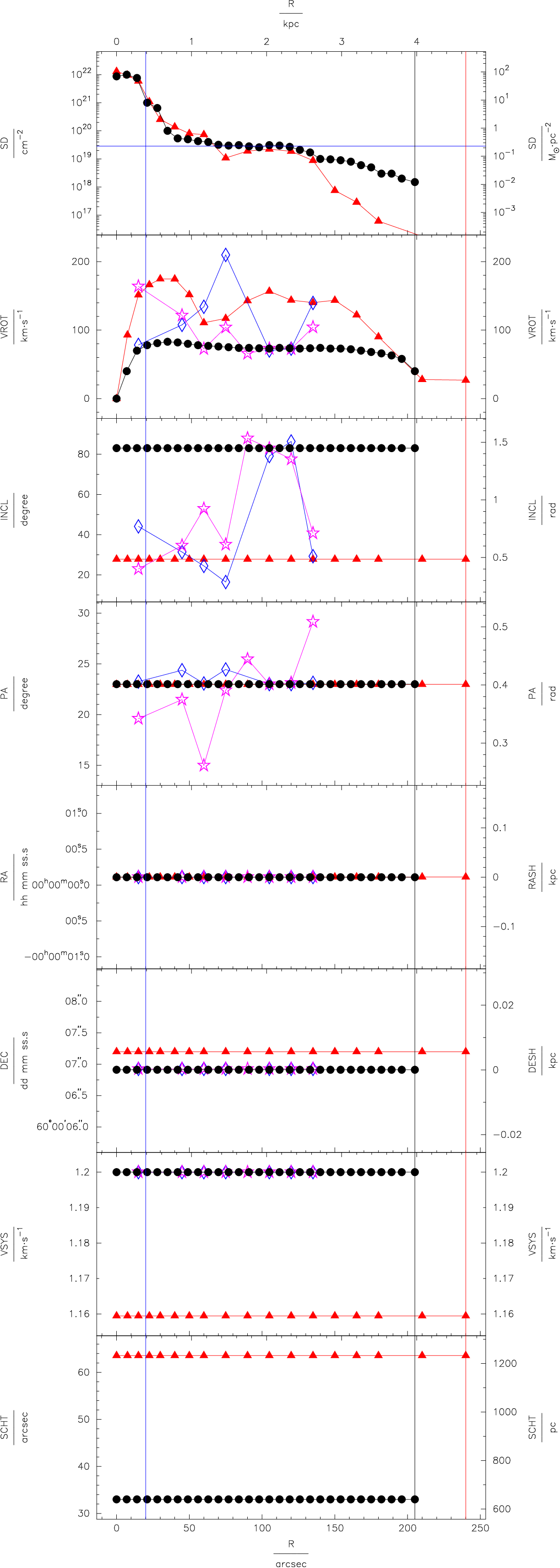}
  \end{center}
}
\parbox[c]{7.5cm} {
{
    \begin{center}
\includegraphics[width=8.5cm,height=22cm]{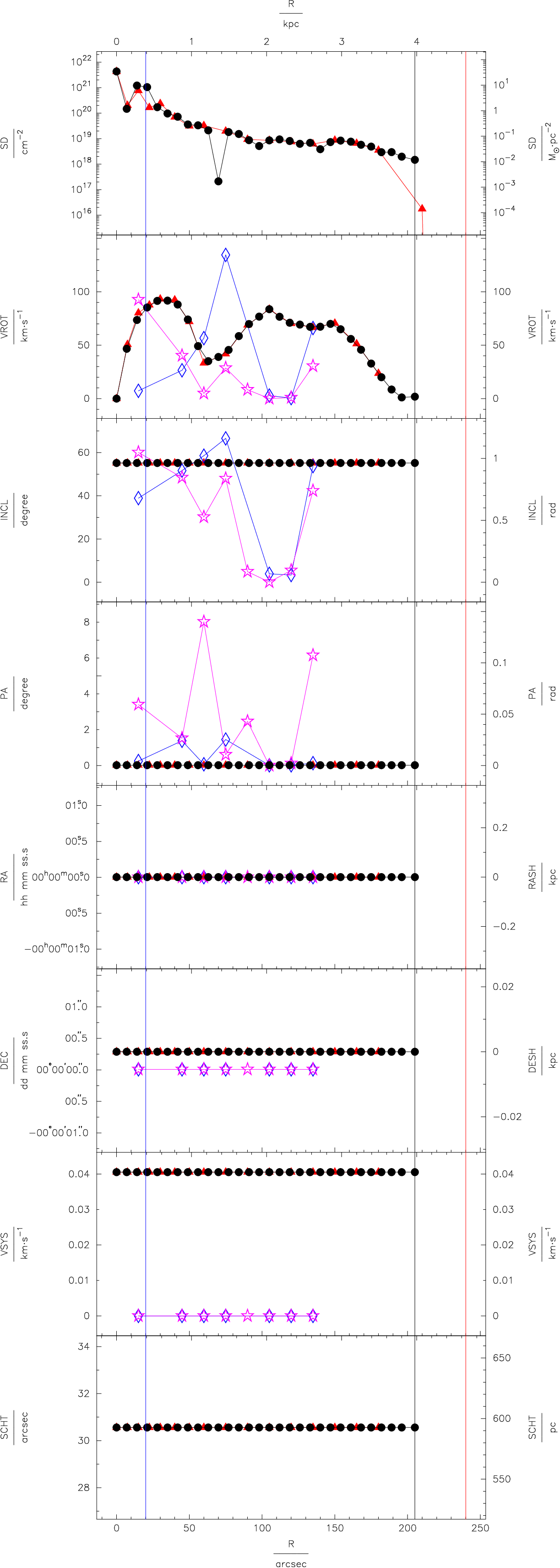}
    \end{center}
}
}
\input{fig_cap_10}
\label{Fig02_add}
\end{figure*}\begin{center}
\begin{table*}

\caption{Test 2: Flat disk with constant large scale height; noise added to the cube. Orientation parameters, centre, systemic velocity and scale height fitted "as one" with TiRiFiC. In a first attempt the observer failed to achieve a good fit. The large scale height enhances drastically the ambiguity of inclination and rotation velocity. Deviation in global dispersion: 0.57 ${\rm km}\,{\rm s}^{-1}$. ROTCUR: input model with optimal guesses. Only the rotation velocity and the orientation parameters were fitted.
}

\begin{center}
\begin{tabular}{rrrrrrrrrrrr}
\hline
\hline
Quantity & Range & method & $ N_{\ion{H}{i}} $ & $ \frac{N_{\ion{H}{i}}}{N_{\ion{H}{i}},exp}$ & $v_{\rm rot}$ & $ i $ & $ pa $ & $ {\rm RA} $ & $ {\rm DEC} $ & $ v_{\rm sys} $ & $ z_0 $ \\
 (1)     & (2)   & (3) & (4)                & (5)                                          & (6)           & (7)   & (8)    & (9)          & (10)          & (11)            & (12)\\ 
\hline\\
$ \overline{\Delta} $ & all & tir &  327.61 &0.536 &61.6 &55.2 &0.0 &0.1 &0.3 &0.0 &30.6 \\
$\sqrt{\overline{\Delta^2}}$ & all & tir &  1038.29 &0.624 &66.7 &55.2 &0.0 &0.1 &0.3 &0.0 &30.6 \\
$\Delta_{\rm max}$ & all & tir &  4323.01 &1.052 &93.2 &55.2 &0.0 &0.1 &0.3 &0.0 &30.6 \\
\hline
$ \overline{\Delta} $ & 70\% & tir &  11.14 &0.353 &49.9 &55.2 &0.0 &0.1 &0.3 &0.0 &30.6 \\
$\sqrt{\overline{\Delta^2}}$ & 70\% & tir &  15.28 &0.403 &54.7 &55.2 &0.0 &0.1 &0.3 &0.0 &30.6 \\
$\Delta_{\rm max}$ & 70\% & tir &  32.42 &0.646 &72.0 &55.2 &0.0 &0.1 &0.3 &0.0 &30.6 \\
\hline
$ \overline{\Delta} $ & hi & tir &  80.80 &0.579 &72.0 &55.2 &0.0 &0.1 &0.3 &0.0 &30.6 \\
$\sqrt{\overline{\Delta^2}}$ & hi & tir &  114.15 &0.642 &75.6 &55.2 &0.0 &0.1 &0.3 &0.0 &30.6 \\
$\Delta_{\rm max}$ & hi & tir &  237.01 &1.052 &93.2 &55.2 &0.0 &0.1 &0.3 &0.0 &30.6 \\
\hline
$ \overline{\Delta} $ & $8-\sigma$ & tir &  15.72 &0.662 &61.3 &55.2 &0.0 &0.1 &0.3 &0.0 &30.6 \\
$\sqrt{\overline{\Delta^2}}$ & $8-\sigma$ & tir &  24.48 &0.721 &64.6 &55.2 &0.0 &0.1 &0.3 &0.0 &30.6 \\
$\Delta_{\rm max}$ & $8-\sigma$ & tir &  70.63 &1.052 &92.4 &55.2 &0.0 &0.1 &0.3 &0.0 &30.6 \\
\hline
$ \overline{\Delta} $ & $2-\sigma$ & tir &  7.01 &0.601 &59.6 &55.2 &0.0 &0.1 &0.3 &0.0 &30.6 \\
$\sqrt{\overline{\Delta^2}}$ & $2-\sigma$ & tir &  8.80 &0.669 &62.3 &55.2 &0.0 &0.1 &0.3 &0.0 &30.6 \\
$\Delta_{\rm max}$ & $2-\sigma$ & tir &  19.74 &0.983 &83.7 &55.2 &0.0 &0.1 &0.3 &0.0 &30.6 \\
\hline
$ \overline{\Delta} $ & $0.5-\sigma$ & tir &  3.75 &0.953 &48.5 &55.2 &0.0 &0.1 &0.3 &0.0 &30.6 \\
$\sqrt{\overline{\Delta^2}}$ & $0.5-\sigma$ & tir &  5.10 &0.954 &52.2 &55.2 &0.0 &0.1 &0.3 &0.0 &30.6 \\
$\Delta_{\rm max}$ & $0.5-\sigma$ & tir &  8.64 &0.983 &70.6 &55.2 &0.0 &0.1 &0.3 &0.0 &30.6 \\
\hline
$ \overline{\Delta} $ & excs  & tir &  0.01 \\
$\sqrt{\overline{\Delta^2}}$ & excs & tir &  0.01 \\
$\Delta_{\rm max}$ & excs & tir &  0.02 \\
\hline
$ \overline{\Delta} $ & all & pk &  & & 25.7 &29.9 &2.8 &0.0 &0.0 &0.0 & \\
$\sqrt{\overline{\Delta^2}}$ & all & pk &  & & 38.8 &37.1 &3.9 &0.0 &0.0 &0.0 & \\
$\Delta_{\rm max}$ & all & pk &  & & 92.6 &60.0 &8.0 &0.0 &0.0 &0.0 & \\
\hline
$ \overline{\Delta} $ & 70\% & pk &  & & 8.5 &16.6 &0.9 &0.0 &0.0 &0.0 & \\
$\sqrt{\overline{\Delta^2}}$ & 70\% & pk &  & & 13.5 &23.5 &1.3 &0.0 &0.0 &0.0 & \\
$\Delta_{\rm max}$ & 70\% & pk &  & & 28.6 &42.3 &2.5 &0.0 &0.0 &0.0 & \\
\hline
$ \overline{\Delta} $ & hi & pk &  & & 18.4 &31.7 &2.5 &0.0 &0.0 &0.0 & \\
$\sqrt{\overline{\Delta^2}}$ & hi & pk &  & & 24.8 &37.3 &4.1 &0.0 &0.0 &0.0 & \\
$\Delta_{\rm max}$ & hi & pk &  & & 40.2 &48.4 &8.0 &0.0 &0.0 &0.0 & \\
\hline
$ \overline{\Delta} $ & all & m1 &  & & 42.0 &39.5 &0.5 &0.0 &0.0 &0.0 & \\
$\sqrt{\overline{\Delta^2}}$ & all & m1 &  & & 61.5 &46.2 &0.8 &0.0 &0.0 &0.0 & \\
$\Delta_{\rm max}$ & all & m1 &  & & 134.5 &66.6 &1.4 &0.0 &0.0 &0.0 & \\
\hline
$ \overline{\Delta} $ & 70\% & m1 &  & & 9.3 &24.5 &0.1 &0.0 &0.0 &0.0 & \\
$\sqrt{\overline{\Delta^2}}$ & 70\% & m1 &  & & 13.8 &32.4 &0.1 &0.0 &0.0 &0.0 & \\
$\Delta_{\rm max}$ & 70\% & m1 &  & & 26.5 &51.6 &0.1 &0.0 &0.0 &0.0 & \\
\hline
$ \overline{\Delta} $ & hi & m1 &  & & 55.1 &45.2 &0.7 &0.0 &0.0 &0.0 & \\
$\sqrt{\overline{\Delta^2}}$ & hi & m1 &  & & 74.2 &51.3 &1.0 &0.0 &0.0 &0.0 & \\
$\Delta_{\rm max}$ & hi & m1 &  & & 134.5 &66.6 &1.4 &0.0 &0.0 &0.0 & \\
\hline
\hline
\end{tabular}
\end{center}
\end{table*}
\end{center}
\clearpage
\begin{figure*}
\parbox[c]{8.5cm}{
  \begin{center}
\includegraphics[width=8.5cm,height=22cm]{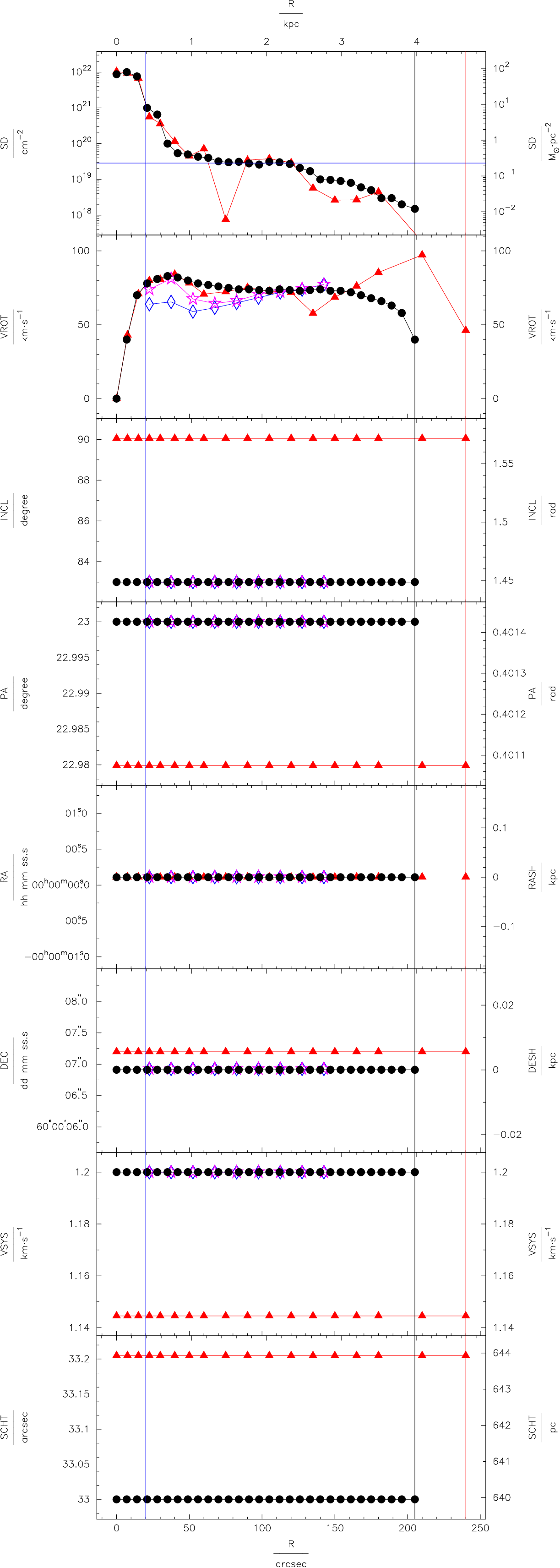}
  \end{center}
}
\parbox[c]{7.5cm} {
{
    \begin{center}
\includegraphics[width=8.5cm,height=22cm]{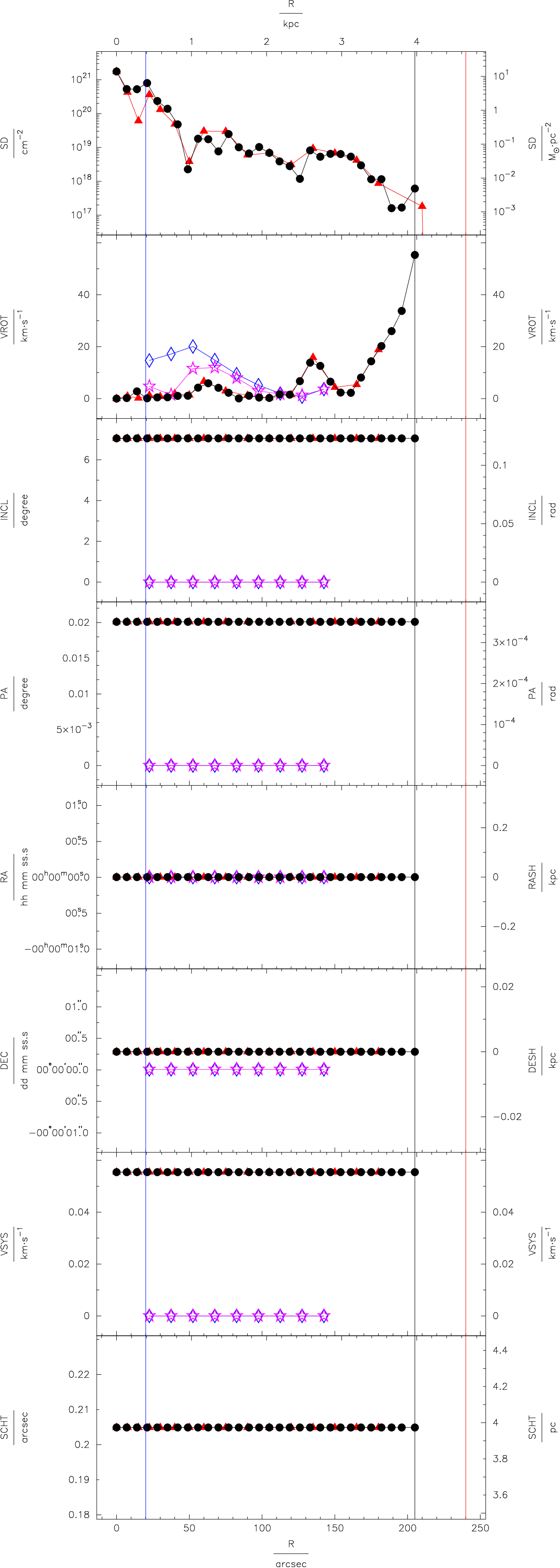}
    \end{center}
}
}
\input{fig_cap_11}
\label{Fig02b_add}
\end{figure*}\begin{center}
\begin{table*}

\caption{Test 2, revised: same as 2, but this time started tear to the correct result. The observer achieved a reasonable result, while TiRiFiC approached the $\chi^2$ minimum in parameter space very slowly. It is to be hoped that this problem will vanish after the implementation of a more effective $\chi^2$ minimiser. Deviation in global dispersion: 0.39 ${\rm km}\,{\rm s}^{-1}$. ROTCUR: input model with optimal guesses. Only the rotation velocity was fitted.
}

\begin{center}
\begin{tabular}{rrrrrrrrrrrr}
\hline
\hline
Quantity & Range & method & $ N_{\ion{H}{i}} $ & $ \frac{N_{\ion{H}{i}}}{N_{\ion{H}{i}},exp}$ & $v_{\rm rot}$ & $ i $ & $ pa $ & $ {\rm RA} $ & $ {\rm DEC} $ & $ v_{\rm sys} $ & $ z_0 $ \\
 (1)     & (2)   & (3) & (4)                & (5)                                          & (6)           & (7)   & (8)    & (9)          & (10)          & (11)            & (12)\\ 
\hline\\
$ \overline{\Delta} $ & all & tir &  160.22 &0.386 &4.0 &7.1 &0.0 &0.1 &0.3 &0.1 &0.2 \\
$\sqrt{\overline{\Delta^2}}$ & all & tir &  435.21 &0.483 &6.7 &7.1 &0.0 &0.1 &0.3 &0.1 &0.2 \\
$\Delta_{\rm max}$ & all & tir &  1752.18 &0.975 &18.8 &7.1 &0.0 &0.1 &0.3 &0.1 &0.2 \\
\hline
$ \overline{\Delta} $ & 70\% & tir &  8.43 &0.217 &1.1 &7.1 &0.0 &0.1 &0.3 &0.1 &0.2 \\
$\sqrt{\overline{\Delta^2}}$ & 70\% & tir &  13.12 &0.271 &1.4 &7.1 &0.0 &0.1 &0.3 &0.1 &0.2 \\
$\Delta_{\rm max}$ & 70\% & tir &  30.33 &0.608 &2.9 &7.1 &0.0 &0.1 &0.3 &0.1 &0.2 \\
\hline
$ \overline{\Delta} $ & hi & tir &  87.62 &0.485 &2.2 &7.1 &0.0 &0.1 &0.3 &0.1 &0.2 \\
$\sqrt{\overline{\Delta^2}}$ & hi & tir &  147.64 &0.573 &2.9 &7.1 &0.0 &0.1 &0.3 &0.1 &0.2 \\
$\Delta_{\rm max}$ & hi & tir &  361.90 &0.975 &6.6 &7.1 &0.0 &0.1 &0.3 &0.1 &0.2 \\
\hline
$ \overline{\Delta} $ & $8-\sigma$ & tir &  11.54 &0.478 &5.5 &7.1 &0.0 &0.1 &0.3 &0.1 &0.2 \\
$\sqrt{\overline{\Delta^2}}$ & $8-\sigma$ & tir &  18.51 &0.561 &8.1 &7.1 &0.0 &0.1 &0.3 &0.1 &0.2 \\
$\Delta_{\rm max}$ & $8-\sigma$ & tir &  48.88 &0.975 &18.8 &7.1 &0.0 &0.1 &0.3 &0.1 &0.2 \\
\hline
$ \overline{\Delta} $ & $2-\sigma$ & tir &  6.70 &0.464 &6.3 &7.1 &0.0 &0.1 &0.3 &0.1 &0.2 \\
$\sqrt{\overline{\Delta^2}}$ & $2-\sigma$ & tir &  10.64 &0.546 &9.1 &7.1 &0.0 &0.1 &0.3 &0.1 &0.2 \\
$\Delta_{\rm max}$ & $2-\sigma$ & tir &  29.82 &0.975 &18.8 &7.1 &0.0 &0.1 &0.3 &0.1 &0.2 \\
\hline
$ \overline{\Delta} $ & $0.5-\sigma$ & tir &  2.40 &0.523 &9.5 &7.1 &0.0 &0.1 &0.3 &0.1 &0.2 \\
$\sqrt{\overline{\Delta^2}}$ & $0.5-\sigma$ & tir &  3.58 &0.562 &11.6 &7.1 &0.0 &0.1 &0.3 &0.1 &0.2 \\
$\Delta_{\rm max}$ & $0.5-\sigma$ & tir &  6.77 &0.720 &18.8 &7.1 &0.0 &0.1 &0.3 &0.1 &0.2 \\
\hline
$ \overline{\Delta} $ & excs  & tir &  0.09 \\
$\sqrt{\overline{\Delta^2}}$ & excs & tir &  0.13 \\
$\Delta_{\rm max}$ & excs & tir &  0.18 \\
\hline
$ \overline{\Delta} $ & all & pk &  & & 5.2 &0.0 &0.0 &0.0 &0.0 &0.0 & \\
$\sqrt{\overline{\Delta^2}}$ & all & pk &  & & 6.6 &0.0 &0.0 &0.0 &0.0 &0.0 & \\
$\Delta_{\rm max}$ & all & pk &  & & 0.0 &0.0 &0.0 &0.0 &0.0 &0.0 & \\
\hline
$ \overline{\Delta} $ & 70\% & pk &  & & 2.6 &0.0 &0.0 &0.0 &0.0 &0.0 & \\
$\sqrt{\overline{\Delta^2}}$ & 70\% & pk &  & & 2.9 &0.0 &0.0 &0.0 &0.0 &0.0 & \\
$\Delta_{\rm max}$ & 70\% & pk &  & & 4.7 &0.0 &0.0 &0.0 &0.0 &0.0 & \\
\hline
$ \overline{\Delta} $ & hi & pk &  & & 6.6 &0.0 &0.0 &0.0 &0.0 &0.0 & \\
$\sqrt{\overline{\Delta^2}}$ & hi & pk &  & & 7.8 &0.0 &0.0 &0.0 &0.0 &0.0 & \\
$\Delta_{\rm max}$ & hi & pk &  & & 12.0 &0.0 &0.0 &0.0 &0.0 &0.0 & \\
\hline
$ \overline{\Delta} $ & all & m1 &  & & 9.7 &0.0 &0.0 &0.0 &0.0 &0.0 & \\
$\sqrt{\overline{\Delta^2}}$ & all & m1 &  & & 11.9 &0.0 &0.0 &0.0 &0.0 &0.0 & \\
$\Delta_{\rm max}$ & all & m1 &  & & 0.0 &0.0 &0.0 &0.0 &0.0 &0.0 & \\
\hline
$ \overline{\Delta} $ & 70\% & m1 &  & & 5.9 &0.0 &0.0 &0.0 &0.0 &0.0 & \\
$\sqrt{\overline{\Delta^2}}$ & 70\% & m1 &  & & 7.6 &0.0 &0.0 &0.0 &0.0 &0.0 & \\
$\Delta_{\rm max}$ & 70\% & m1 &  & & 14.7 &0.0 &0.0 &0.0 &0.0 &0.0 & \\
\hline
$ \overline{\Delta} $ & hi & m1 &  & & 13.0 &0.0 &0.0 &0.0 &0.0 &0.0 & \\
$\sqrt{\overline{\Delta^2}}$ & hi & m1 &  & & 14.3 &0.0 &0.0 &0.0 &0.0 &0.0 & \\
$\Delta_{\rm max}$ & hi & m1 &  & & 20.1 &0.0 &0.0 &0.0 &0.0 &0.0 & \\
\hline
\hline
\end{tabular}
\end{center}
\end{table*}
\end{center}
\clearpage
\begin{figure*}
\parbox[c]{8.5cm}{
  \begin{center}
\includegraphics[width=8.5cm,height=22cm]{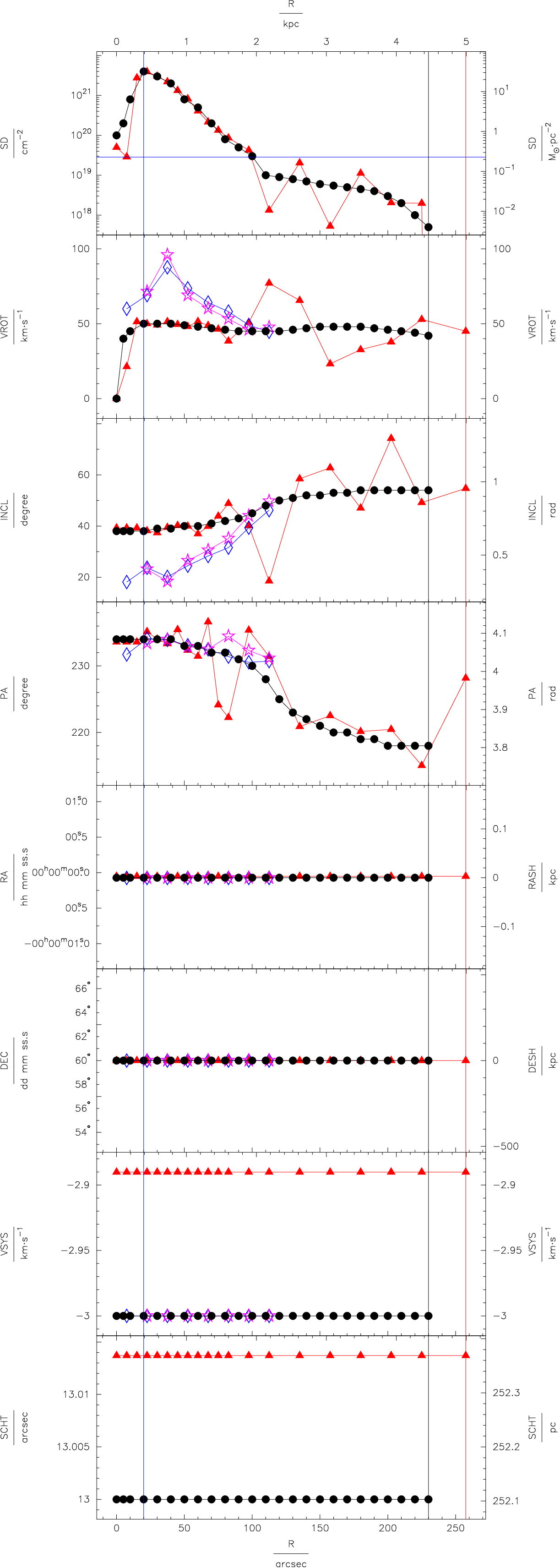}
  \end{center}
}
\parbox[c]{7.5cm} {
{
    \begin{center}
\includegraphics[width=8.5cm,height=22cm]{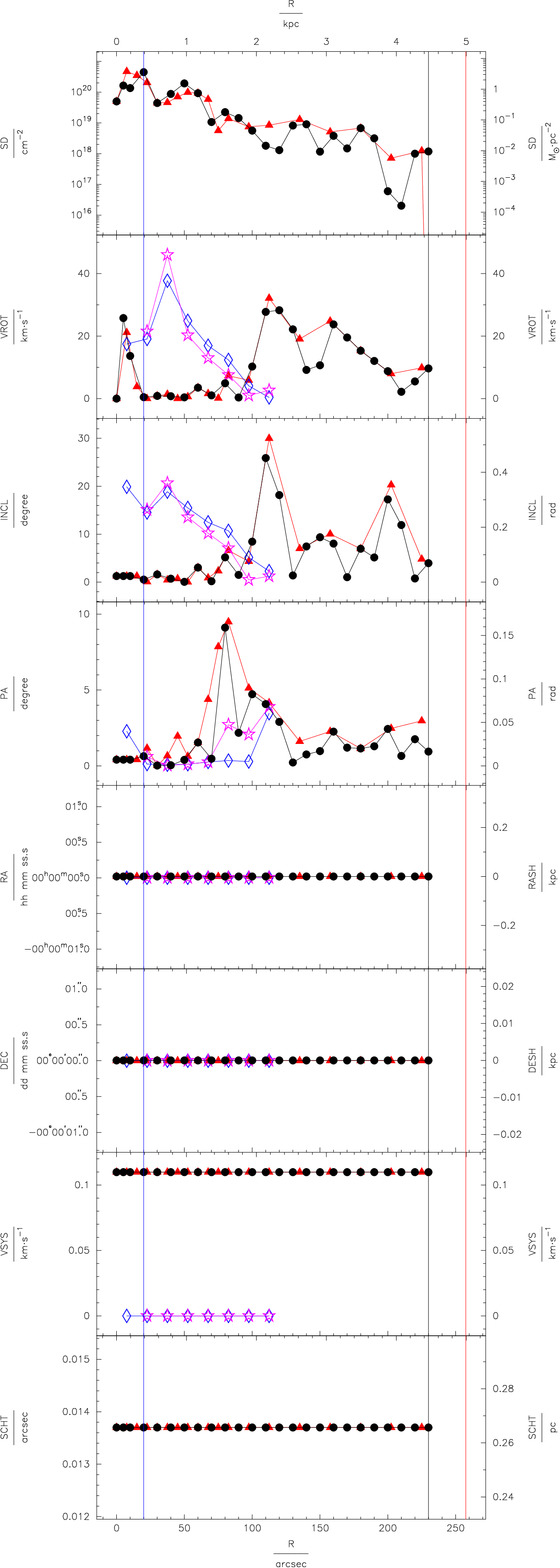}
    \end{center}
}
}
\input{fig_cap_12}
\end{figure*}\begin{center}
\begin{table*}

\caption{Test 3: symmetric warp without central shift and without shift in systemic velocity. Noise was added to the data cube. Deviation in global dispersion: 0.03 ${\rm km}\,{\rm s}^{-1}$. ROTCUR: input model with optimal guesses. Only the rotation velocity and the orientation parameters were fitted.
}

\begin{center}
\begin{tabular}{rrrrrrrrrrrr}
\hline
\hline
Quantity & Range & method & $ N_{\ion{H}{i}} $ & $ \frac{N_{\ion{H}{i}}}{N_{\ion{H}{i}},exp}$ & $v_{\rm rot}$ & $ i $ & $ pa $ & $ {\rm RA} $ & $ {\rm DEC} $ & $ v_{\rm sys} $ & $ z_0 $ \\
 (1)     & (2)   & (3) & (4)                & (5)                                          & (6)           & (7)   & (8)    & (9)          & (10)          & (11)            & (12)\\ 
\hline\\
$ \overline{\Delta} $ & all & tir &  77.34 &0.505 &8.2 &5.4 &2.6 &0.3 &0.0 &0.1 &0.0 \\
$\sqrt{\overline{\Delta^2}}$ & all & tir &  144.86 &0.757 &12.5 &9.2 &3.6 &0.3 &0.0 &0.1 &0.0 \\
$\Delta_{\rm max}$ & all & tir &  471.18 &1.724 &32.1 &30.0 &9.5 &0.3 &0.0 &0.1 &0.0 \\
\hline
$ \overline{\Delta} $ & 70\% & tir &  18.61 &0.155 &2.5 &1.7 &1.1 &0.3 &0.0 &0.1 &0.0 \\
$\sqrt{\overline{\Delta^2}}$ & 70\% & tir &  27.45 &0.201 &3.8 &2.2 &1.4 &0.3 &0.0 &0.1 &0.0 \\
$\Delta_{\rm max}$ & 70\% & tir &  58.22 &0.499 &7.9 &4.8 &2.5 &0.3 &0.0 &0.1 &0.0 \\
\hline
$ \overline{\Delta} $ & hi & tir &  64.09 &0.112 &2.1 &2.0 &3.3 &0.3 &0.0 &0.1 &0.0 \\
$\sqrt{\overline{\Delta^2}}$ & hi & tir &  85.23 &0.137 &3.2 &2.8 &4.5 &0.3 &0.0 &0.1 &0.0 \\
$\Delta_{\rm max}$ & hi & tir &  204.22 &0.213 &7.3 &6.6 &9.5 &0.3 &0.0 &0.1 &0.0 \\
\hline
$ \overline{\Delta} $ & $8-\sigma$ & tir &  6.20 &0.815 &13.6 &10.3 &4.1 &0.3 &0.0 &0.1 &0.0 \\
$\sqrt{\overline{\Delta^2}}$ & $8-\sigma$ & tir &  7.68 &1.035 &16.6 &13.3 &4.9 &0.3 &0.0 &0.1 &0.0 \\
$\Delta_{\rm max}$ & $8-\sigma$ & tir &  13.83 &1.724 &32.1 &30.0 &9.5 &0.3 &0.0 &0.1 &0.0 \\
\hline
$ \overline{\Delta} $ & $2-\sigma$ & tir &  5.32 &1.015 &16.4 &11.9 &2.8 &0.3 &0.0 &0.1 &0.0 \\
$\sqrt{\overline{\Delta^2}}$ & $2-\sigma$ & tir &  6.77 &1.171 &18.7 &14.9 &3.1 &0.3 &0.0 &0.1 &0.0 \\
$\Delta_{\rm max}$ & $2-\sigma$ & tir &  12.93 &1.724 &32.1 &30.0 &5.1 &0.3 &0.0 &0.1 &0.0 \\
\hline
$ \overline{\Delta} $ & $0.5-\sigma$ & tir &  5.02 &1.149 &18.2 &13.2 &2.4 &0.3 &0.0 &0.1 &0.0 \\
$\sqrt{\overline{\Delta^2}}$ & $0.5-\sigma$ & tir &  6.67 &1.262 &20.0 &16.0 &2.6 &0.3 &0.0 &0.1 &0.0 \\
$\Delta_{\rm max}$ & $0.5-\sigma$ & tir &  12.93 &1.724 &32.1 &30.0 &4.2 &0.3 &0.0 &0.1 &0.0 \\
\hline
$ \overline{\Delta} $ & excs  & tir &  0.00 \\
$\sqrt{\overline{\Delta^2}}$ & excs & tir &  0.00 \\
$\Delta_{\rm max}$ & excs & tir &  0.00 \\
\hline
$ \overline{\Delta} $ & all & pk &  & & 16.0 &9.8 &1.4 &0.0 &0.0 &0.0 & \\
$\sqrt{\overline{\Delta^2}}$ & all & pk &  & & 21.5 &11.9 &2.0 &0.0 &0.0 &0.0 & \\
$\Delta_{\rm max}$ & all & pk &  & & 21.5 &20.7 &3.9 &0.0 &0.0 &0.0 & \\
\hline
$ \overline{\Delta} $ & 70\% & pk &  & & 6.0 &4.7 &0.2 &0.0 &0.0 &0.0 & \\
$\sqrt{\overline{\Delta^2}}$ & 70\% & pk &  & & 7.6 &6.2 &0.3 &0.0 &0.0 &0.0 & \\
$\Delta_{\rm max}$ & 70\% & pk &  & & 13.0 &10.2 &0.6 &0.0 &0.0 &0.0 & \\
\hline
$ \overline{\Delta} $ & hi & pk &  & & 18.2 &11.2 &1.0 &0.0 &0.0 &0.0 & \\
$\sqrt{\overline{\Delta^2}}$ & hi & pk &  & & 23.1 &12.9 &1.4 &0.0 &0.0 &0.0 & \\
$\Delta_{\rm max}$ & hi & pk &  & & 46.0 &20.7 &2.7 &0.0 &0.0 &0.0 & \\
\hline
$ \overline{\Delta} $ & all & m1 &  & & 16.6 &12.4 &0.9 &0.0 &0.0 &0.0 & \\
$\sqrt{\overline{\Delta^2}}$ & all & m1 &  & & 19.9 &13.7 &1.5 &0.0 &0.0 &0.0 & \\
$\Delta_{\rm max}$ & all & m1 &  & & 37.7 &19.9 &3.5 &0.0 &0.0 &0.0 & \\
\hline
$ \overline{\Delta} $ & 70\% & m1 &  & & 10.2 &9.0 &0.2 &0.0 &0.0 &0.0 & \\
$\sqrt{\overline{\Delta^2}}$ & 70\% & m1 &  & & 12.3 &10.1 &0.2 &0.0 &0.0 &0.0 & \\
$\Delta_{\rm max}$ & 70\% & m1 &  & & 17.5 &14.5 &0.3 &0.0 &0.0 &0.0 & \\
\hline
$ \overline{\Delta} $ & hi & m1 &  & & 19.2 &12.9 &0.2 &0.0 &0.0 &0.0 & \\
$\sqrt{\overline{\Delta^2}}$ & hi & m1 &  & & 21.8 &13.6 &0.2 &0.0 &0.0 &0.0 & \\
$\Delta_{\rm max}$ & hi & m1 &  & & 37.7 &18.9 &0.3 &0.0 &0.0 &0.0 & \\
\hline
\hline
\end{tabular}
\end{center}
\end{table*}
\end{center}
\clearpage
\begin{figure*}
\parbox[c]{8.5cm}{
  \begin{center}
\includegraphics[width=8.5cm,height=22cm]{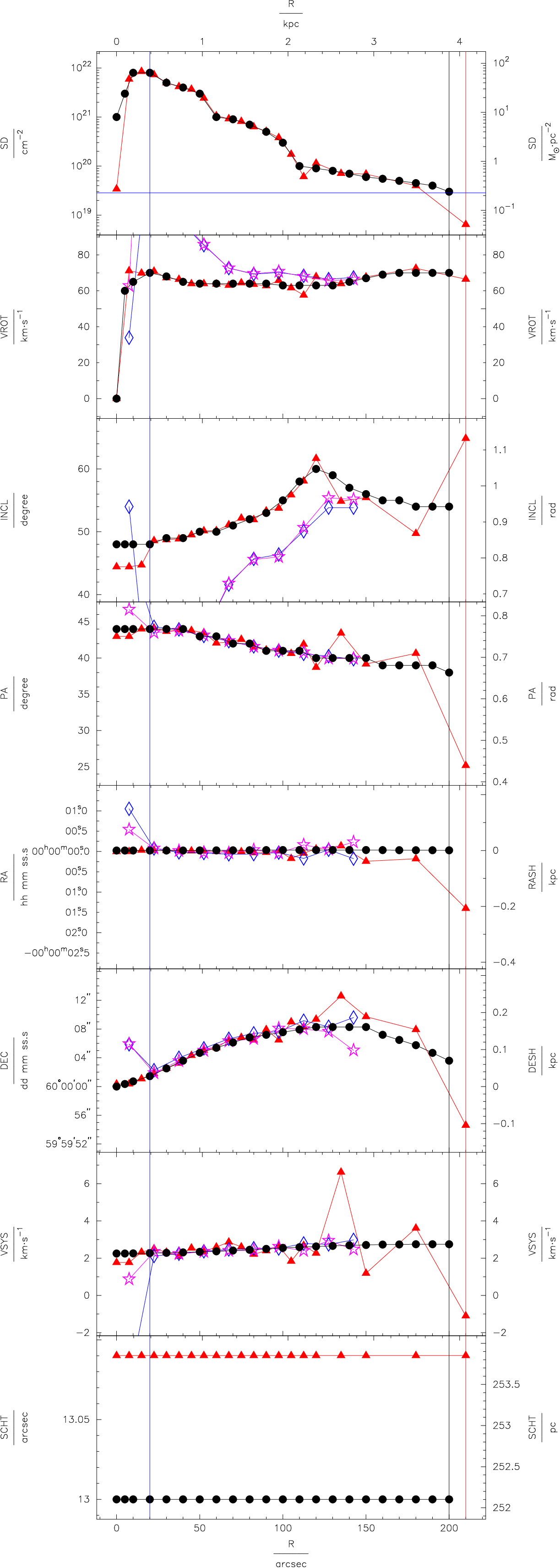}
  \end{center}
}
\parbox[c]{7.5cm} {
{
    \begin{center}
\includegraphics[width=8.5cm,height=22cm]{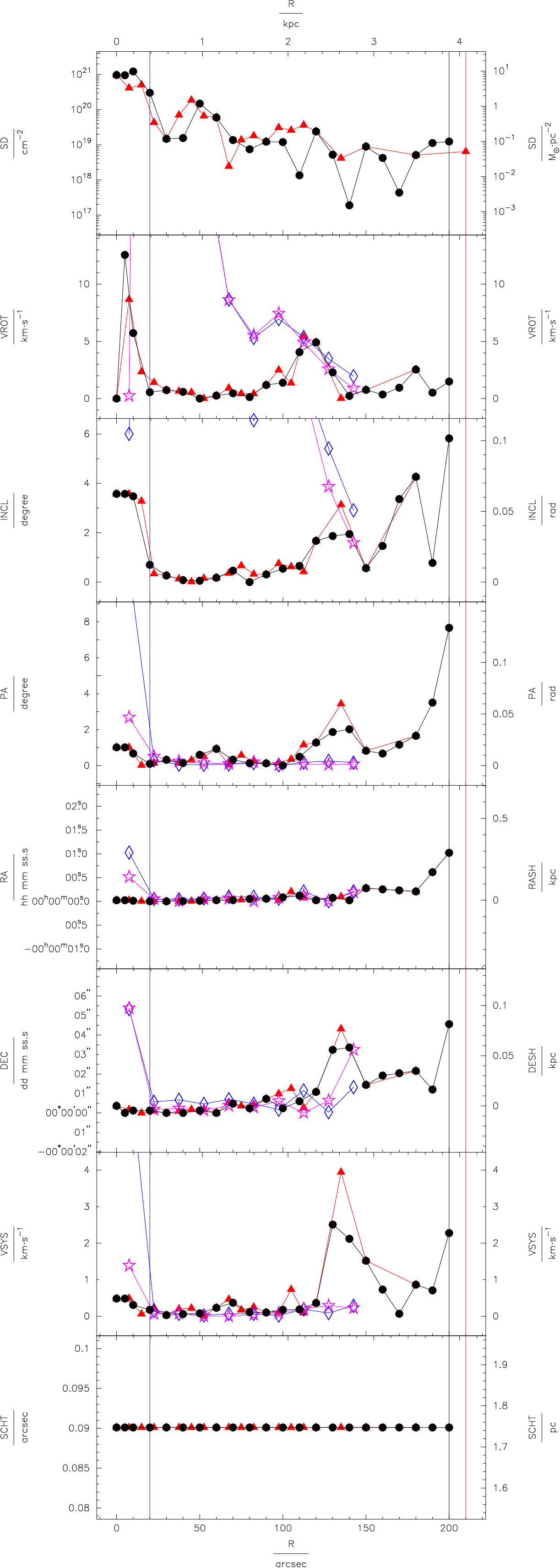}
    \end{center}
}
}
\input{fig_cap_13}
\end{figure*}\begin{center}
\begin{table*}

\caption{Test 4: asymmetric warp with shift of centre and systemic velocity; the scale height is kept constant with radius and noise was added to the data cube. Deviation in global dispersion: below 0.01 ${\rm km}\,{\rm s}^{-1}$. ROTCUR: input model with optimal guesses. All parameters except expansion velocity were left variable.}

\begin{center}
\begin{tabular}{rrrrrrrrrrrr}
\hline
\hline
Quantity & Range & method & $ N_{\ion{H}{i}} $ & $ \frac{N_{\ion{H}{i}}}{N_{\ion{H}{i}},exp}$ & $v_{\rm rot}$ & $ i $ & $ pa $ & $ {\rm RA} $ & $ {\rm DEC} $ & $ v_{\rm sys} $ & $ z_0 $ \\
 (1)     & (2)   & (3) & (4)                & (5)                                          & (6)           & (7)   & (8)    & (9)          & (10)          & (11)            & (12)\\ 
\hline\\
$ \overline{\Delta} $ & all & tir &  119.82 &0.126 &1.8 &1.2 &0.7 &0.9 &0.7 &0.5 &0.1 \\
$\sqrt{\overline{\Delta^2}}$ & all & tir &  259.53 &0.248 &2.8 &1.9 &1.1 &1.5 &1.2 &1.0 &0.1 \\
$\Delta_{\rm max}$ & all & tir &  965.85 &0.966 &8.7 &4.3 &3.4 &4.2 &4.3 &3.9 &0.1 \\
\hline
$ \overline{\Delta} $ & 70\% & tir &  17.63 &0.037 &0.6 &0.4 &0.3 &0.3 &0.2 &0.2 &0.1 \\
$\sqrt{\overline{\Delta^2}}$ & 70\% & tir &  21.51 &0.046 &0.8 &0.4 &0.4 &0.4 &0.3 &0.2 &0.1 \\
$\Delta_{\rm max}$ & 70\% & tir &  43.33 &0.088 &1.4 &0.8 &0.9 &0.9 &0.7 &0.5 &0.1 \\
\hline
$ \overline{\Delta} $ & hi & tir &  36.71 &0.083 &1.4 &0.8 &0.7 &1.0 &0.8 &0.6 &0.1 \\
$\sqrt{\overline{\Delta^2}}$ & hi & tir &  56.58 &0.129 &2.1 &1.4 &1.1 &1.6 &1.3 &1.1 &0.1 \\
$\Delta_{\rm max}$ & hi & tir &  187.26 &0.375 &5.4 &4.3 &3.4 &4.2 &4.3 &3.9 &0.1 \\
\hline
$ \overline{\Delta} $ & $8-\sigma$ & tir &  14.20 &0.192 &2.7 &2.0 &1.7 &2.1 &1.9 &1.4 &0.1 \\
$\sqrt{\overline{\Delta^2}}$ & $8-\sigma$ & tir &  18.60 &0.223 &3.5 &2.5 &1.9 &2.5 &2.3 &1.9 &0.1 \\
$\Delta_{\rm max}$ & $8-\sigma$ & tir &  36.53 &0.375 &5.4 &4.3 &3.4 &4.2 &4.3 &3.9 &0.1 \\
\hline
$ \overline{\Delta} $ & excs  & tir &  6.45 \\
$\sqrt{\overline{\Delta^2}}$ & excs & tir &  6.45 \\
$\Delta_{\rm max}$ & excs & tir &  6.45 \\
\hline
$ \overline{\Delta} $ & all & pk &  & & 34.0 &12.8 &0.4 &1.7 &1.1 &0.2 & \\
$\sqrt{\overline{\Delta^2}}$ & all & pk &  & & 81.3 &16.6 &0.9 &2.7 &2.0 &0.5 & \\
$\Delta_{\rm max}$ & all & pk &  & & 253.2 &39.9 &2.7 &7.7 &5.4 &1.4 & \\
\hline
$ \overline{\Delta} $ & 70\% & pk &  & & 4.3 &7.2 &0.1 &0.6 &0.3 &0.1 & \\
$\sqrt{\overline{\Delta^2}}$ & 70\% & pk &  & & 5.2 &8.0 &0.1 &0.7 &0.3 &0.1 & \\
$\Delta_{\rm max}$ & 70\% & pk &  & & 8.6 &12.9 &0.2 &1.2 &0.6 &0.2 & \\
\hline
$ \overline{\Delta} $ & hi & pk &  & & 37.8 &12.7 &0.2 &1.0 &0.6 &0.1 & \\
$\sqrt{\overline{\Delta^2}}$ & hi & pk &  & & 85.6 &16.9 &0.2 &1.3 &1.1 &0.1 & \\
$\Delta_{\rm max}$ & hi & pk &  & & 253.2 &39.9 &0.5 &2.9 &3.2 &0.3 & \\
\hline
$ \overline{\Delta} $ & all & m1 &  & & 21.3 &11.7 &1.2 &2.8 &1.1 &0.8 & \\
$\sqrt{\overline{\Delta^2}}$ & all & m1 &  & & 34.9 &14.6 &3.5 &5.1 &1.8 &2.2 & \\
$\Delta_{\rm max}$ & all & m1 &  & & 97.9 &33.0 &11.0 &15.4 &5.3 &6.9 & \\
\hline
$ \overline{\Delta} $ & 70\% & m1 &  & & 7.6 &6.7 &0.1 &0.9 &0.4 &0.1 & \\
$\sqrt{\overline{\Delta^2}}$ & 70\% & m1 &  & & 9.7 &6.9 &0.1 &1.0 &0.5 &0.1 & \\
$\Delta_{\rm max}$ & 70\% & m1 &  & & 21.4 &9.2 &0.2 &1.5 &0.7 &0.2 & \\
\hline
$ \overline{\Delta} $ & hi & m1 &  & & 20.5 &12.3 &0.1 &1.4 &0.6 &0.1 & \\
$\sqrt{\overline{\Delta^2}}$ & hi & m1 &  & & 35.5 &15.3 &0.2 &1.7 &0.7 &0.1 & \\
$\Delta_{\rm max}$ & hi & m1 &  & & 97.9 &33.0 &0.3 &3.2 &1.3 &0.3 & \\
\hline
\hline
\end{tabular}
\end{center}
\end{table*}
\end{center}
\clearpage
\begin{figure*}
\parbox[c]{8.5cm}{
  \begin{center}
\includegraphics[width=8.5cm,height=22cm]{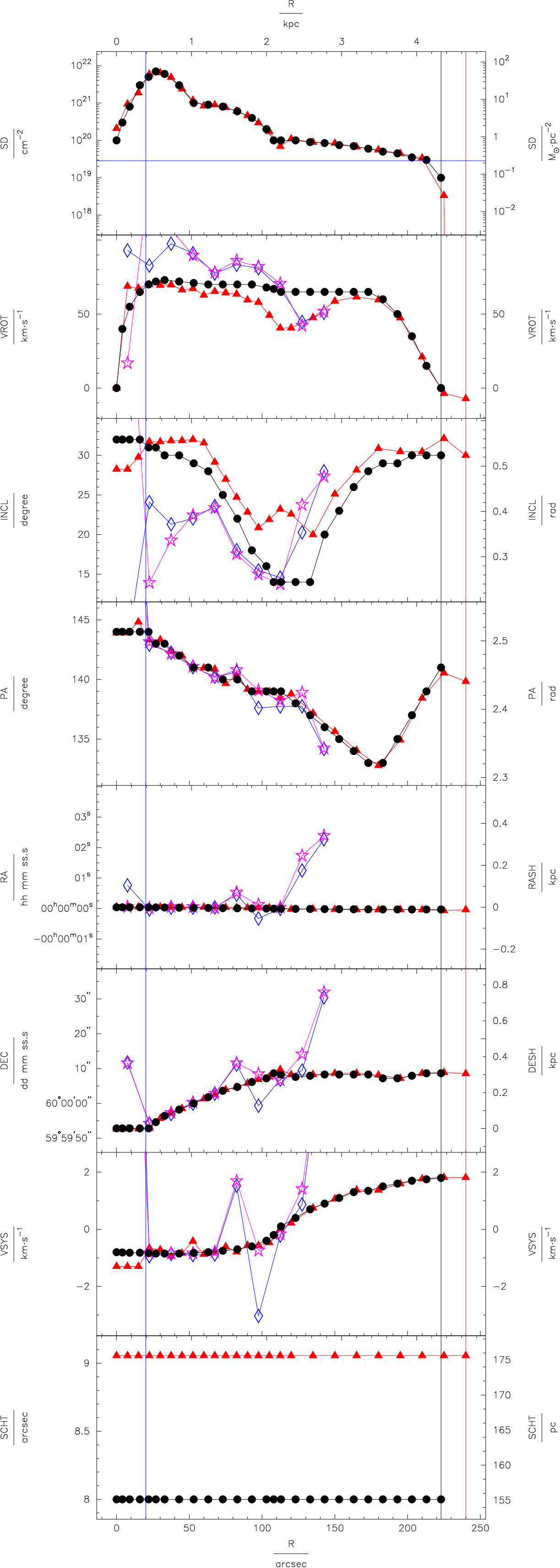}
  \end{center}
}
\parbox[c]{7.5cm} {
{
    \begin{center}
\includegraphics[width=8.5cm,height=22cm]{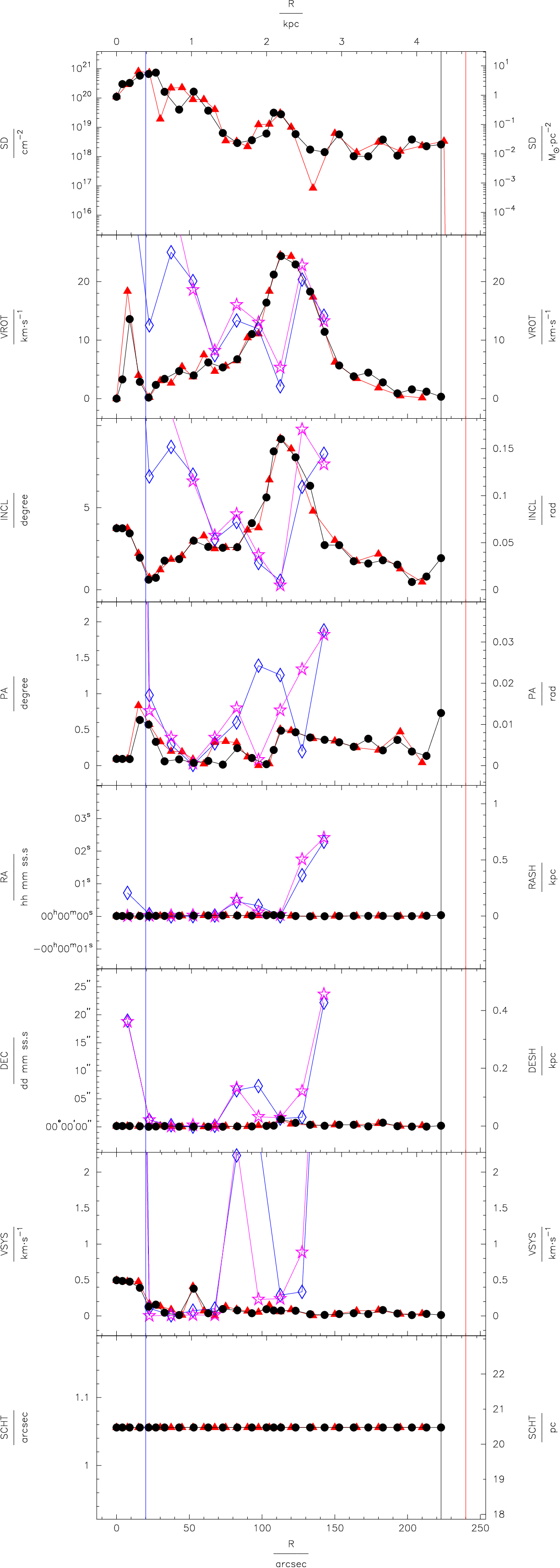}
    \end{center}
}
}
\input{fig_cap_14}
\end{figure*}\begin{center}
\begin{table*}

\caption{Test 5: asymmetric warp with central shift and shift in sytemic velocity. The scale height is constant with radius, the data cube contains no noise.
The scale height is constant. Deviation in global dispersion: 0.06 ${\rm km}\,{\rm s}^{-1}$.
The results get unreliable at low inclination ($i<25\deg$). ROTCUR: first guess deviates by $10\,{\rm km}\,{\rm s}^{-1}$ in rotation velocity, by $10\deg$ in inclination and position angle, by $10\,{\rm km}\,{\rm s}^{-1}$ in systemic velocity, and by $10\arcsec$ in both central coordinates from the optimal guess. All parameters except expansion velocity were left variable.
}

\begin{center}
\begin{tabular}{rrrrrrrrrrrr}
\hline
\hline
Quantity & Range & method & $ N_{\ion{H}{i}} $ & $ \frac{N_{\ion{H}{i}}}{N_{\ion{H}{i}},exp}$ & $v_{\rm rot}$ & $ i $ & $ pa $ & $ {\rm RA} $ & $ {\rm DEC} $ & $ v_{\rm sys} $ & $ z_0 $ \\
 (1)     & (2)   & (3) & (4)                & (5)                                          & (6)           & (7)   & (8)    & (9)          & (10)          & (11)            & (12)\\ 
\hline\\
$ \overline{\Delta} $ & all & tir &  108.87 &0.138 &7.8 &3.3 &0.3 &0.3 &0.2 &0.1 &1.1 \\
$\sqrt{\overline{\Delta^2}}$ & all & tir &  237.61 &0.270 &10.8 &3.9 &0.3 &0.3 &0.4 &0.2 &1.1 \\
$\Delta_{\rm max}$ & all & tir &  808.64 &1.103 &24.6 &9.2 &0.8 &0.6 &1.4 &0.5 &1.1 \\
\hline
$ \overline{\Delta} $ & 70\% & tir &  8.97 &0.042 &3.5 &2.1 &0.2 &0.2 &0.1 &0.1 &1.1 \\
$\sqrt{\overline{\Delta^2}}$ & 70\% & tir &  14.27 &0.052 &4.2 &2.3 &0.2 &0.2 &0.1 &0.1 &1.1 \\
$\Delta_{\rm max}$ & 70\% & tir &  40.55 &0.087 &7.5 &3.6 &0.3 &0.3 &0.2 &0.1 &1.1 \\
\hline
$ \overline{\Delta} $ & hi & tir &  75.50 &0.066 &7.9 &3.3 &0.3 &0.3 &0.3 &0.1 &1.1 \\
$\sqrt{\overline{\Delta^2}}$ & hi & tir &  181.88 &0.094 &10.8 &4.0 &0.3 &0.3 &0.4 &0.1 &1.1 \\
$\Delta_{\rm max}$ & hi & tir &  736.63 &0.305 &24.6 &9.2 &0.6 &0.6 &1.4 &0.4 &1.1 \\
\hline
$ \overline{\Delta} $ & $8-\sigma$ & tir &  5.86 &0.084 &9.8 &3.9 &0.3 &0.2 &0.5 &0.0 &1.1 \\
$\sqrt{\overline{\Delta^2}}$ & $8-\sigma$ & tir &  10.51 &0.123 &13.9 &5.0 &0.4 &0.3 &0.6 &0.1 &1.1 \\
$\Delta_{\rm max}$ & $8-\sigma$ & tir &  30.54 &0.305 &24.6 &9.2 &0.5 &0.6 &1.4 &0.1 &1.1 \\
\hline
$ \overline{\Delta} $ & $2-\sigma$ & tir &  1.90 &0.074 &0.1 &0.5 &0.0 &0.2 &0.2 &0.0 &1.1 \\
$\sqrt{\overline{\Delta^2}}$ & $2-\sigma$ & tir &  2.36 &0.074 &0.1 &0.5 &0.0 &0.2 &0.2 &0.0 &1.1 \\
$\Delta_{\rm max}$ & $2-\sigma$ & tir &  3.35 &0.074 &0.1 &0.5 &0.0 &0.2 &0.2 &0.0 &1.1 \\
\hline
$ \overline{\Delta} $ & excs  & tir &  1.67 \\
$\sqrt{\overline{\Delta^2}}$ & excs & tir &  2.37 \\
$\Delta_{\rm max}$ & excs & tir &  3.35 \\
\hline
$ \overline{\Delta} $ & all & pk &  & & 23.8 &8.5 &2.3 &7.4 &6.1 &2.7 & \\
$\sqrt{\overline{\Delta^2}}$ & all & pk &  & & 30.3 &10.8 &5.2 &14.3 &10.0 &5.7 & \\
$\Delta_{\rm max}$ & all & pk &  & & 33.6 &22.9 &16.2 &36.1 &23.7 &16.7 & \\
\hline
$ \overline{\Delta} $ & 70\% & pk &  & & 13.9 &4.9 &0.5 &0.5 &1.7 &0.2 & \\
$\sqrt{\overline{\Delta^2}}$ & 70\% & pk &  & & 14.9 &5.8 &0.6 &0.8 &2.6 &0.4 & \\
$\Delta_{\rm max}$ & 70\% & pk &  & & 22.8 &9.8 &0.8 &1.9 &6.4 &0.9 & \\
\hline
$ \overline{\Delta} $ & hi & pk &  & & 22.7 &6.9 &0.7 &8.2 &4.7 &1.1 & \\
$\sqrt{\overline{\Delta^2}}$ & hi & pk &  & & 29.9 &8.4 &0.9 &15.1 &8.5 &2.2 & \\
$\Delta_{\rm max}$ & hi & pk &  & & 72.7 &17.1 &1.8 &36.1 &23.7 &6.1 & \\
\hline
$ \overline{\Delta} $ & all & m1 &  & & 16.9 &7.3 &3.7 &7.7 &6.0 &3.6 & \\
$\sqrt{\overline{\Delta^2}}$ & all & m1 &  & & 20.0 &9.9 &9.6 &13.1 &9.8 &7.9 & \\
$\Delta_{\rm max}$ & all & m1 &  & & 42.6 &26.0 &30.1 &34.3 &22.2 &24.0 & \\
\hline
$ \overline{\Delta} $ & 70\% & m1 &  & & 11.7 &4.2 &0.5 &1.9 &1.6 &0.5 & \\
$\sqrt{\overline{\Delta^2}}$ & 70\% & m1 &  & & 12.8 &4.9 &0.7 &3.2 &2.7 &0.9 & \\
$\Delta_{\rm max}$ & 70\% & m1 &  & & 20.1 &7.0 &1.3 &6.7 &6.5 &2.2 & \\
\hline
$ \overline{\Delta} $ & hi & m1 &  & & 14.1 &5.2 &0.8 &7.4 &4.5 &1.3 & \\
$\sqrt{\overline{\Delta^2}}$ & hi & m1 &  & & 15.6 &5.9 &1.0 &13.4 &8.1 &2.3 & \\
$\Delta_{\rm max}$ & hi & m1 &  & & 25.0 &8.7 &1.9 &34.3 &22.2 &6.0 & \\
\hline
\hline
\end{tabular}
\end{center}
\end{table*}
\end{center}
\clearpage
\begin{figure*}
\parbox[c]{8.5cm}{
  \begin{center}
\includegraphics[width=8.5cm,height=22cm]{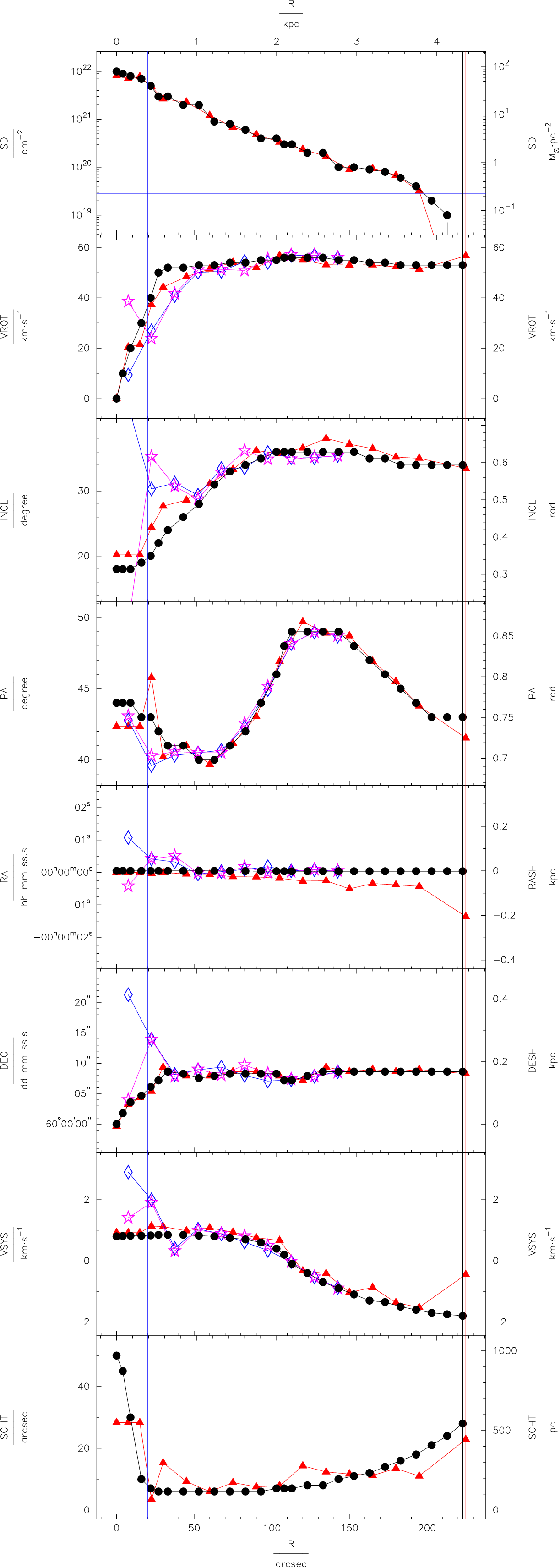}
  \end{center}
}
\parbox[c]{7.5cm} {
{
    \begin{center}
\includegraphics[width=8.5cm,height=22cm]{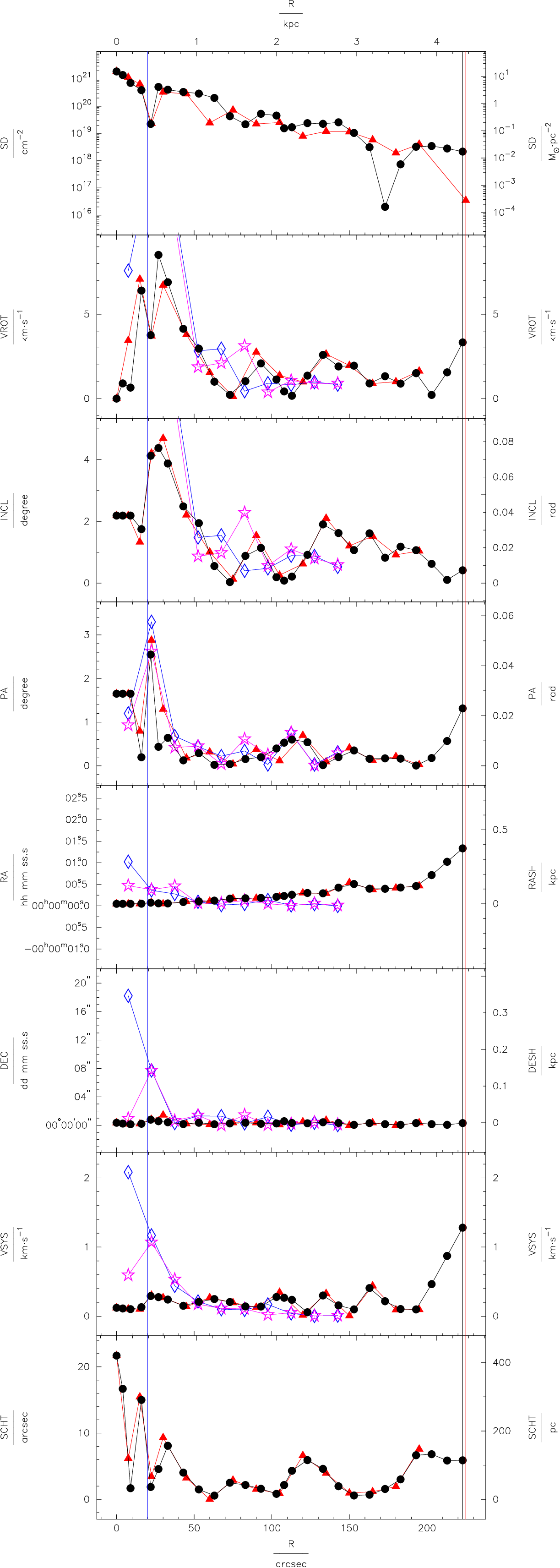}
    \end{center}
}
}
\input{fig_cap_15}
\end{figure*}\begin{center}
\begin{table*}

\caption{Test 6: warp with central/sytemic velocity shift and variable scale height without noise added to the data cube. Deviation in global dispersion: 0.06 ${\rm km}\,{\rm s}^{-1}$. ROTCUR: first guess deviates by $5\,{\rm km}\,{\rm s}^{-1}$ in rotation velocity, by $5\deg$ in inclination and position angle, by $5\,{\rm km}\,{\rm s}^{-1}$ in systemic velocity, and by $5\arcsec$ in both central coordinates from the optimal guess. All parameters except expansion velocity were left variable.
}

\begin{center}
\begin{tabular}{rrrrrrrrrrrr}
\hline
\hline
Quantity & Range & method & $ N_{\ion{H}{i}} $ & $ \frac{N_{\ion{H}{i}}}{N_{\ion{H}{i}},exp}$ & $v_{\rm rot}$ & $ i $ & $ pa $ & $ {\rm RA} $ & $ {\rm DEC} $ & $ v_{\rm sys} $ & $ z_0 $ \\
 (1)     & (2)   & (3) & (4)                & (5)                                          & (6)           & (7)   & (8)    & (9)          & (10)          & (11)            & (12)\\ 
\hline\\
$ \overline{\Delta} $ & all & tir &  265.41 &0.083 &2.5 &1.7 &0.7 &3.2 &0.4 &0.2 &5.4 \\
$\sqrt{\overline{\Delta^2}}$ & all & tir &  569.74 &0.096 &3.2 &2.1 &1.0 &4.0 &0.5 &0.2 &7.9 \\
$\Delta_{\rm max}$ & all & tir &  1895.32 &0.190 &7.1 &4.7 &2.9 &8.1 &1.4 &0.4 &21.7 \\
\hline
$ \overline{\Delta} $ & 70\% & tir &  12.61 &0.058 &1.4 &1.1 &0.2 &1.8 &0.2 &0.1 &2.4 \\
$\sqrt{\overline{\Delta^2}}$ & 70\% & tir &  15.59 &0.066 &1.6 &1.2 &0.3 &2.2 &0.2 &0.1 &2.9 \\
$\Delta_{\rm max}$ & 70\% & tir &  25.18 &0.110 &2.7 &2.1 &0.7 &4.4 &0.4 &0.3 &6.2 \\
\hline
$ \overline{\Delta} $ & hi & tir &  63.43 &0.070 &2.2 &1.7 &0.5 &3.8 &0.4 &0.2 &3.3 \\
$\sqrt{\overline{\Delta^2}}$ & hi & tir &  123.19 &0.081 &2.8 &2.1 &0.9 &4.4 &0.6 &0.2 &4.3 \\
$\Delta_{\rm max}$ & hi & tir &  329.48 &0.142 &6.7 &4.7 &2.9 &8.1 &1.4 &0.4 &9.3 \\
\hline
$ \overline{\Delta} $ & $8-\sigma$ & tir &  4.67 &0.081 &1.4 &1.2 &0.2 &6.8 &0.2 &0.2 &2.9 \\
$\sqrt{\overline{\Delta^2}}$ & $8-\sigma$ & tir &  6.10 &0.088 &1.4 &1.2 &0.2 &6.8 &0.3 &0.2 &4.0 \\
$\Delta_{\rm max}$ & $8-\sigma$ & tir &  11.50 &0.115 &2.0 &1.5 &0.4 &8.1 &0.4 &0.4 &7.6 \\
\hline
$ \overline{\Delta} $ & $2-\sigma$ & tir &  2.03 &0.112 &1.6 &1.0 &0.0 &7.0 &0.4 &0.1 &7.6 \\
$\sqrt{\overline{\Delta^2}}$ & $2-\sigma$ & tir &  2.85 &0.112 &1.6 &1.0 &0.0 &7.0 &0.4 &0.1 &7.6 \\
$\Delta_{\rm max}$ & $2-\sigma$ & tir &  4.03 &0.112 &1.6 &1.0 &0.0 &7.0 &0.4 &0.1 &7.6 \\
\hline
$ \overline{\Delta} $ & excs  & tir &  0.03 \\
$\sqrt{\overline{\Delta^2}}$ & excs & tir &  0.03 \\
$\Delta_{\rm max}$ & excs & tir &  0.03 \\
\hline
$ \overline{\Delta} $ & all & pk &  & & 5.9 &3.7 &0.6 &2.5 &1.3 &0.3 & \\
$\sqrt{\overline{\Delta^2}}$ & all & pk &  & & 9.4 &5.9 &1.0 &3.7 &2.5 &0.4 & \\
$\Delta_{\rm max}$ & all & pk &  & & 21.6 &15.1 &2.6 &7.0 &7.7 &1.1 & \\
\hline
$ \overline{\Delta} $ & 70\% & pk &  & & 1.5 &1.0 &0.3 &0.8 &0.3 &0.1 & \\
$\sqrt{\overline{\Delta^2}}$ & 70\% & pk &  & & 1.7 &1.2 &0.4 &1.0 &0.5 &0.1 & \\
$\Delta_{\rm max}$ & 70\% & pk &  & & 3.1 &2.3 &0.6 &1.9 &0.9 &0.2 & \\
\hline
$ \overline{\Delta} $ & hi & pk &  & & 4.2 &3.1 &0.6 &2.0 &1.3 &0.2 & \\
$\sqrt{\overline{\Delta^2}}$ & hi & pk &  & & 6.8 &5.5 &1.0 &3.1 &2.7 &0.4 & \\
$\Delta_{\rm max}$ & hi & pk &  & & 17.1 &15.1 &2.6 &6.9 &7.7 &1.1 & \\
\hline
$ \overline{\Delta} $ & all & m1 &  & & 4.3 &4.8 &0.7 &3.0 &3.1 &0.4 & \\
$\sqrt{\overline{\Delta^2}}$ & all & m1 &  & & 6.3 &8.9 &1.2 &5.4 &6.3 &0.8 & \\
$\Delta_{\rm max}$ & all & m1 &  & & 14.0 &25.2 &3.3 &15.4 &18.2 &2.1 & \\
\hline
$ \overline{\Delta} $ & 70\% & m1 &  & & 1.4 &0.9 &0.3 &0.8 &0.5 &0.1 & \\
$\sqrt{\overline{\Delta^2}}$ & 70\% & m1 &  & & 1.7 &1.0 &0.4 &1.0 &0.7 &0.1 & \\
$\Delta_{\rm max}$ & 70\% & m1 &  & & 3.0 &1.6 &0.7 &2.0 &1.3 &0.2 & \\
\hline
$ \overline{\Delta} $ & hi & m1 &  & & 3.9 &2.5 &0.7 &1.7 &1.4 &0.2 & \\
$\sqrt{\overline{\Delta^2}}$ & hi & m1 &  & & 6.2 &4.1 &1.2 &2.5 &2.7 &0.4 & \\
$\Delta_{\rm max}$ & hi & m1 &  & & 14.0 &10.1 &3.3 &5.4 &7.7 &1.2 & \\
\hline
\hline
\end{tabular}
\end{center}
\end{table*}
\end{center}
\clearpage
\begin{figure*}
\parbox[c]{8.5cm}{
  \begin{center}
\includegraphics[width=8.5cm,height=22cm]{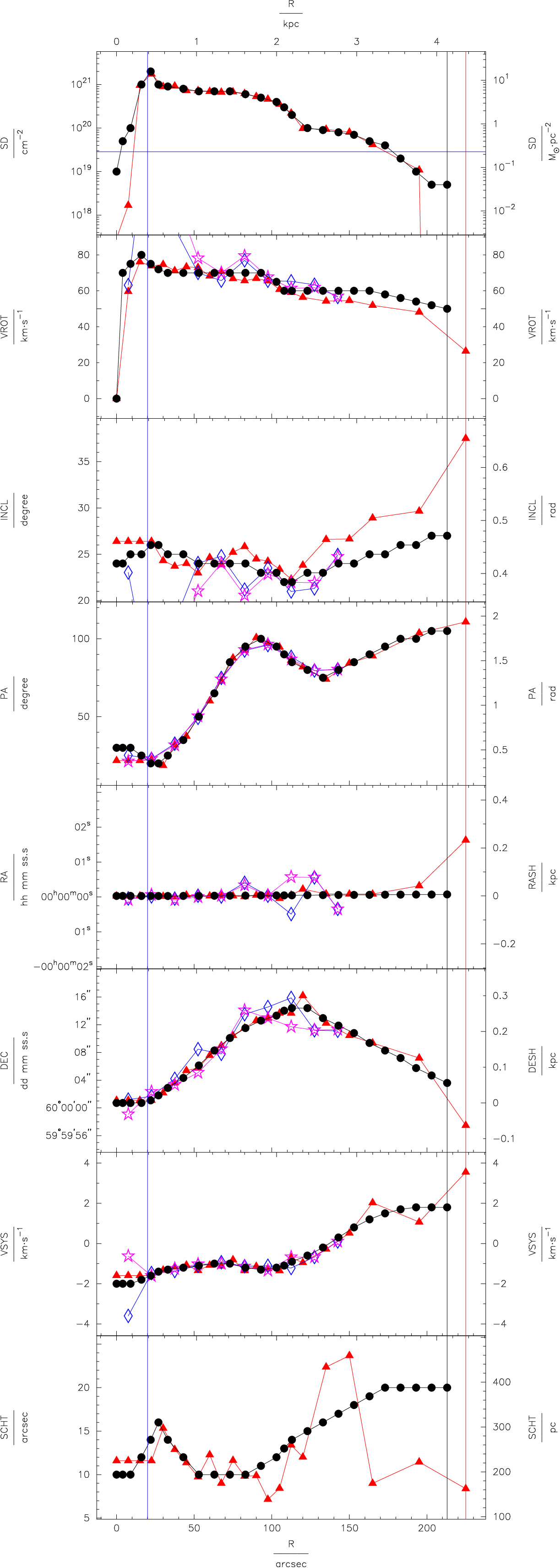}
  \end{center}
}
\parbox[c]{7.5cm} {
{
    \begin{center}
\includegraphics[width=8.5cm,height=22cm]{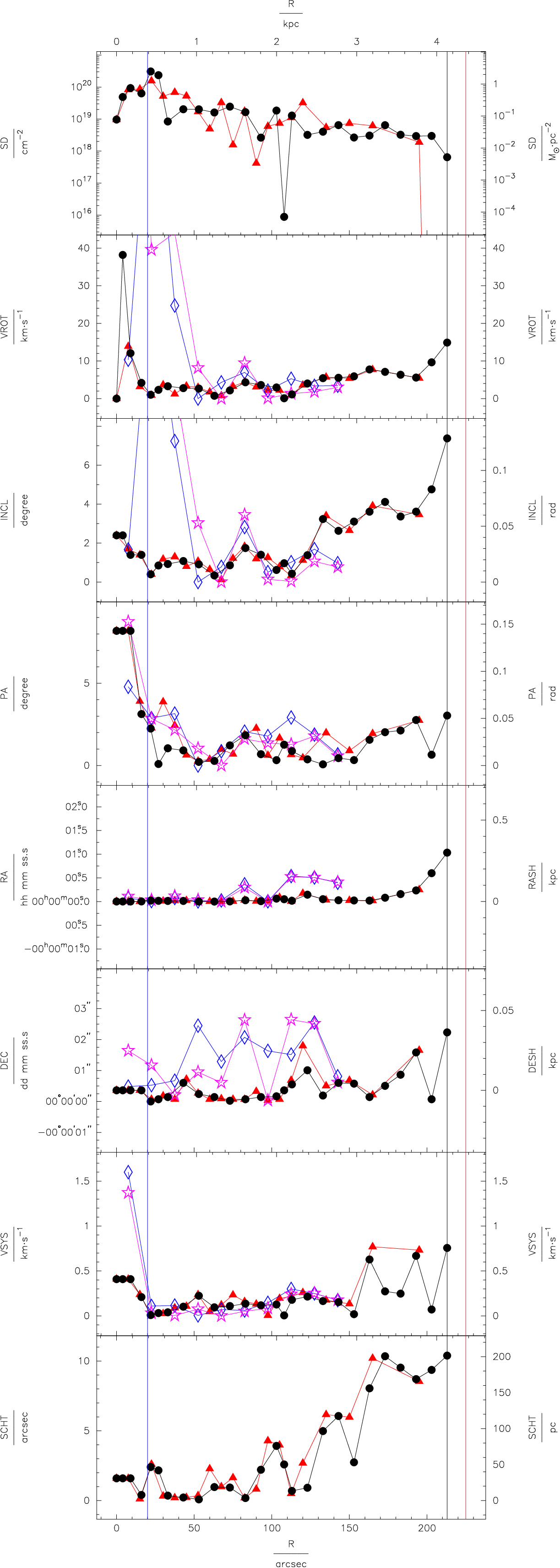}
    \end{center}
}
}
\input{fig_cap_16}
\end{figure*}\begin{center}
\begin{table*}

\caption{Test 7: asymmetric warp with central/sytemic velocity shift and variable scale height, noise added to the data cube. Deviation in global dispersion: below 0.01 ${\rm km}\,{\rm s}^{-1}$. ROTCUR: optimal first guesses. All parameters except expansion velocity were left variable.}

\begin{center}
\begin{tabular}{rrrrrrrrrrrr}
\hline
\hline
Quantity & Range & method & $ N_{\ion{H}{i}} $ & $ \frac{N_{\ion{H}{i}}}{N_{\ion{H}{i}},exp}$ & $v_{\rm rot}$ & $ i $ & $ pa $ & $ {\rm RA} $ & $ {\rm DEC} $ & $ v_{\rm sys} $ & $ z_0 $ \\
 (1)     & (2)   & (3) & (4)                & (5)                                          & (6)           & (7)   & (8)    & (9)          & (10)          & (11)            & (12)\\ 
\hline\\
$ \overline{\Delta} $ & all & tir &  30.11 &0.156 &3.6 &1.5 &2.2 &0.5 &0.4 &0.2 &2.6 \\
$\sqrt{\overline{\Delta^2}}$ & all & tir &  48.80 &0.315 &4.7 &1.8 &3.1 &1.1 &0.6 &0.3 &3.9 \\
$\Delta_{\rm max}$ & all & tir &  156.23 &0.980 &13.9 &3.9 &8.2 &3.8 &1.8 &0.8 &10.2 \\
\hline
$ \overline{\Delta} $ & 70\% & tir &  8.62 &0.039 &2.1 &0.9 &1.1 &0.1 &0.2 &0.1 &1.0 \\
$\sqrt{\overline{\Delta^2}}$ & 70\% & tir &  11.87 &0.048 &2.4 &1.0 &1.3 &0.2 &0.2 &0.1 &1.3 \\
$\Delta_{\rm max}$ & 70\% & tir &  32.38 &0.082 &3.6 &1.4 &2.3 &0.3 &0.4 &0.2 &2.6 \\
\hline
$ \overline{\Delta} $ & hi & tir &  28.24 &0.060 &3.1 &1.4 &1.4 &0.4 &0.3 &0.2 &2.6 \\
$\sqrt{\overline{\Delta^2}}$ & hi & tir &  47.10 &0.084 &3.6 &1.7 &1.7 &0.8 &0.6 &0.2 &3.7 \\
$\Delta_{\rm max}$ & hi & tir &  156.23 &0.249 &7.7 &3.9 &3.9 &2.6 &1.8 &0.8 &10.2 \\
\hline
$ \overline{\Delta} $ & $8-\sigma$ & tir &  8.89 &0.150 &5.6 &2.9 &1.6 &1.5 &1.0 &0.4 &6.7 \\
$\sqrt{\overline{\Delta^2}}$ & $8-\sigma$ & tir &  14.00 &0.166 &5.7 &3.1 &1.8 &2.1 &1.2 &0.5 &7.2 \\
$\Delta_{\rm max}$ & $8-\sigma$ & tir &  32.38 &0.249 &7.7 &3.9 &2.7 &3.8 &1.8 &0.8 &10.2 \\
\hline
$ \overline{\Delta} $ & $2-\sigma$ & tir &  0.95 &0.210 &5.4 &3.5 &2.7 &3.8 &1.7 &0.7 &8.5 \\
$\sqrt{\overline{\Delta^2}}$ & $2-\sigma$ & tir &  1.34 &0.210 &5.4 &3.5 &2.7 &3.8 &1.7 &0.7 &8.5 \\
$\Delta_{\rm max}$ & $2-\sigma$ & tir &  1.89 &0.210 &5.4 &3.5 &2.7 &3.8 &1.7 &0.7 &8.5 \\
\hline
$ \overline{\Delta} $ & $0.5-\sigma$ & tir &  0.95 &0.210 &5.4 &3.5 &2.7 &3.8 &1.7 &0.7 &8.5 \\
$\sqrt{\overline{\Delta^2}}$ & $0.5-\sigma$ & tir &  1.34 &0.210 &5.4 &3.5 &2.7 &3.8 &1.7 &0.7 &8.5 \\
$\Delta_{\rm max}$ & $0.5-\sigma$ & tir &  1.89 &0.210 &5.4 &3.5 &2.7 &3.8 &1.7 &0.7 &8.5 \\
\hline
$ \overline{\Delta} $ & excs  & tir &  0.00 \\
$\sqrt{\overline{\Delta^2}}$ & excs & tir &  0.00 \\
$\Delta_{\rm max}$ & excs & tir &  0.00 \\
\hline
$ \overline{\Delta} $ & all & pk &  & & 28.6 &4.6 &2.1 &3.0 &1.3 &0.2 & \\
$\sqrt{\overline{\Delta^2}}$ & all & pk &  & & 59.5 &7.3 &3.2 &4.3 &1.6 &0.5 & \\
$\Delta_{\rm max}$ & all & pk &  & & 178.2 &17.3 &8.7 &7.8 &2.6 &1.4 & \\
\hline
$ \overline{\Delta} $ & 70\% & pk &  & & 3.4 &1.2 &1.1 &1.3 &0.7 &0.1 & \\
$\sqrt{\overline{\Delta^2}}$ & 70\% & pk &  & & 4.9 &1.8 &1.2 &1.9 &0.9 &0.1 & \\
$\Delta_{\rm max}$ & 70\% & pk &  & & 9.4 &3.4 &1.8 &4.4 &1.6 &0.2 & \\
\hline
$ \overline{\Delta} $ & hi & pk &  & & 11.9 &3.2 &1.4 &3.2 &1.3 &0.1 & \\
$\sqrt{\overline{\Delta^2}}$ & hi & pk &  & & 20.2 &5.1 &1.6 &4.5 &1.6 &0.1 & \\
$\Delta_{\rm max}$ & hi & pk &  & & 44.1 &10.2 &2.8 &7.8 &2.6 &0.3 & \\
\hline
$ \overline{\Delta} $ & all & m1 &  & & 15.3 &3.2 &2.1 &3.0 &1.4 &0.3 & \\
$\sqrt{\overline{\Delta^2}}$ & all & m1 &  & & 30.7 &5.6 &2.5 &4.4 &1.6 &0.5 & \\
$\Delta_{\rm max}$ & all & m1 &  & & 92.5 &15.7 &4.8 &8.1 &2.6 &1.6 & \\
\hline
$ \overline{\Delta} $ & 70\% & m1 &  & & 3.6 &0.9 &1.5 &1.2 &1.0 &0.1 & \\
$\sqrt{\overline{\Delta^2}}$ & 70\% & m1 &  & & 4.2 &1.1 &1.7 &2.2 &1.1 &0.1 & \\
$\Delta_{\rm max}$ & 70\% & m1 &  & & 7.0 &1.7 &2.9 &5.5 &1.6 &0.2 & \\
\hline
$ \overline{\Delta} $ & hi & m1 &  & & 15.9 &3.4 &1.8 &3.2 &1.5 &0.1 & \\
$\sqrt{\overline{\Delta^2}}$ & hi & m1 &  & & 32.1 &5.9 &2.1 &4.6 &1.7 &0.2 & \\
$\Delta_{\rm max}$ & hi & m1 &  & & 92.5 &15.7 &3.2 &8.1 &2.6 &0.3 & \\
\hline
\hline
\end{tabular}
\end{center}
\end{table*}
\end{center}
\clearpage
\begin{figure*}
\parbox[c]{8.5cm}{
  \begin{center}
\includegraphics[width=8.5cm,height=22cm]{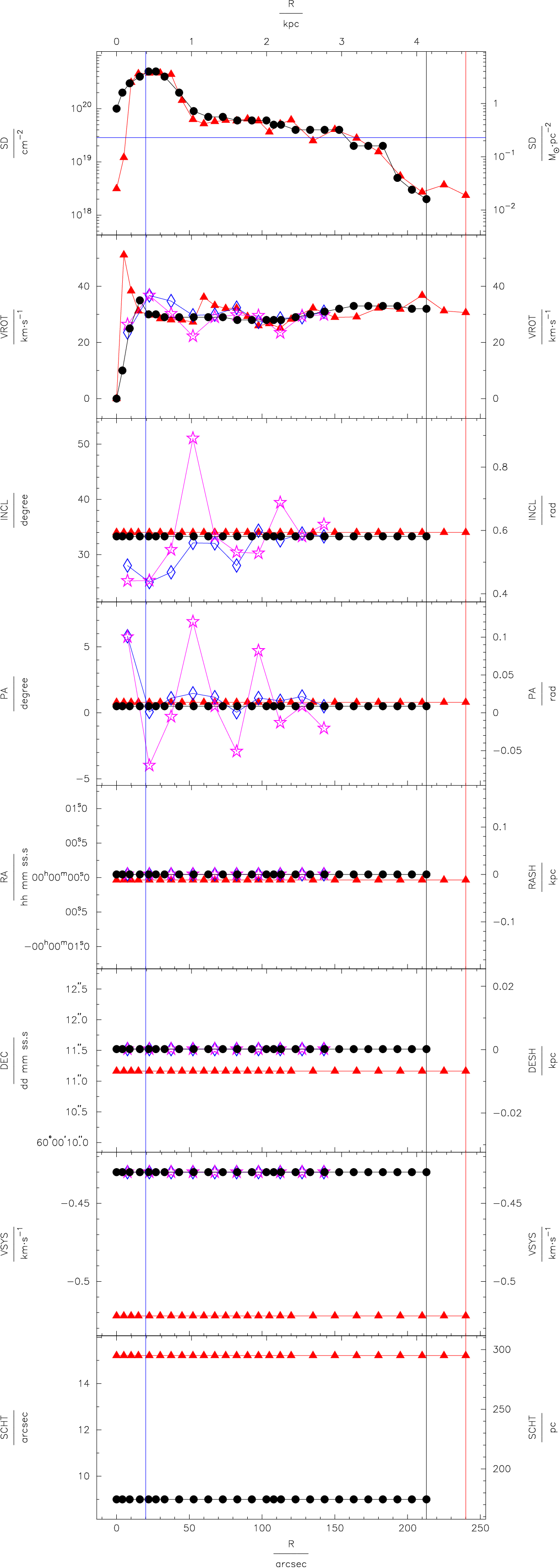}
  \end{center}
}
\parbox[c]{7.5cm} {
{
    \begin{center}
\includegraphics[width=8.5cm,height=22cm]{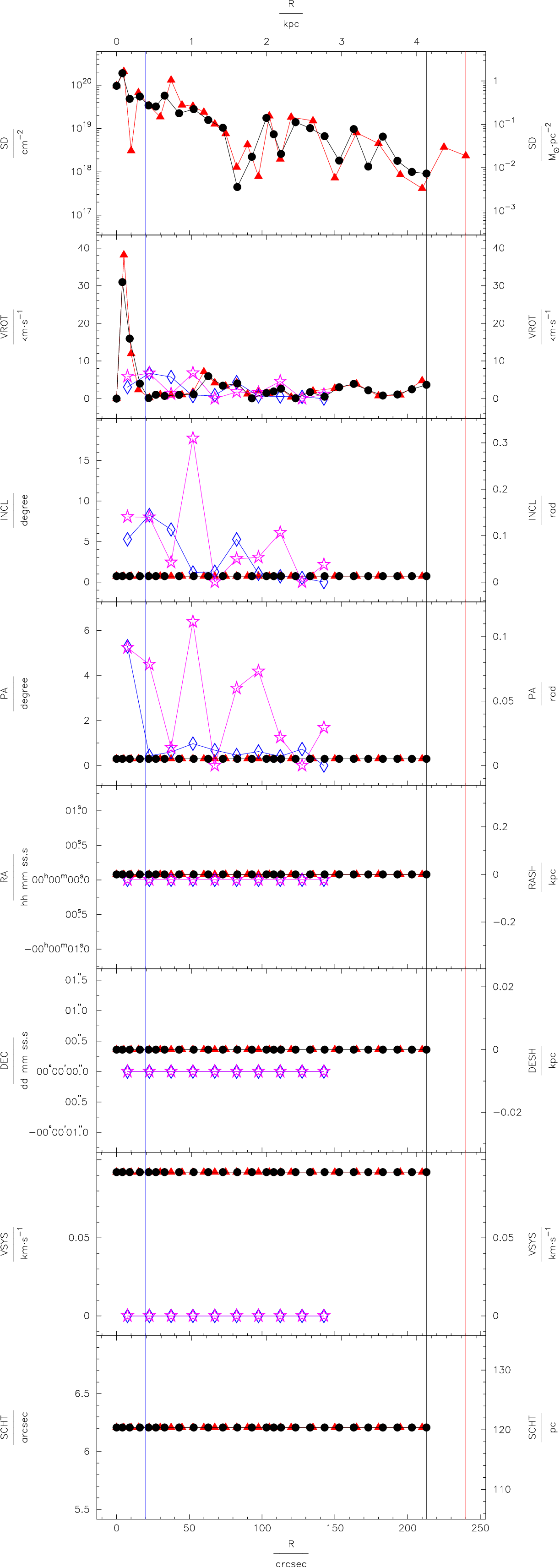}
    \end{center}
}
}
\input{fig_cap_17}
\end{figure*}\begin{center}
\begin{table*}

\caption{Test 8: flat disk with constant scale height, centre, and systemic velocity. Noise was added to the data cube. Deviation in global dispersion: 0.12 ${\rm km}\,{\rm s}^{-1}$. ROTCUR: input model with optimal guesses. Only the rotation velocity and the orientation parameters were fitted.}

\begin{center}
\begin{tabular}{rrrrrrrrrrrr}
\hline
\hline
Quantity & Range & method & $ N_{\ion{H}{i}} $ & $ \frac{N_{\ion{H}{i}}}{N_{\ion{H}{i}},exp}$ & $v_{\rm rot}$ & $ i $ & $ pa $ & $ {\rm RA} $ & $ {\rm DEC} $ & $ v_{\rm sys} $ & $ z_0 $ \\
 (1)     & (2)   & (3) & (4)                & (5)                                          & (6)           & (7)   & (8)    & (9)          & (10)          & (11)            & (12)\\ 
\hline\\
$ \overline{\Delta} $ & all & tir &  28.90 &0.253 &4.1 &0.7 &0.3 &1.2 &0.4 &0.1 &6.2 \\
$\sqrt{\overline{\Delta^2}}$ & all & tir &  55.43 &0.358 &8.6 &0.7 &0.3 &1.2 &0.4 &0.1 &6.2 \\
$\Delta_{\rm max}$ & all & tir &  207.90 &0.969 &38.2 &0.7 &0.3 &1.2 &0.4 &0.1 &6.2 \\
\hline
$ \overline{\Delta} $ & 70\% & tir &  6.87 &0.116 &1.4 &0.7 &0.3 &1.2 &0.4 &0.1 &6.2 \\
$\sqrt{\overline{\Delta^2}}$ & 70\% & tir &  9.57 &0.147 &1.6 &0.7 &0.3 &1.2 &0.4 &0.1 &6.2 \\
$\Delta_{\rm max}$ & 70\% & tir &  19.57 &0.313 &2.8 &0.7 &0.3 &1.2 &0.4 &0.1 &6.2 \\
\hline
$ \overline{\Delta} $ & hi & tir &  22.29 &0.187 &2.3 &0.7 &0.3 &1.2 &0.4 &0.1 &6.2 \\
$\sqrt{\overline{\Delta^2}}$ & hi & tir &  37.63 &0.242 &2.9 &0.7 &0.3 &1.2 &0.4 &0.1 &6.2 \\
$\Delta_{\rm max}$ & hi & tir &  130.94 &0.426 &7.1 &0.7 &0.3 &1.2 &0.4 &0.1 &6.2 \\
\hline
$ \overline{\Delta} $ & $8-\sigma$ & tir &  8.82 &0.203 &2.7 &0.7 &0.3 &1.2 &0.4 &0.1 &6.2 \\
$\sqrt{\overline{\Delta^2}}$ & $8-\sigma$ & tir &  12.77 &0.249 &3.2 &0.7 &0.3 &1.2 &0.4 &0.1 &6.2 \\
$\Delta_{\rm max}$ & $8-\sigma$ & tir &  32.78 &0.426 &7.1 &0.7 &0.3 &1.2 &0.4 &0.1 &6.2 \\
\hline
$ \overline{\Delta} $ & $2-\sigma$ & tir &  3.31 &0.248 &2.6 &0.7 &0.3 &1.2 &0.4 &0.1 &6.2 \\
$\sqrt{\overline{\Delta^2}}$ & $2-\sigma$ & tir &  4.17 &0.263 &3.1 &0.7 &0.3 &1.2 &0.4 &0.1 &6.2 \\
$\Delta_{\rm max}$ & $2-\sigma$ & tir &  7.99 &0.400 &4.8 &0.7 &0.3 &1.2 &0.4 &0.1 &6.2 \\
\hline
$ \overline{\Delta} $ & $0.5-\sigma$ & tir &  1.83 &0.183 &2.9 &0.7 &0.3 &1.2 &0.4 &0.1 &6.2 \\
$\sqrt{\overline{\Delta^2}}$ & $0.5-\sigma$ & tir &  2.25 &0.183 &3.4 &0.7 &0.3 &1.2 &0.4 &0.1 &6.2 \\
$\Delta_{\rm max}$ & $0.5-\sigma$ & tir &  3.71 &0.186 &4.8 &0.7 &0.3 &1.2 &0.4 &0.1 &6.2 \\
\hline
$ \overline{\Delta} $ & excs  & tir &  3.03 \\
$\sqrt{\overline{\Delta^2}}$ & excs & tir &  3.11 \\
$\Delta_{\rm max}$ & excs & tir &  3.71 \\
\hline
$ \overline{\Delta} $ & all & pk &  & & 3.0 &5.0 &2.7 &0.0 &0.0 &0.0 & \\
$\sqrt{\overline{\Delta^2}}$ & all & pk &  & & 3.9 &7.1 &3.5 &0.0 &0.0 &0.0 & \\
$\Delta_{\rm max}$ & all & pk &  & & 6.8 &17.7 &6.4 &0.0 &0.0 &0.0 & \\
\hline
$ \overline{\Delta} $ & 70\% & pk &  & & 1.5 &2.4 &1.6 &0.0 &0.0 &0.0 & \\
$\sqrt{\overline{\Delta^2}}$ & 70\% & pk &  & & 2.0 &3.0 &2.2 &0.0 &0.0 &0.0 & \\
$\Delta_{\rm max}$ & 70\% & pk &  & & 4.6 &6.1 &4.2 &0.0 &0.0 &0.0 & \\
\hline
$ \overline{\Delta} $ & hi & pk &  & & 2.6 &4.7 &2.5 &0.0 &0.0 &0.0 & \\
$\sqrt{\overline{\Delta^2}}$ & hi & pk &  & & 3.7 &7.0 &3.3 &0.0 &0.0 &0.0 & \\
$\Delta_{\rm max}$ & hi & pk &  & & 6.8 &17.7 &6.4 &0.0 &0.0 &0.0 & \\
\hline
$ \overline{\Delta} $ & all & m1 &  & & 2.3 &3.0 &1.0 &0.0 &0.0 &0.0 & \\
$\sqrt{\overline{\Delta^2}}$ & all & m1 &  & & 3.3 &4.1 &1.8 &0.0 &0.0 &0.0 & \\
$\Delta_{\rm max}$ & all & m1 &  & & 6.7 &8.3 &5.3 &0.0 &0.0 &0.0 & \\
\hline
$ \overline{\Delta} $ & 70\% & m1 &  & & 0.9 &1.4 &0.5 &0.0 &0.0 &0.0 & \\
$\sqrt{\overline{\Delta^2}}$ & 70\% & m1 &  & & 1.3 &2.2 &0.5 &0.0 &0.0 &0.0 & \\
$\Delta_{\rm max}$ & 70\% & m1 &  & & 3.1 &5.3 &0.7 &0.0 &0.0 &0.0 & \\
\hline
$ \overline{\Delta} $ & hi & m1 &  & & 2.2 &2.7 &0.5 &0.0 &0.0 &0.0 & \\
$\sqrt{\overline{\Delta^2}}$ & hi & m1 &  & & 3.3 &4.0 &0.6 &0.0 &0.0 &0.0 & \\
$\Delta_{\rm max}$ & hi & m1 &  & & 6.7 &8.3 &1.0 &0.0 &0.0 &0.0 & \\
\hline
\hline
\end{tabular}
\end{center}
\end{table*}
\end{center}
\clearpage
\begin{figure*}
\parbox[c]{8.5cm}{
  \begin{center}
\includegraphics[width=8.5cm,height=22cm]{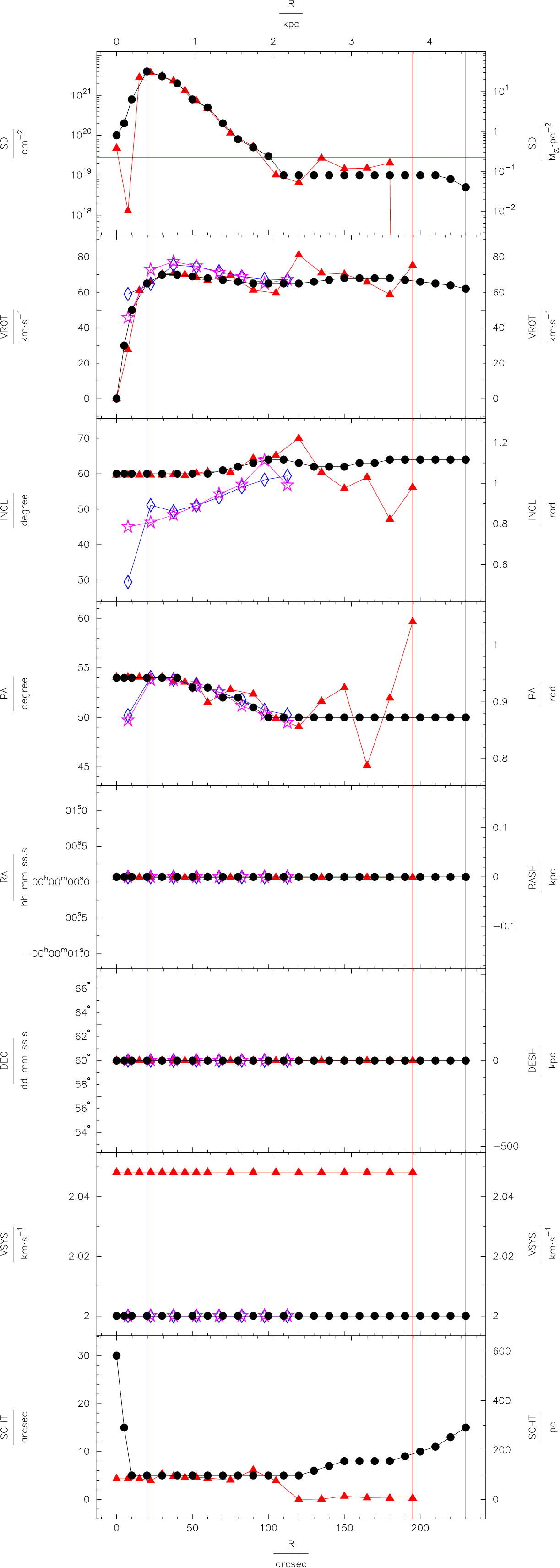}
  \end{center}
}
\parbox[c]{7.5cm} {
{
    \begin{center}
\includegraphics[width=8.5cm,height=22cm]{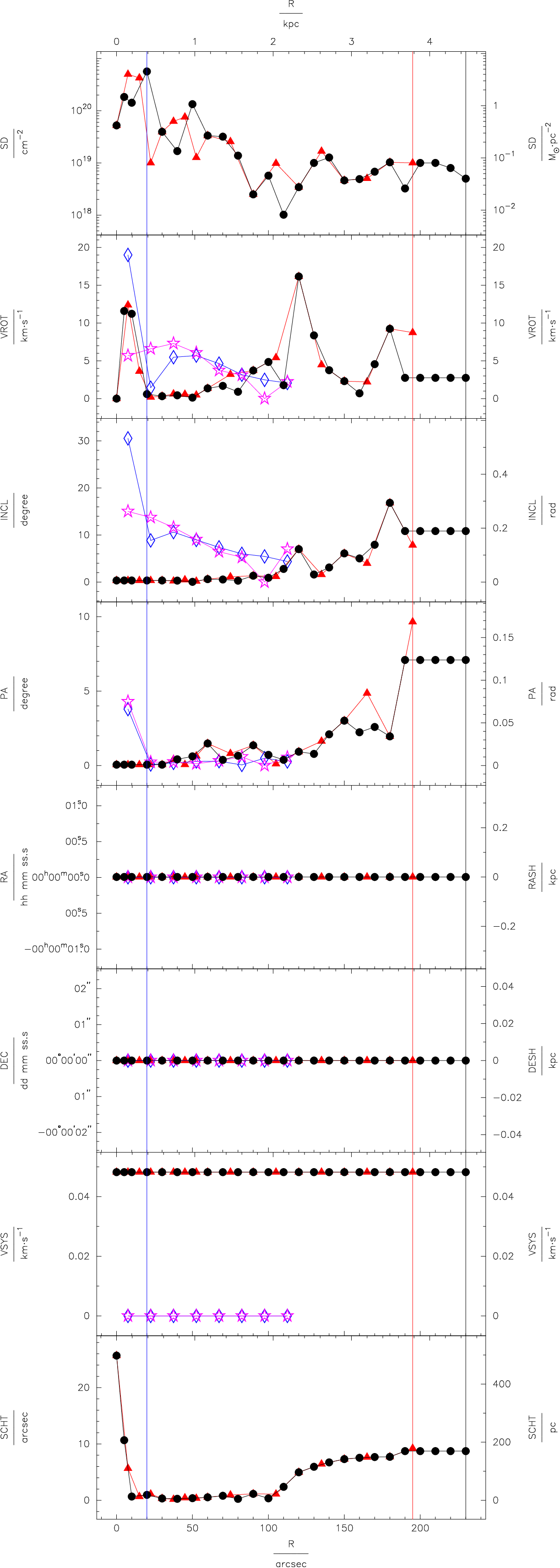}
    \end{center}
}
}
\input{fig_cap_18}
\end{figure*}\begin{center}
\begin{table*}

\caption{Test 9: slight symmetric warp without central/sytemic velocity shift and variable scale height, noise added. Deviation in global dispersion: 0.09 ${\rm km}\,{\rm s}^{-1}$. ROTCUR: input model with optimal guesses. Only the rotation velocity  and the orientation parameters were fitted.}

\begin{center}
\begin{tabular}{rrrrrrrrrrrr}
\hline
\hline
Quantity & Range & method & $ N_{\ion{H}{i}} $ & $ \frac{N_{\ion{H}{i}}}{N_{\ion{H}{i}},exp}$ & $v_{\rm rot}$ & $ i $ & $ pa $ & $ {\rm RA} $ & $ {\rm DEC} $ & $ v_{\rm sys} $ & $ z_0 $ \\
 (1)     & (2)   & (3) & (4)                & (5)                                          & (6)           & (7)   & (8)    & (9)          & (10)          & (11)            & (12)\\ 
\hline\\
$ \overline{\Delta} $ & all & tir &  72.06 &0.423 &4.2 &2.8 &1.5 &0.1 &0.0 &0.0 &4.5 \\
$\sqrt{\overline{\Delta^2}}$ & all & tir &  157.30 &0.625 &6.1 &5.0 &2.8 &0.1 &0.0 &0.0 &7.5 \\
$\Delta_{\rm max}$ & all & tir &  498.73 &1.674 &16.2 &16.8 &9.7 &0.1 &0.0 &0.0 &25.7 \\
\hline
$ \overline{\Delta} $ & 70\% & tir &  12.01 &0.157 &1.5 &0.6 &0.4 &0.1 &0.0 &0.0 &1.5 \\
$\sqrt{\overline{\Delta^2}}$ & 70\% & tir &  14.93 &0.232 &2.0 &0.7 &0.6 &0.1 &0.0 &0.0 &2.3 \\
$\Delta_{\rm max}$ & 70\% & tir &  33.38 &0.491 &3.7 &1.3 &1.4 &0.1 &0.0 &0.0 &5.7 \\
\hline
$ \overline{\Delta} $ & hi & tir &  32.72 &0.052 &1.3 &0.6 &0.6 &0.1 &0.0 &0.0 &0.6 \\
$\sqrt{\overline{\Delta^2}}$ & hi & tir &  40.64 &0.075 &1.8 &0.7 &0.8 &0.1 &0.0 &0.0 &0.7 \\
$\Delta_{\rm max}$ & hi & tir &  75.55 &0.183 &3.7 &1.3 &1.5 &0.1 &0.0 &0.0 &1.1 \\
\hline
$ \overline{\Delta} $ & $8-\sigma$ & tir &  9.78 &0.637 &6.2 &5.2 &2.7 &0.1 &0.0 &0.0 &5.2 \\
$\sqrt{\overline{\Delta^2}}$ & $8-\sigma$ & tir &  12.02 &0.797 &7.5 &7.1 &3.9 &0.1 &0.0 &0.0 &6.0 \\
$\Delta_{\rm max}$ & $8-\sigma$ & tir &  25.56 &1.674 &16.2 &16.8 &9.7 &0.1 &0.0 &0.0 &9.2 \\
\hline
$ \overline{\Delta} $ & $2-\sigma$ & tir &  8.56 &0.786 &6.9 &6.4 &3.2 &0.1 &0.0 &0.0 &6.3 \\
$\sqrt{\overline{\Delta^2}}$ & $2-\sigma$ & tir &  9.57 &0.901 &8.3 &8.0 &4.4 &0.1 &0.0 &0.0 &6.8 \\
$\Delta_{\rm max}$ & $2-\sigma$ & tir &  16.74 &1.674 &16.2 &16.8 &9.7 &0.1 &0.0 &0.0 &9.2 \\
\hline
$ \overline{\Delta} $ & excs  & tir &  8.25 \\
$\sqrt{\overline{\Delta^2}}$ & excs & tir &  8.50 \\
$\Delta_{\rm max}$ & excs & tir &  10.00 \\
\hline
$ \overline{\Delta} $ & all & pk &  & & 4.3 &8.5 &0.8 &0.0 &0.0 &0.0 & \\
$\sqrt{\overline{\Delta^2}}$ & all & pk &  & & 4.9 &9.7 &1.6 &0.0 &0.0 &0.0 & \\
$\Delta_{\rm max}$ & all & pk &  & & 2.2 &15.0 &4.3 &0.0 &0.0 &0.0 & \\
\hline
$ \overline{\Delta} $ & 70\% & pk &  & & 2.9 &5.6 &0.2 &0.0 &0.0 &0.0 & \\
$\sqrt{\overline{\Delta^2}}$ & 70\% & pk &  & & 3.5 &6.3 &0.2 &0.0 &0.0 &0.0 & \\
$\Delta_{\rm max}$ & 70\% & pk &  & & 5.7 &9.1 &0.3 &0.0 &0.0 &0.0 & \\
\hline
$ \overline{\Delta} $ & hi & pk &  & & 4.5 &7.7 &0.2 &0.0 &0.0 &0.0 & \\
$\sqrt{\overline{\Delta^2}}$ & hi & pk &  & & 5.1 &8.9 &0.3 &0.0 &0.0 &0.0 & \\
$\Delta_{\rm max}$ & hi & pk &  & & 7.3 &13.7 &0.6 &0.0 &0.0 &0.0 & \\
\hline
$ \overline{\Delta} $ & all & m1 &  & & 5.5 &10.3 &0.7 &0.0 &0.0 &0.0 & \\
$\sqrt{\overline{\Delta^2}}$ & all & m1 &  & & 7.6 &13.0 &1.4 &0.0 &0.0 &0.0 & \\
$\Delta_{\rm max}$ & all & m1 &  & & 19.0 &30.6 &3.8 &0.0 &0.0 &0.0 & \\
\hline
$ \overline{\Delta} $ & 70\% & m1 &  & & 2.8 &6.4 &0.2 &0.0 &0.0 &0.0 & \\
$\sqrt{\overline{\Delta^2}}$ & 70\% & m1 &  & & 3.0 &6.6 &0.2 &0.0 &0.0 &0.0 & \\
$\Delta_{\rm max}$ & 70\% & m1 &  & & 4.6 &8.9 &0.3 &0.0 &0.0 &0.0 & \\
\hline
$ \overline{\Delta} $ & hi & m1 &  & & 3.8 &7.9 &0.2 &0.0 &0.0 &0.0 & \\
$\sqrt{\overline{\Delta^2}}$ & hi & m1 &  & & 4.1 &8.1 &0.3 &0.0 &0.0 &0.0 & \\
$\Delta_{\rm max}$ & hi & m1 &  & & 5.7 &10.6 &0.5 &0.0 &0.0 &0.0 & \\
\hline
\hline
\end{tabular}
\end{center}
\end{table*}
\end{center}
\clearpage
\begin{figure*}
\parbox[c]{8.5cm}{
  \begin{center}
\includegraphics[width=8.5cm,height=22cm]{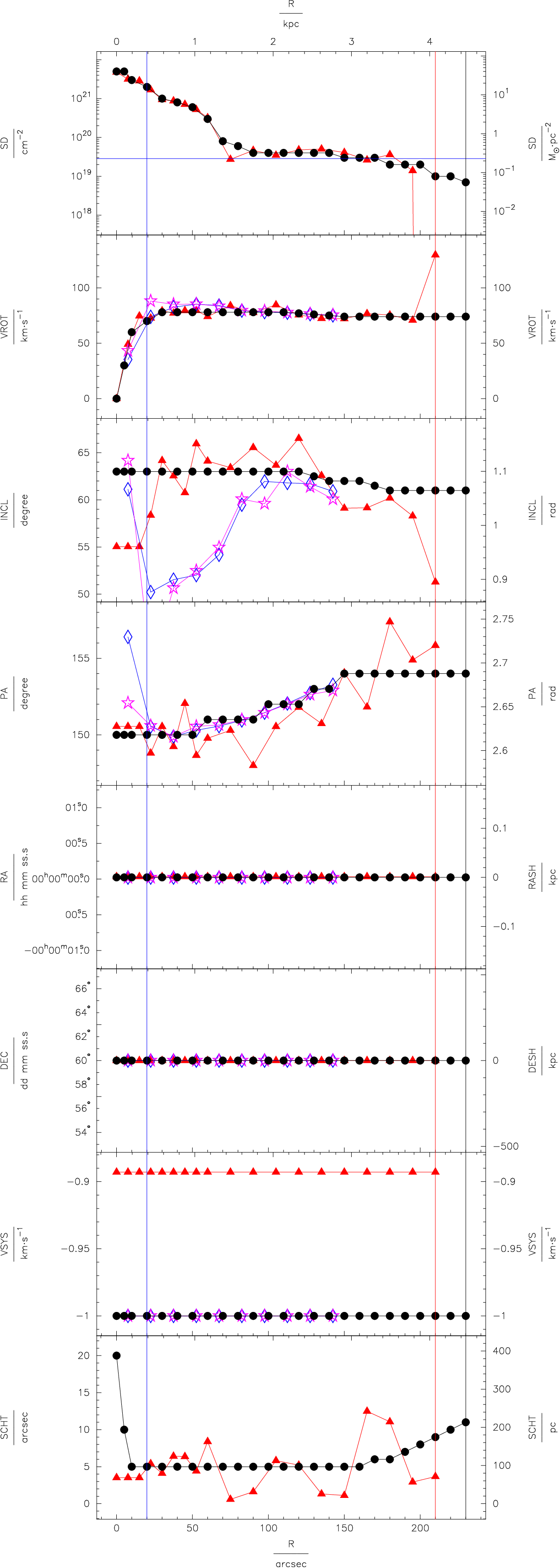}
  \end{center}
}
\parbox[c]{7.5cm} {
{
    \begin{center}
\includegraphics[width=8.5cm,height=22cm]{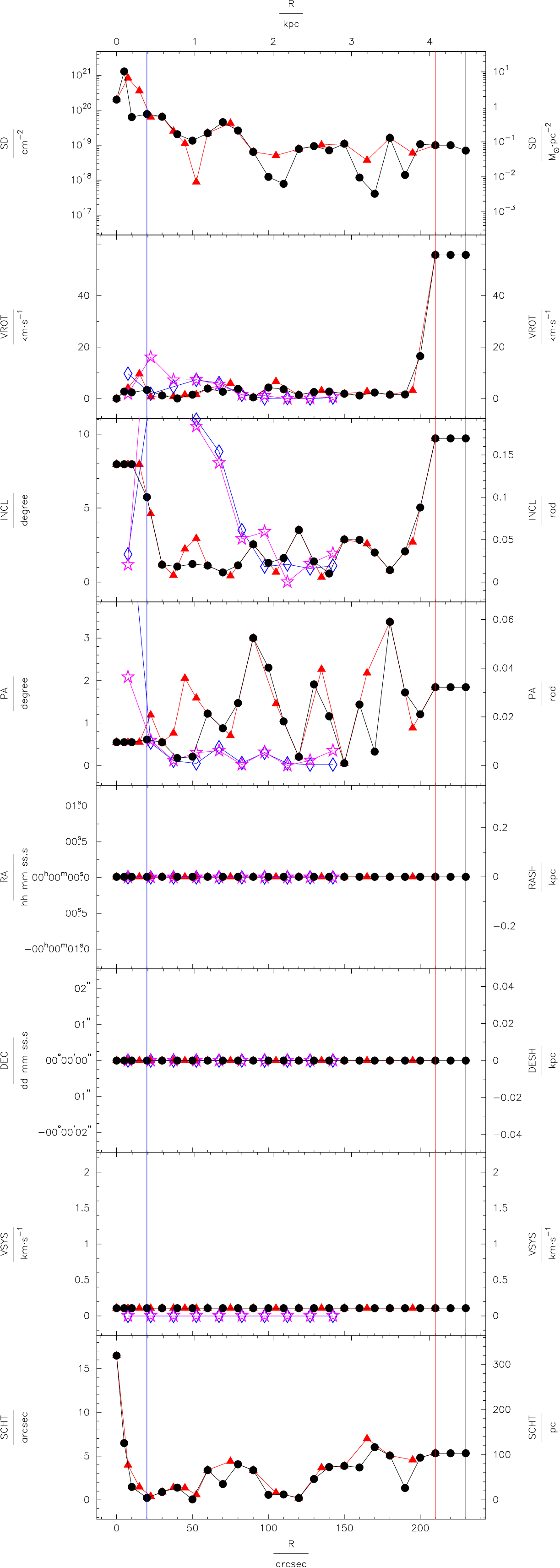}
    \end{center}
}
}
\input{fig_cap_19}
\end{figure*}\begin{center}
\begin{table*}

\caption{Test 10: slight warp without central/sytemic velocity shift and variable scale height, noise added to the data cube. Deviation in global dispersion: 0.12 ${\rm km}\,{\rm s}^{-1}$. ROTCUR: input model with optimal guesses. Only rotation velocity, position angle, and inclination were fitted.}

\begin{center}
\begin{tabular}{rrrrrrrrrrrr}
\hline
\hline
Quantity & Range & method & $ N_{\ion{H}{i}} $ & $ \frac{N_{\ion{H}{i}}}{N_{\ion{H}{i}},exp}$ & $v_{\rm rot}$ & $ i $ & $ pa $ & $ {\rm RA} $ & $ {\rm DEC} $ & $ v_{\rm sys} $ & $ z_0 $ \\
 (1)     & (2)   & (3) & (4)                & (5)                                          & (6)           & (7)   & (8)    & (9)          & (10)          & (11)            & (12)\\ 
\hline\\
$ \overline{\Delta} $ & all & tir &  89.66 &0.240 &5.6 &3.3 &1.3 &0.2 &0.0 &0.1 &3.6 \\
$\sqrt{\overline{\Delta^2}}$ & all & tir &  215.03 &0.362 &13.3 &4.4 &1.6 &0.2 &0.0 &0.1 &5.1 \\
$\Delta_{\rm max}$ & all & tir &  834.24 &1.000 &55.8 &9.7 &3.4 &0.2 &0.0 &0.1 &16.5 \\
\hline
$ \overline{\Delta} $ & 70\% & tir &  10.46 &0.094 &1.5 &1.6 &0.8 &0.2 &0.0 &0.1 &2.0 \\
$\sqrt{\overline{\Delta^2}}$ & 70\% & tir &  12.46 &0.116 &1.8 &1.9 &0.9 &0.2 &0.0 &0.1 &2.4 \\
$\Delta_{\rm max}$ & 70\% & tir &  25.22 &0.209 &3.2 &2.9 &1.6 &0.2 &0.0 &0.1 &4.0 \\
\hline
$ \overline{\Delta} $ & hi & tir &  21.19 &0.159 &2.5 &2.0 &1.3 &0.2 &0.0 &0.1 &2.4 \\
$\sqrt{\overline{\Delta^2}}$ & hi & tir &  30.10 &0.229 &3.1 &2.4 &1.6 &0.2 &0.0 &0.1 &3.1 \\
$\Delta_{\rm max}$ & hi & tir &  65.39 &0.610 &6.7 &4.6 &3.0 &0.2 &0.0 &0.1 &7.0 \\
\hline
$ \overline{\Delta} $ & $8-\sigma$ & tir &  11.92 &0.395 &8.3 &2.6 &1.6 &0.2 &0.0 &0.1 &3.8 \\
$\sqrt{\overline{\Delta^2}}$ & $8-\sigma$ & tir &  16.09 &0.491 &18.0 &3.7 &1.9 &0.2 &0.0 &0.1 &4.3 \\
$\Delta_{\rm max}$ & $8-\sigma$ & tir &  42.69 &1.000 &55.8 &9.7 &3.4 &0.2 &0.0 &0.1 &7.0 \\
\hline
$ \overline{\Delta} $ & $2-\sigma$ & tir &  9.38 &0.520 &13.0 &3.7 &1.7 &0.2 &0.0 &0.1 &5.2 \\
$\sqrt{\overline{\Delta^2}}$ & $2-\sigma$ & tir &  10.33 &0.616 &25.0 &4.8 &2.0 &0.2 &0.0 &0.1 &5.3 \\
$\Delta_{\rm max}$ & $2-\sigma$ & tir &  16.21 &1.000 &55.8 &9.7 &3.4 &0.2 &0.0 &0.1 &7.0 \\
\hline
$ \overline{\Delta} $ & excs  & tir &  8.50 \\
$\sqrt{\overline{\Delta^2}}$ & excs & tir &  8.63 \\
$\Delta_{\rm max}$ & excs & tir &  10.00 \\
\hline
$ \overline{\Delta} $ & all & pk &  & & 4.1 &6.2 &0.4 &0.0 &0.0 &0.0 & \\
$\sqrt{\overline{\Delta^2}}$ & all & pk &  & & 6.3 &8.9 &0.7 &0.0 &0.0 &0.0 & \\
$\Delta_{\rm max}$ & all & pk &  & & 16.0 &20.7 &2.1 &0.0 &0.0 &0.0 & \\
\hline
$ \overline{\Delta} $ & 70\% & pk &  & & 1.5 &2.7 &0.2 &0.0 &0.0 &0.0 & \\
$\sqrt{\overline{\Delta^2}}$ & 70\% & pk &  & & 2.3 &3.6 &0.2 &0.0 &0.0 &0.0 & \\
$\Delta_{\rm max}$ & 70\% & pk &  & & 5.5 &8.0 &0.3 &0.0 &0.0 &0.0 & \\
\hline
$ \overline{\Delta} $ & hi & pk &  & & 4.4 &6.8 &0.2 &0.0 &0.0 &0.0 & \\
$\sqrt{\overline{\Delta^2}}$ & hi & pk &  & & 6.6 &9.3 &0.3 &0.0 &0.0 &0.0 & \\
$\Delta_{\rm max}$ & hi & pk &  & & 16.0 &20.7 &0.6 &0.0 &0.0 &0.0 & \\
\hline
$ \overline{\Delta} $ & all & m1 &  & & 3.2 &5.4 &0.8 &0.0 &0.0 &0.0 & \\
$\sqrt{\overline{\Delta^2}}$ & all & m1 &  & & 4.6 &7.2 &2.0 &0.0 &0.0 &0.0 & \\
$\Delta_{\rm max}$ & all & m1 &  & & 6.1 &12.8 &6.4 &0.0 &0.0 &0.0 & \\
\hline
$ \overline{\Delta} $ & 70\% & m1 &  & & 1.2 &2.6 &0.1 &0.0 &0.0 &0.0 & \\
$\sqrt{\overline{\Delta^2}}$ & 70\% & m1 &  & & 2.0 &3.7 &0.1 &0.0 &0.0 &0.0 & \\
$\Delta_{\rm max}$ & 70\% & m1 &  & & 4.6 &8.8 &0.3 &0.0 &0.0 &0.0 & \\
\hline
$ \overline{\Delta} $ & hi & m1 &  & & 2.4 &5.8 &0.2 &0.0 &0.0 &0.0 & \\
$\sqrt{\overline{\Delta^2}}$ & hi & m1 &  & & 3.6 &7.5 &0.3 &0.0 &0.0 &0.0 & \\
$\Delta_{\rm max}$ & hi & m1 &  & & 7.3 &12.8 &0.5 &0.0 &0.0 &0.0 & \\
\hline
\hline
\end{tabular}
\end{center}
\end{table*}
\end{center}
\clearpage
\begin{figure*}
\parbox[c]{8.5cm}{
  \begin{center}
\includegraphics[width=8.5cm,height=22cm]{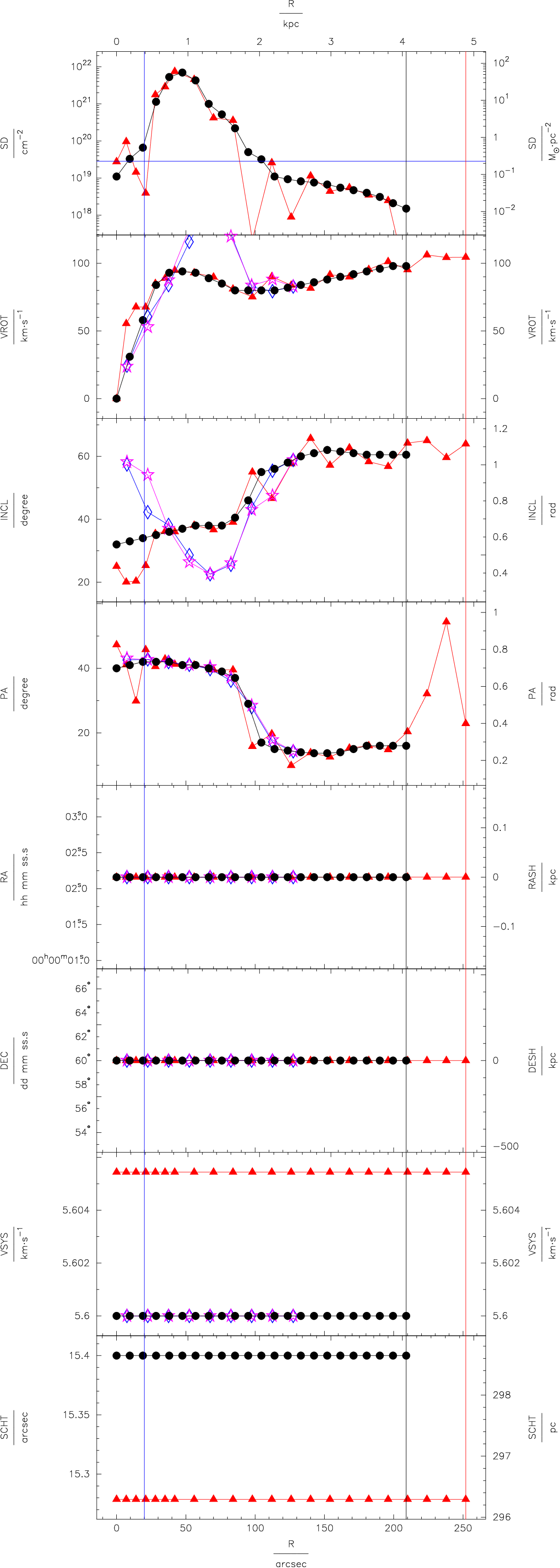}
  \end{center}
}
\parbox[c]{7.5cm} {
{
    \begin{center}
\includegraphics[width=8.5cm,height=22cm]{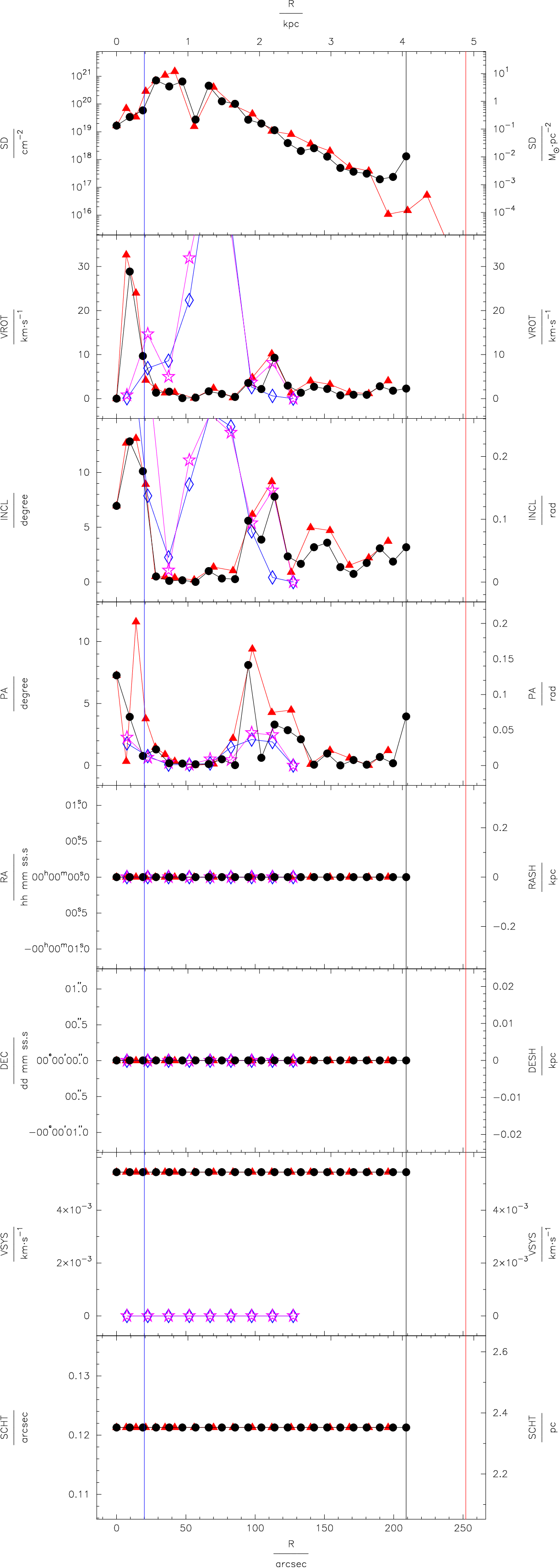}
    \end{center}
}
}
\input{fig_cap_20}
\end{figure*}\begin{center}
\begin{table*}

\caption{Test 11: symmetric warp without central/sytemic velocity shift, constant scale height. The data cube contains no noise. Deviation in global dispersion: 0.01 ${\rm km}\,{\rm s}^{-1}$. ROTCUR: input model with optimal guesses. Only the rotation velocity and the orientation parameters were fitted.}

\begin{center}
\begin{tabular}{rrrrrrrrrrrr}
\hline
\hline
Quantity & Range & method & $ N_{\ion{H}{i}} $ & $ \frac{N_{\ion{H}{i}}}{N_{\ion{H}{i}},exp}$ & $v_{\rm rot}$ & $ i $ & $ pa $ & $ {\rm RA} $ & $ {\rm DEC} $ & $ v_{\rm sys} $ & $ z_0 $ \\
 (1)     & (2)   & (3) & (4)                & (5)                                          & (6)           & (7)   & (8)    & (9)          & (10)          & (11)            & (12)\\ 
\hline\\
$ \overline{\Delta} $ & all & tir &  193.54 &0.629 &5.5 &4.4 &2.7 &0.0 &0.0 &0.0 &0.1 \\
$\sqrt{\overline{\Delta^2}}$ & all & tir &  434.19 &0.874 &10.1 &6.0 &4.3 &0.0 &0.0 &0.0 &0.1 \\
$\Delta_{\rm max}$ & all & tir &  1489.84 &2.540 &32.6 &13.1 &11.6 &0.0 &0.0 &0.0 &0.1 \\
\hline
$ \overline{\Delta} $ & 70\% & tir &  9.06 &0.307 &1.6 &1.8 &0.7 &0.0 &0.0 &0.0 &0.1 \\
$\sqrt{\overline{\Delta^2}}$ & 70\% & tir &  15.98 &0.378 &2.0 &2.5 &1.0 &0.0 &0.0 &0.0 &0.1 \\
$\Delta_{\rm max}$ & 70\% & tir &  44.10 &0.685 &4.0 &5.0 &2.2 &0.0 &0.0 &0.0 &0.1 \\
\hline
$ \overline{\Delta} $ & hi & tir &  514.06 &0.497 &2.1 &2.4 &2.3 &0.0 &0.0 &0.0 &0.1 \\
$\sqrt{\overline{\Delta^2}}$ & hi & tir &  719.46 &0.598 &2.6 &3.9 &3.7 &0.0 &0.0 &0.0 &0.1 \\
$\Delta_{\rm max}$ & hi & tir &  1489.84 &0.995 &4.8 &8.9 &9.4 &0.0 &0.0 &0.0 &0.1 \\
\hline
$ \overline{\Delta} $ & $8-\sigma$ & tir &  5.79 &0.448 &3.8 &4.2 &2.7 &0.0 &0.0 &0.0 &0.1 \\
$\sqrt{\overline{\Delta^2}}$ & $8-\sigma$ & tir &  13.35 &0.572 &4.7 &4.9 &4.0 &0.0 &0.0 &0.0 &0.1 \\
$\Delta_{\rm max}$ & $8-\sigma$ & tir &  44.10 &0.995 &10.2 &9.2 &9.4 &0.0 &0.0 &0.0 &0.1 \\
\hline
$ \overline{\Delta} $ & $2-\sigma$ & tir &  2.31 &0.370 &3.6 &3.9 &1.7 &0.0 &0.0 &0.0 &0.1 \\
$\sqrt{\overline{\Delta^2}}$ & $2-\sigma$ & tir &  4.22 &0.482 &4.6 &4.7 &2.4 &0.0 &0.0 &0.0 &0.1 \\
$\Delta_{\rm max}$ & $2-\sigma$ & tir &  10.57 &0.902 &10.2 &9.2 &4.5 &0.0 &0.0 &0.0 &0.1 \\
\hline
$ \overline{\Delta} $ & $0.5-\sigma$ & tir &  1.48 &0.317 &2.5 &3.0 &1.3 &0.0 &0.0 &0.0 &0.1 \\
$\sqrt{\overline{\Delta^2}}$ & $0.5-\sigma$ & tir &  2.90 &0.439 &2.8 &3.4 &2.0 &0.0 &0.0 &0.0 &0.1 \\
$\Delta_{\rm max}$ & $0.5-\sigma$ & tir &  8.12 &0.902 &4.0 &5.0 &4.5 &0.0 &0.0 &0.0 &0.1 \\
\hline
$ \overline{\Delta} $ & excs  & tir &  0.02 \\
$\sqrt{\overline{\Delta^2}}$ & excs & tir &  0.03 \\
$\Delta_{\rm max}$ & excs & tir &  0.05 \\
\hline
$ \overline{\Delta} $ & all & pk &  & & 17.3 &11.1 &1.0 &0.0 &0.0 &0.0 & \\
$\sqrt{\overline{\Delta^2}}$ & all & pk &  & & 25.2 &13.7 &1.5 &0.0 &0.0 &0.0 & \\
$\Delta_{\rm max}$ & all & pk &  & & 53.8 &25.4 &2.6 &0.0 &0.0 &0.0 & \\
\hline
$ \overline{\Delta} $ & 70\% & pk &  & & 5.4 &6.6 &0.3 &0.0 &0.0 &0.0 & \\
$\sqrt{\overline{\Delta^2}}$ & 70\% & pk &  & & 7.3 &8.3 &0.4 &0.0 &0.0 &0.0 & \\
$\Delta_{\rm max}$ & 70\% & pk &  & & 14.6 &13.6 &0.6 &0.0 &0.0 &0.0 & \\
\hline
$ \overline{\Delta} $ & hi & pk &  & & 24.5 &11.0 &0.7 &0.0 &0.0 &0.0 & \\
$\sqrt{\overline{\Delta^2}}$ & hi & pk &  & & 30.6 &12.7 &1.1 &0.0 &0.0 &0.0 & \\
$\Delta_{\rm max}$ & hi & pk &  & & 53.8 &19.7 &2.6 &0.0 &0.0 &0.0 & \\
\hline
$ \overline{\Delta} $ & all & m1 &  & & 15.0 &8.7 &0.9 &0.0 &0.0 &0.0 & \\
$\sqrt{\overline{\Delta^2}}$ & all & m1 &  & & 23.8 &11.6 &1.2 &0.0 &0.0 &0.0 & \\
$\Delta_{\rm max}$ & all & m1 &  & & 22.4 &24.6 &2.1 &0.0 &0.0 &0.0 & \\
\hline
$ \overline{\Delta} $ & 70\% & m1 &  & & 3.1 &4.0 &0.4 &0.0 &0.0 &0.0 & \\
$\sqrt{\overline{\Delta^2}}$ & 70\% & m1 &  & & 4.6 &5.3 &0.7 &0.0 &0.0 &0.0 & \\
$\Delta_{\rm max}$ & 70\% & m1 &  & & 8.6 &8.9 &1.5 &0.0 &0.0 &0.0 & \\
\hline
$ \overline{\Delta} $ & hi & m1 &  & & 22.4 &8.9 &0.8 &0.0 &0.0 &0.0 & \\
$\sqrt{\overline{\Delta^2}}$ & hi & m1 &  & & 29.2 &10.1 &1.1 &0.0 &0.0 &0.0 & \\
$\Delta_{\rm max}$ & hi & m1 &  & & 53.5 &15.5 &2.1 &0.0 &0.0 &0.0 & \\
\hline
\hline
\end{tabular}
\end{center}
\end{table*}
\end{center}
\clearpage
\begin{figure*}
\parbox[c]{8.5cm}{
  \begin{center}
\includegraphics[width=8.5cm,height=22cm]{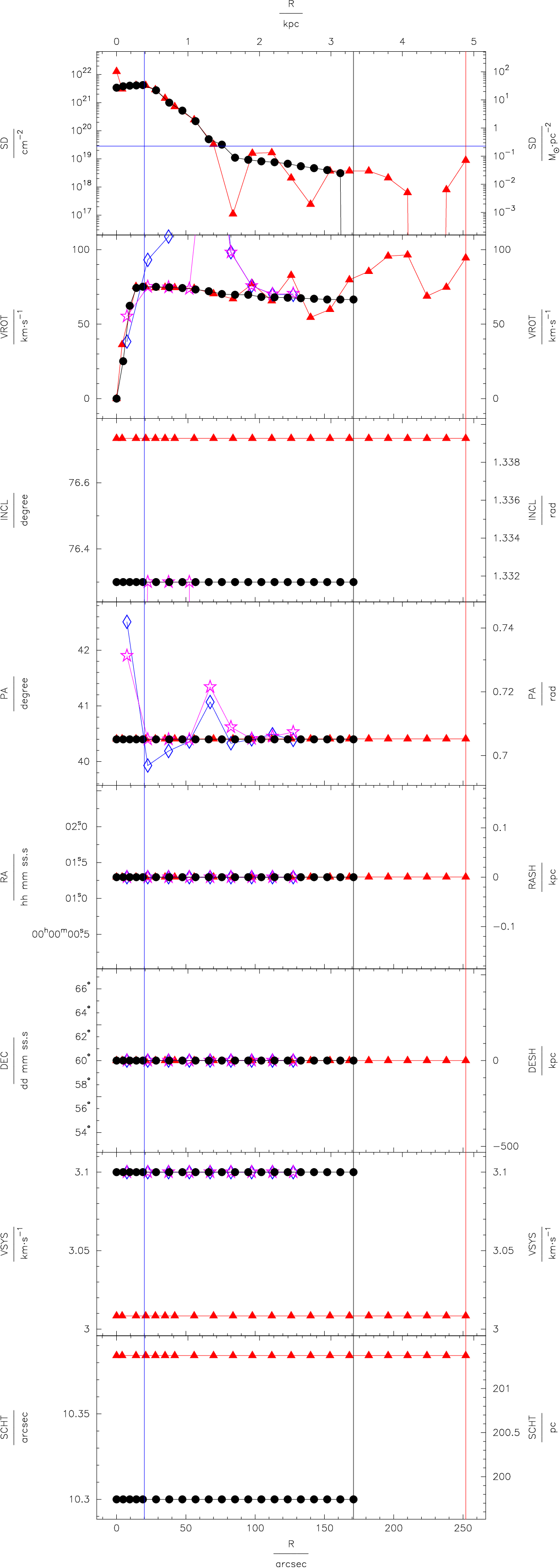}
  \end{center}
}
\parbox[c]{7.5cm} {
{
    \begin{center}
\includegraphics[width=8.5cm,height=22cm]{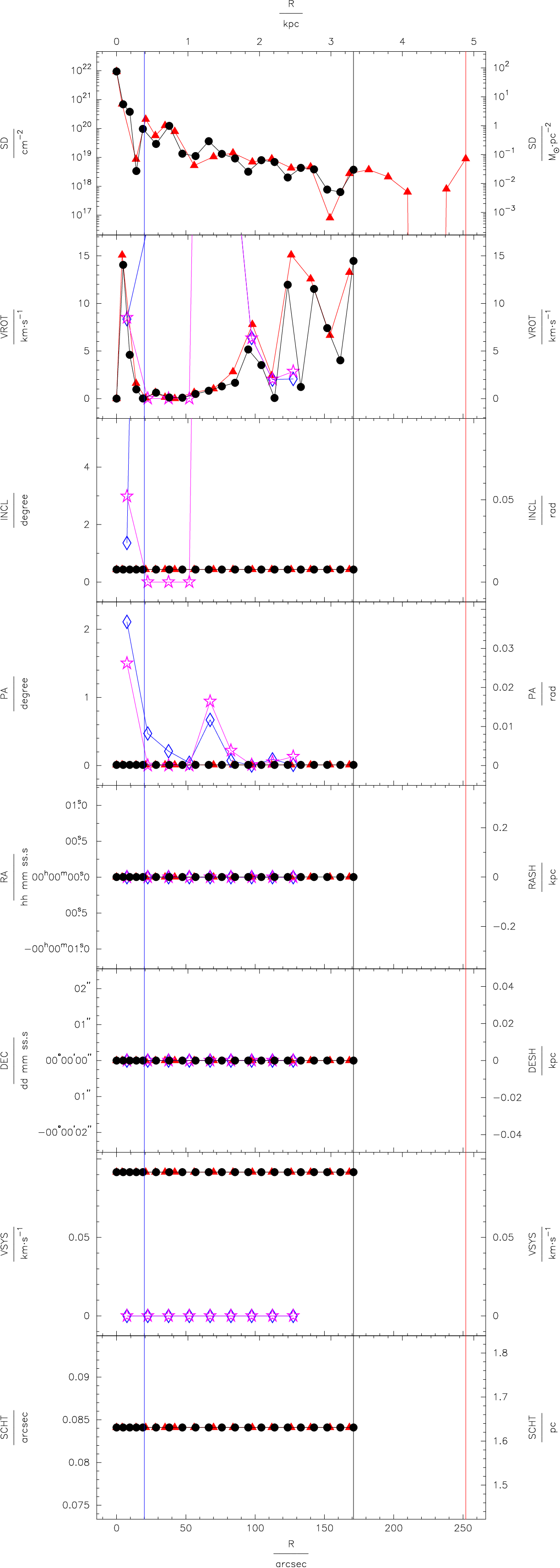}
    \end{center}
}
}
\input{fig_cap_21}
\end{figure*}\begin{center}
\begin{table*}

\caption{Test 12: flat disk with constant scale height, noise added to the data cube. Deviation in global dispersion: 0.08 ${\rm km}\,{\rm s}^{-1}$. ROTCUR: input model with optimal guesses.  Only the rotation velocity  was fitted.}

\begin{center}
\begin{tabular}{rrrrrrrrrrrr}
\hline
\hline
Quantity & Range & method & $ N_{\ion{H}{i}} $ & $ \frac{N_{\ion{H}{i}}}{N_{\ion{H}{i}},exp}$ & $v_{\rm rot}$ & $ i $ & $ pa $ & $ {\rm RA} $ & $ {\rm DEC} $ & $ v_{\rm sys} $ & $ z_0 $ \\
 (1)     & (2)   & (3) & (4)                & (5)                                          & (6)           & (7)   & (8)    & (9)          & (10)          & (11)            & (12)\\ 
\hline\\
$ \overline{\Delta} $ & all & tir &  486.53 &0.678 &5.0 &0.4 &0.0 &0.0 &0.0 &0.1 &0.1 \\
$\sqrt{\overline{\Delta^2}}$ & all & tir &  2020.45 &1.118 &7.6 &0.4 &0.0 &0.0 &0.0 &0.1 &0.1 \\
$\Delta_{\rm max}$ & all & tir &  9446.95 &2.810 &15.1 &0.4 &0.0 &0.0 &0.0 &0.1 &0.1 \\
\hline
$ \overline{\Delta} $ & 70\% & tir &  4.54 &0.198 &1.5 &0.4 &0.0 &0.0 &0.0 &0.1 &0.1 \\
$\sqrt{\overline{\Delta^2}}$ & 70\% & tir &  5.69 &0.325 &2.4 &0.4 &0.0 &0.0 &0.0 &0.1 &0.1 \\
$\Delta_{\rm max}$ & 70\% & tir &  10.50 &0.773 &6.6 &0.4 &0.0 &0.0 &0.0 &0.1 &0.1 \\
\hline
$ \overline{\Delta} $ & hi & tir &  81.15 &0.086 &0.4 &0.4 &0.0 &0.0 &0.0 &0.1 &0.1 \\
$\sqrt{\overline{\Delta^2}}$ & hi & tir &  107.39 &0.115 &0.6 &0.4 &0.0 &0.0 &0.0 &0.1 &0.1 \\
$\Delta_{\rm max}$ & hi & tir &  208.61 &0.242 &1.0 &0.4 &0.0 &0.0 &0.0 &0.1 &0.1 \\
\hline
$ \overline{\Delta} $ & $8-\sigma$ & tir &  4.89 &0.951 &7.7 &0.4 &0.0 &0.0 &0.0 &0.1 &0.1 \\
$\sqrt{\overline{\Delta^2}}$ & $8-\sigma$ & tir &  6.46 &1.235 &9.2 &0.4 &0.0 &0.0 &0.0 &0.1 &0.1 \\
$\Delta_{\rm max}$ & $8-\sigma$ & tir &  14.21 &2.810 &15.1 &0.4 &0.0 &0.0 &0.0 &0.1 &0.1 \\
\hline
$ \overline{\Delta} $ & $2-\sigma$ & tir &  4.46 &1.052 &8.7 &0.4 &0.0 &0.0 &0.0 &0.1 &0.1 \\
$\sqrt{\overline{\Delta^2}}$ & $2-\sigma$ & tir &  6.04 &1.317 &9.9 &0.4 &0.0 &0.0 &0.0 &0.1 &0.1 \\
$\Delta_{\rm max}$ & $2-\sigma$ & tir &  14.21 &2.810 &15.1 &0.4 &0.0 &0.0 &0.0 &0.1 &0.1 \\
\hline
$ \overline{\Delta} $ & $0.5-\sigma$ & tir &  3.65 &1.062 &9.6 &0.4 &0.0 &0.0 &0.0 &0.1 &0.1 \\
$\sqrt{\overline{\Delta^2}}$ & $0.5-\sigma$ & tir &  4.77 &1.364 &10.6 &0.4 &0.0 &0.0 &0.0 &0.1 &0.1 \\
$\Delta_{\rm max}$ & $0.5-\sigma$ & tir &  8.92 &2.810 &15.1 &0.4 &0.0 &0.0 &0.0 &0.1 &0.1 \\
\hline
$ \overline{\Delta} $ & excs  & tir &  2.70 \\
$\sqrt{\overline{\Delta^2}}$ & excs & tir &  4.06 \\
$\Delta_{\rm max}$ & excs & tir &  8.92 \\
\hline
$ \overline{\Delta} $ & all & pk &  & & 19.9 &15.2 &0.3 &0.0 &0.0 &0.0 & \\
$\sqrt{\overline{\Delta^2}}$ & all & pk &  & & 44.8 &23.7 &0.6 &0.0 &0.0 &0.0 & \\
$\Delta_{\rm max}$ & all & pk &  & & 131.0 &56.5 &1.5 &0.0 &0.0 &0.0 & \\
\hline
$ \overline{\Delta} $ & 70\% & pk &  & & 1.9 &4.4 &0.0 &0.0 &0.0 &0.0 & \\
$\sqrt{\overline{\Delta^2}}$ & 70\% & pk &  & & 3.0 &6.9 &0.1 &0.0 &0.0 &0.0 & \\
$\Delta_{\rm max}$ & 70\% & pk &  & & 6.3 &12.2 &0.1 &0.0 &0.0 &0.0 & \\
\hline
$ \overline{\Delta} $ & hi & pk &  & & 32.7 &14.1 &0.2 &0.0 &0.0 &0.0 & \\
$\sqrt{\overline{\Delta^2}}$ & hi & pk &  & & 65.5 &28.3 &0.5 &0.0 &0.0 &0.0 & \\
$\Delta_{\rm max}$ & hi & pk &  & & 131.0 &56.5 &0.9 &0.0 &0.0 &0.0 & \\
\hline
$ \overline{\Delta} $ & all & m1 &  & & 42.8 &29.9 &0.4 &0.0 &0.0 &0.0 & \\
$\sqrt{\overline{\Delta^2}}$ & all & m1 &  & & 73.2 &35.5 &0.8 &0.0 &0.0 &0.0 & \\
$\Delta_{\rm max}$ & all & m1 &  & & 93.9 &61.8 &2.1 &0.0 &0.0 &0.0 & \\
\hline
$ \overline{\Delta} $ & 70\% & m1 &  & & 10.9 &19.4 &0.1 &0.0 &0.0 &0.0 & \\
$\sqrt{\overline{\Delta^2}}$ & 70\% & m1 &  & & 14.4 &23.5 &0.1 &0.0 &0.0 &0.0 & \\
$\Delta_{\rm max}$ & 70\% & m1 &  & & 28.4 &37.8 &0.2 &0.0 &0.0 &0.0 & \\
\hline
$ \overline{\Delta} $ & hi & m1 &  & & 84.6 &47.6 &0.3 &0.0 &0.0 &0.0 & \\
$\sqrt{\overline{\Delta^2}}$ & hi & m1 &  & & 108.7 &48.6 &0.4 &0.0 &0.0 &0.0 & \\
$\Delta_{\rm max}$ & hi & m1 &  & & 192.3 &61.8 &0.7 &0.0 &0.0 &0.0 & \\
\hline
\hline
\end{tabular}
\end{center}
\end{table*}
\end{center}
\clearpage
\begin{figure*}
\parbox[c]{8.5cm}{
  \begin{center}
\includegraphics[width=8.5cm,height=22cm]{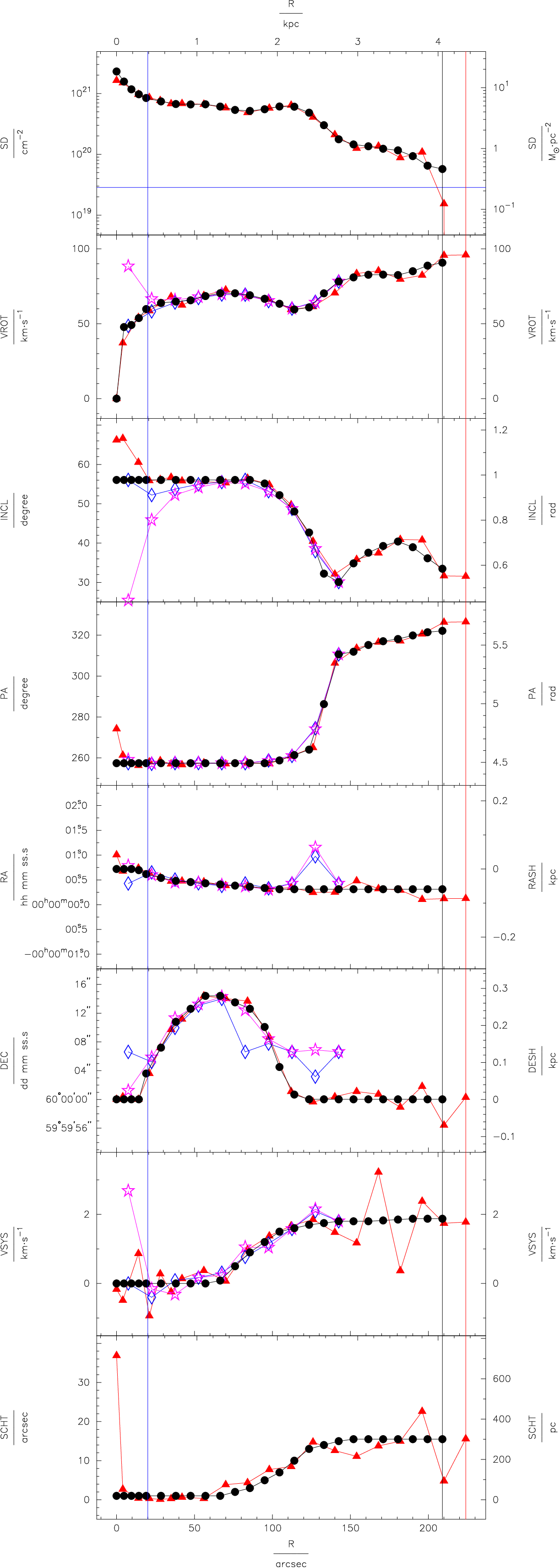}
  \end{center}
}
\parbox[c]{7.5cm} {
{
    \begin{center}
\includegraphics[width=8.5cm,height=22cm]{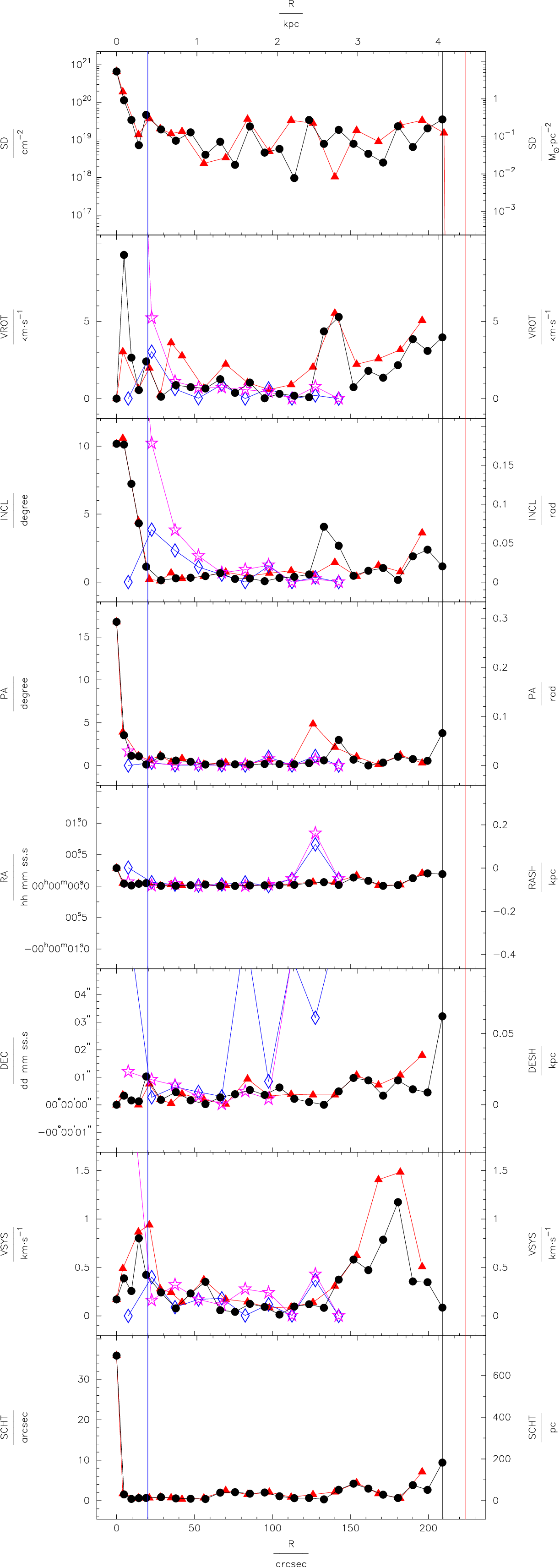}
    \end{center}
}
}
\input{fig_cap_22}
\end{figure*}\begin{center}
\begin{table*}

\caption{Test 13: asymmetric warp with variable scale height, central shift and shift in systemic velocity. Noise added to the data cube. Deviation in global dispersion: 0.08 ${\rm km}\,{\rm s}^{-1}$. ROTCUR: optimal first guess. All parameters except expansion velocity were left variable.}

\begin{center}
\begin{tabular}{rrrrrrrrrrrr}
\hline
\hline
Quantity & Range & method & $ N_{\ion{H}{i}} $ & $ \frac{N_{\ion{H}{i}}}{N_{\ion{H}{i}},exp}$ & $v_{\rm rot}$ & $ i $ & $ pa $ & $ {\rm RA} $ & $ {\rm DEC} $ & $ v_{\rm sys} $ & $ z_0 $ \\
 (1)     & (2)   & (3) & (4)                & (5)                                          & (6)           & (7)   & (8)    & (9)          & (10)          & (11)            & (12)\\ 
\hline\\
$ \overline{\Delta} $ & all & tir &  58.49 &0.089 &2.1 &2.1 &2.0 &0.9 &0.5 &0.5 &3.7 \\
$\sqrt{\overline{\Delta^2}}$ & all & tir &  156.54 &0.145 &2.6 &3.8 &4.3 &1.5 &0.7 &0.6 &8.8 \\
$\Delta_{\rm max}$ & all & tir &  667.63 &0.442 &5.5 &10.6 &16.8 &4.3 &1.8 &1.5 &35.9 \\
\hline
$ \overline{\Delta} $ & 70\% & tir &  12.31 &0.029 &1.3 &0.5 &0.5 &0.3 &0.2 &0.2 &1.0 \\
$\sqrt{\overline{\Delta^2}}$ & 70\% & tir &  15.05 &0.036 &1.5 &0.6 &0.6 &0.4 &0.3 &0.2 &1.1 \\
$\Delta_{\rm max}$ & 70\% & tir &  27.88 &0.069 &2.6 &0.8 &1.1 &0.6 &0.4 &0.5 &1.7 \\
\hline
$ \overline{\Delta} $ & hi & tir &  18.82 &0.079 &2.3 &0.8 &0.9 &0.7 &0.6 &0.5 &1.8 \\
$\sqrt{\overline{\Delta^2}}$ & hi & tir &  22.50 &0.137 &2.8 &1.2 &1.5 &1.1 &0.7 &0.6 &2.5 \\
$\Delta_{\rm max}$ & hi & tir &  36.76 &0.442 &5.5 &3.6 &4.9 &3.1 &1.8 &1.5 &7.1 \\
\hline
$ \overline{\Delta} $ & $8-\sigma$ & tir &  16.76 &0.215 &3.3 &1.5 &0.7 &1.5 &1.2 &1.0 &3.4 \\
$\sqrt{\overline{\Delta^2}}$ & $8-\sigma$ & tir &  19.86 &0.257 &3.4 &2.0 &0.8 &2.0 &1.2 &1.1 &4.3 \\
$\Delta_{\rm max}$ & $8-\sigma$ & tir &  33.35 &0.442 &5.1 &3.6 &1.2 &3.1 &1.8 &1.5 &7.1 \\
\hline
$ \overline{\Delta} $ & excs  & tir &  7.70 \\
$\sqrt{\overline{\Delta^2}}$ & excs & tir &  10.89 \\
$\Delta_{\rm max}$ & excs & tir &  15.40 \\
\hline
$ \overline{\Delta} $ & all & pk &  & & 4.9 &5.0 &0.4 &1.8 &2.3 &0.4 & \\
$\sqrt{\overline{\Delta^2}}$ & all & pk &  & & 12.7 &10.3 &0.6 &4.1 &3.5 &0.9 & \\
$\Delta_{\rm max}$ & all & pk &  & & 39.7 &30.7 &1.7 &12.6 &6.9 &2.7 & \\
\hline
$ \overline{\Delta} $ & 70\% & pk &  & & 0.5 &0.7 &0.1 &0.3 &0.6 &0.1 & \\
$\sqrt{\overline{\Delta^2}}$ & 70\% & pk &  & & 0.5 &1.0 &0.1 &0.4 &0.7 &0.2 & \\
$\Delta_{\rm max}$ & 70\% & pk &  & & 0.8 &1.9 &0.3 &1.0 &1.2 &0.3 & \\
\hline
$ \overline{\Delta} $ & hi & pk &  & & 1.1 &2.1 &0.2 &1.9 &2.4 &0.2 & \\
$\sqrt{\overline{\Delta^2}}$ & hi & pk &  & & 1.8 &3.7 &0.4 &4.3 &3.7 &0.2 & \\
$\Delta_{\rm max}$ & hi & pk &  & & 5.2 &10.2 &0.8 &12.6 &6.9 &0.4 & \\
\hline
$ \overline{\Delta} $ & all & m1 &  & & 0.5 &0.9 &0.3 &2.1 &3.1 &0.1 & \\
$\sqrt{\overline{\Delta^2}}$ & all & m1 &  & & 1.0 &1.5 &0.5 &3.6 &4.1 &0.2 & \\
$\Delta_{\rm max}$ & all & m1 &  & & 3.0 &3.9 &1.1 &10.0 &6.6 &0.4 & \\
\hline
$ \overline{\Delta} $ & 70\% & m1 &  & & 0.1 &0.3 &0.0 &0.6 &1.6 &0.1 & \\
$\sqrt{\overline{\Delta^2}}$ & 70\% & m1 &  & & 0.2 &0.5 &0.0 &0.8 &2.4 &0.1 & \\
$\Delta_{\rm max}$ & 70\% & m1 &  & & 0.6 &1.1 &0.1 &1.7 &5.3 &0.2 & \\
\hline
$ \overline{\Delta} $ & hi & m1 &  & & 0.6 &1.0 &0.3 &1.8 &2.7 &0.1 & \\
$\sqrt{\overline{\Delta^2}}$ & hi & m1 &  & & 1.1 &1.6 &0.5 &3.4 &3.7 &0.2 & \\
$\Delta_{\rm max}$ & hi & m1 &  & & 3.0 &3.9 &1.1 &10.0 &6.6 &0.4 & \\
\hline
\hline
\end{tabular}
\end{center}
\end{table*}
\end{center}
\clearpage
\begin{figure*}
\parbox[c]{8.5cm}{
  \begin{center}
\includegraphics[width=8.5cm,height=22cm]{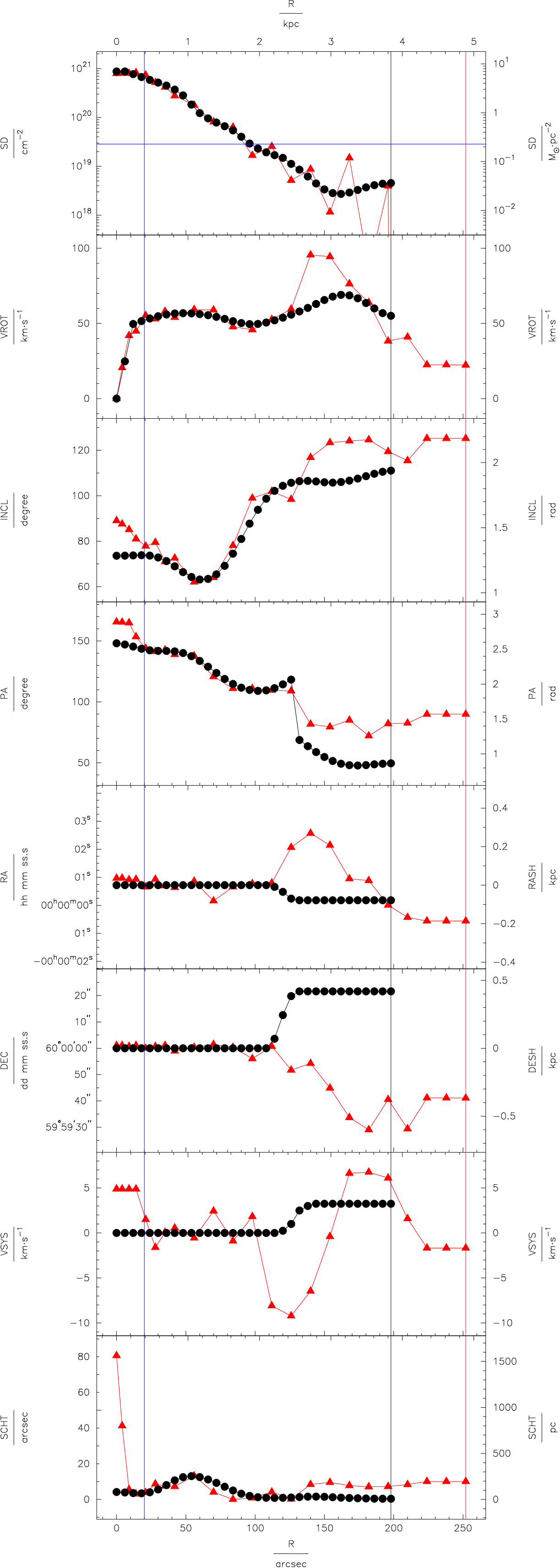}
  \end{center}
}
\parbox[c]{8.5cm} {
{
    \begin{center}
\includegraphics[width=8.5cm,height=22cm]{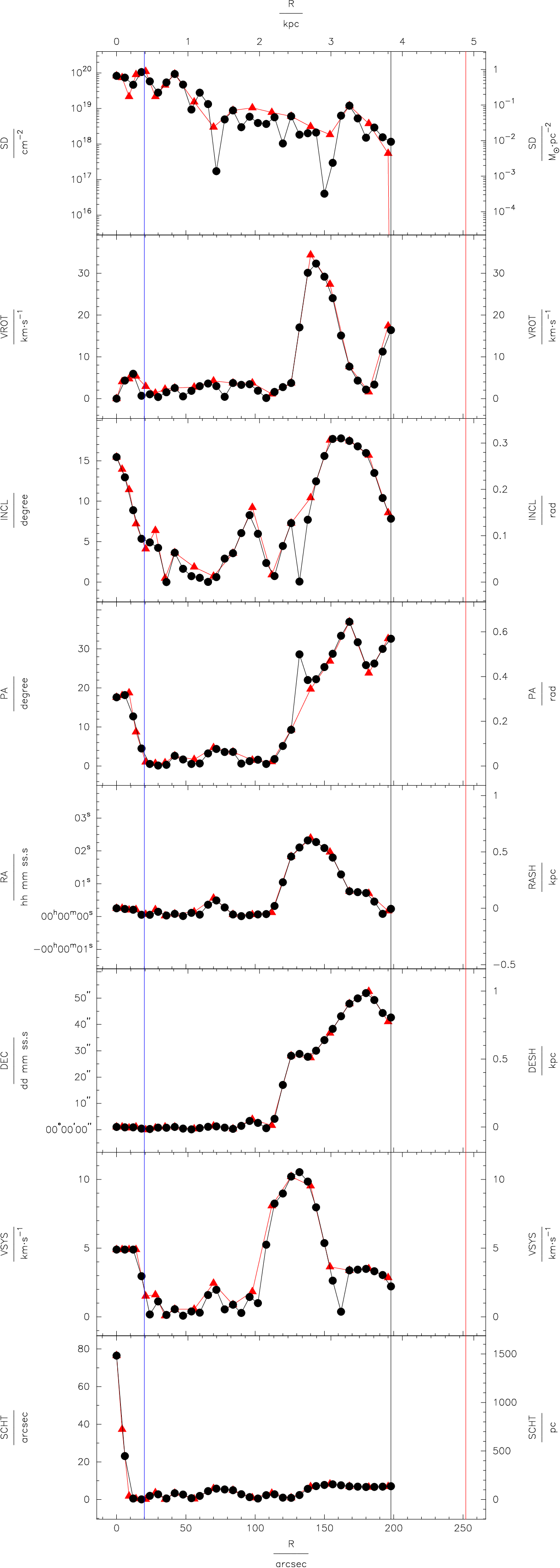}
    \end{center}
}
}
\input{fig_cap_23}
\end{figure*}\begin{center}
\begin{table*}

\caption{Test 14: asymmetric warp with variable scale height, central shift and shift in systemic velocity. Noise added to the data cube. The cloud flux used was $10^{-3}\,{\rm Jy}\,{\rm km}\,{\rm s}^{-1}$ for a radius below $48\arcsec$, $10^{-4}\,{\rm Jy}\,{\rm km}\,{\rm s}^{-1}$ for a radius between $48\arcsec$ and $114\arcsec$, and $10^{-4}\,{\rm Jy}\,{\rm km}\,{\rm s}^{-1}$ beyond $114\arcsec$, correspinding to \ion{H}{i} masses of $3.8\cdot 10^4\,{\rm M}_\odot$, $3.8\cdot 10^3\,{\rm M}_\odot$, and $3.8\cdot 10^2\,{\rm M}_\odot$, respectively, for a galaxy at a distance of $4\,{\rm Mpc}$, the cloud masses scaling with the distance squared. With this setup, an inhomogeneous distibution of the \ion{H}{i} was simulated. Since the orientation of the galactic disk is near to edge-on, no attempt was made to perform a fit to the velocity field.}

\begin{center}
\begin{tabular}{rrrrrrrrrrrr}
\hline
\hline
Quantity & Range & method & $ N_{\ion{H}{i}} $ & $ \frac{N_{\ion{H}{i}}}{N_{\ion{H}{i}},exp}$ & $v_{\rm rot}$ & $ i $ & $ pa $ & $ {\rm RA} $ & $ {\rm DEC} $ & $ v_{\rm sys} $ & $ z_0 $ \\
 (1)     & (2)   & (3) & (4)                & (5)                                          & (6)           & (7)   & (8)    & (9)          & (10)          & (11)            & (12)\\ 
\hline\\
$ \overline{\Delta} $ & all & tir &  26.60 &0.471 &6.9 &8.2 &12.1 &7.9 &13.1 &3.7 &9.2 \\
$\sqrt{\overline{\Delta^2}}$ & all & tir &  44.24 &1.013 &11.4 &10.0 &16.7 &13.2 &22.5 &4.7 &20.0 \\
$\Delta_{\rm max}$ & all & tir &  110.27 &4.126 &34.4 &17.5 &36.9 &35.9 &52.6 &10.2 &76.4 \\
\hline
$ \overline{\Delta} $ & 70\% & tir &  5.88 &0.130 &2.6 &5.0 &5.5 &2.1 &1.1 &2.1 &2.5 \\
$\sqrt{\overline{\Delta^2}}$ & 70\% & tir &  8.52 &0.161 &2.9 &6.0 &8.1 &2.4 &1.5 &2.6 &3.3 \\
$\Delta_{\rm max}$ & 70\% & tir &  21.57 &0.385 &4.2 &10.4 &18.0 &3.7 &4.0 &4.9 &6.6 \\
\hline
$ \overline{\Delta} $ & hi & tir &  42.58 &0.122 &2.8 &3.0 &2.2 &2.5 &0.7 &1.1 &2.6 \\
$\sqrt{\overline{\Delta^2}}$ & hi & tir &  58.27 &0.142 &3.0 &3.5 &2.6 &3.6 &0.9 &1.3 &3.4 \\
$\Delta_{\rm max}$ & hi & tir &  110.27 &0.252 &4.2 &6.4 &4.7 &8.4 &1.4 &2.4 &5.8 \\
\hline
$ \overline{\Delta} $ & $8-\sigma$ & tir &  4.10 &0.795 &10.5 &9.1 &16.0 &12.9 &24.1 &4.6 &5.1 \\
$\sqrt{\overline{\Delta^2}}$ & $8-\sigma$ & tir &  5.76 &1.390 &15.4 &10.9 &20.6 &18.0 &31.1 &5.6 &5.7 \\
$\Delta_{\rm max}$ & $8-\sigma$ & tir &  11.99 &4.126 &34.4 &17.5 &36.9 &35.9 &52.6 &10.2 &8.2 \\
\hline
$ \overline{\Delta} $ & $2-\sigma$ & tir &  3.79 &0.969 &12.1 &10.9 &19.0 &15.0 &29.9 &5.4 &5.1 \\
$\sqrt{\overline{\Delta^2}}$ & $2-\sigma$ & tir &  5.61 &1.554 &17.0 &12.1 &22.9 &19.9 &34.7 &6.2 &5.8 \\
$\Delta_{\rm max}$ & $2-\sigma$ & tir &  11.99 &4.126 &34.4 &17.5 &36.9 &35.9 &52.6 &10.2 &8.2 \\
\hline
$ \overline{\Delta} $ & $0.5-\sigma$ & tir &  2.36 &1.279 &17.7 &13.9 &28.0 &18.0 &41.1 &4.6 &7.1 \\
$\sqrt{\overline{\Delta^2}}$ & $0.5-\sigma$ & tir &  4.36 &1.933 &21.4 &14.4 &28.7 &21.9 &42.0 &5.2 &7.1 \\
$\Delta_{\rm max}$ & $0.5-\sigma$ & tir &  11.99 &4.126 &34.4 &17.5 &36.9 &35.9 &52.6 &9.5 &8.2 \\
\hline
$ \overline{\Delta} $ & excs  & tir &  0.00 \\
$\sqrt{\overline{\Delta^2}}$ & excs & tir &  0.00 \\
$\Delta_{\rm max}$ & excs & tir &  0.00 \\
\hline
\hline
\end{tabular}
\end{center}
\end{table*}
\end{center}
\clearpage
\begin{figure*}
\parbox[c]{8.5cm}{
  \begin{center}
\includegraphics[width=8.5cm,height=22cm]{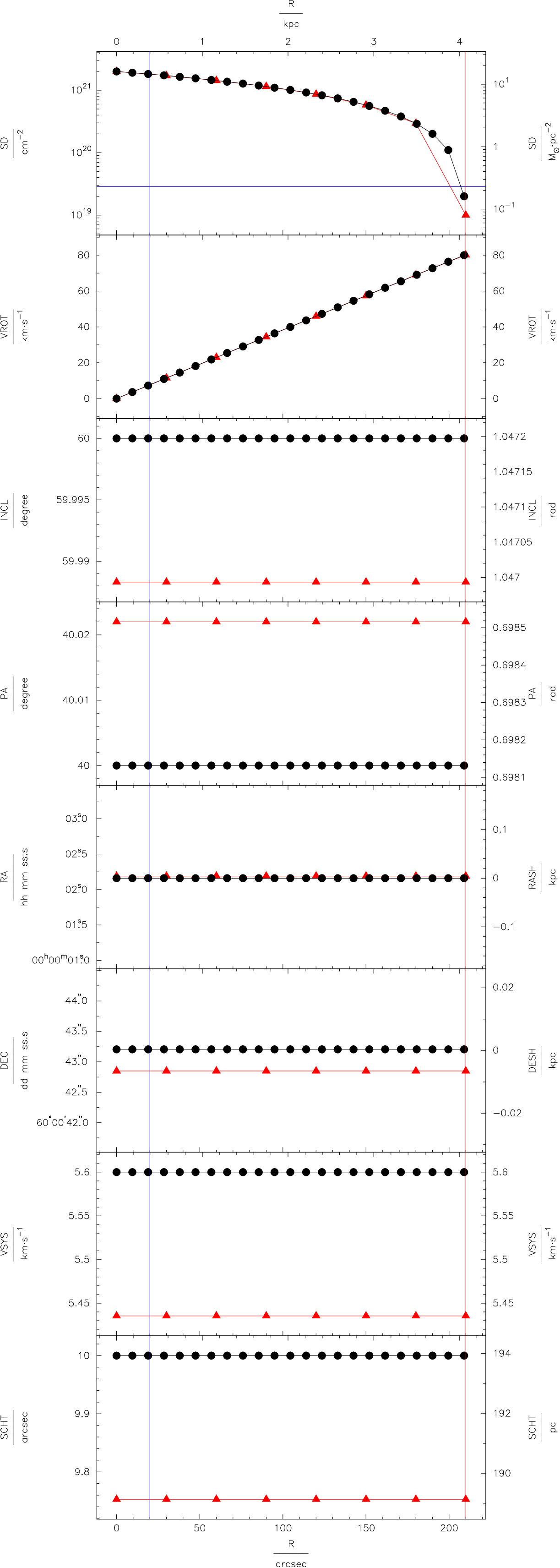}
  \end{center}
}
\parbox[c]{7.5cm} {
{
    \begin{center}
\includegraphics[width=8.5cm,height=22cm]{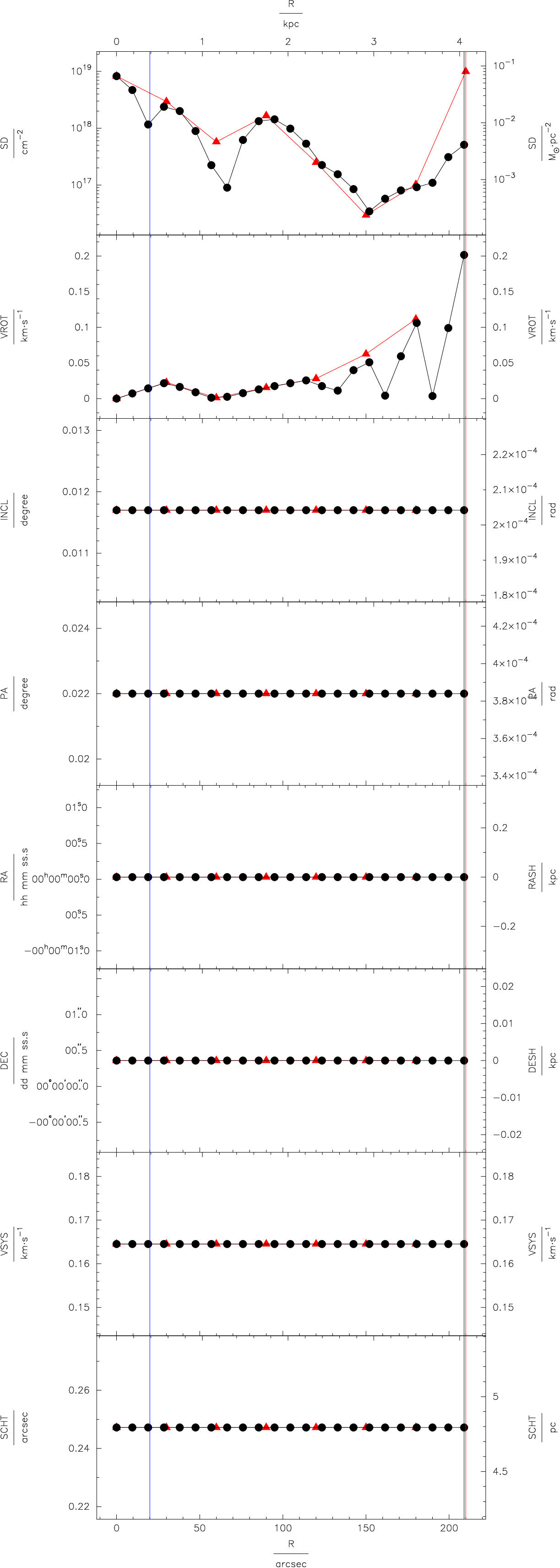}
    \end{center}
}
}
\input{fig_cap_24}
\end{figure*}\begin{center}
\begin{table*}

\caption{Test 15: flat disk with constant scale height and solid-body rotation, no noise added to the data cube. Deviation in global dispersion: 0.01 ${\rm km}\,{\rm s}^{-1}$. While the fitting was in this case not performed without the knowledge of the parametrisation of the fake observation, convergence was reached very quickly: since TiRiFiC fits the surface brightness profile, the degeneracy of inclination and rotation velocity is broken.}

\begin{center}
\begin{tabular}{rrrrrrrrrrrr}
\hline
\hline
Quantity & Range & method & $ N_{\ion{H}{i}} $ & $ \frac{N_{\ion{H}{i}}}{N_{\ion{H}{i}},exp}$ & $v_{\rm rot}$ & $ i $ & $ pa $ & $ {\rm RA} $ & $ {\rm DEC} $ & $ v_{\rm sys} $ & $ z_0 $ \\
 (1)     & (2)   & (3) & (4)                & (5)                                          & (6)           & (7)   & (8)    & (9)          & (10)          & (11)            & (12)\\ 
\hline\\
$ \overline{\Delta} $ & all & tir &  2.98 &0.001 &0.0 &0.0 &0.0 &0.5 &0.4 &0.2 &0.2 \\
$\sqrt{\overline{\Delta^2}}$ & all & tir &  4.74 &0.002 &0.1 &0.0 &0.0 &0.5 &0.4 &0.2 &0.2 \\
$\Delta_{\rm max}$ & all & tir &  9.99 &0.004 &0.1 &0.0 &0.0 &0.5 &0.4 &0.2 &0.2 \\
\hline
$ \overline{\Delta} $ & 70\% & tir &  0.52 &0.000 &0.0 &0.0 &0.0 &0.5 &0.4 &0.2 &0.2 \\
$\sqrt{\overline{\Delta^2}}$ & 70\% & tir &  0.80 &0.000 &0.0 &0.0 &0.0 &0.5 &0.4 &0.2 &0.2 \\
$\Delta_{\rm max}$ & 70\% & tir &  1.67 &0.000 &0.0 &0.0 &0.0 &0.5 &0.4 &0.2 &0.2 \\
\hline
$ \overline{\Delta} $ & hi & tir &  0.93 &0.001 &0.0 &0.0 &0.0 &0.5 &0.4 &0.2 &0.2 \\
$\sqrt{\overline{\Delta^2}}$ & hi & tir &  1.41 &0.001 &0.1 &0.0 &0.0 &0.5 &0.4 &0.2 &0.2 \\
$\Delta_{\rm max}$ & hi & tir &  2.95 &0.002 &0.1 &0.0 &0.0 &0.5 &0.4 &0.2 &0.2 \\
\hline
$ \overline{\Delta} $ & excs  & tir &  9.99 \\
$\sqrt{\overline{\Delta^2}}$ & excs & tir &  9.99 \\
$\Delta_{\rm max}$ & excs & tir &  9.99 \\
\hline
\hline
\end{tabular}
\end{center}
\end{table*}
\end{center}
\end{document}